\documentclass[
  twocolumn,
  prb,
  showpacs,
  amsmath,
  amssymb,
  superscriptaddress,
  floatfix
]{revtex4}

\usepackage{bm}
\usepackage{graphicx}
\usepackage{color}

\DeclareMathOperator{\Tr}{Tr}
\DeclareMathOperator{\sgn}{sgn}
\DeclareMathOperator{\tr}{tr}

\begin{document}

\title{Superconductor-insulator transitions: Phase diagram and magnetoresistance}

\author{I.S.~Burmistrov}

\affiliation{ L.D. Landau Institute for Theoretical Physics, Kosygina
  street 2, 119334 Moscow, Russia}
\affiliation{Moscow
Institute of Physics and Technology, 141700 Moscow, Russia}

\author{I.V.~Gornyi}
\affiliation{
 Institut f\"ur Nanotechnologie, Karlsruhe Institute of Technology,
 76021 Karlsruhe, Germany
}
\affiliation{
\mbox{Institut f\"ur Theorie der kondensierten Materie,
 Karlsruhe Institute of Technology, 76128 Karlsruhe, Germany}
}
\affiliation{
 A.F.~Ioffe Physico-Technical Institute,
 194021 St.~Petersburg, Russia.
}
\affiliation{ L.D. Landau Institute for Theoretical Physics, Kosygina
  street 2, 119334 Moscow, Russia}

\author{A.D.~Mirlin}
\affiliation{
 Institut f\"ur Nanotechnologie, Karlsruhe Institute of Technology,
 76021 Karlsruhe, Germany
 }
\affiliation{
 \mbox{Institut f\"ur Theorie der kondensierten Materie,
 Karlsruhe Institute of Technology, 76128 Karlsruhe, Germany}
}
\affiliation{
 Petersburg Nuclear Physics Institute,
 188300 St.~Petersburg, Russia.
}
\affiliation{ L.D. Landau Institute for Theoretical Physics, Kosygina
  street 2, 119334 Moscow, Russia}

\begin{abstract}
Influence of disorder-induced Anderson localization and of electron-electron interaction on superconductivity in two-dimensional systems is explored. We determine the superconducting transition temperature $T_c$, the temperature dependence of the resistivity,
the phase diagram, as well as  the magnetoresistance. The analysis is based on the renormalization group (RG) for a nonlinear sigma model. Derived RG equations are valid to the lowest order in disorder but for arbitrary electron-electron interaction strength in particle-hole and Cooper channels. Systems with preserved and broken spin-rotational symmetry are considered, both with short-range and with long-range (Coulomb) interaction. In the cases of short-range interaction, we identify parameter regions where the superconductivity is enhanced by localization effects.  Our RG analysis indicates that the superconductor-insulator transition is controlled by a fixed point with a resistivity $R_c$ of the order of the quantum resistance $R_q = h/ 4e^2$. When a transverse magnetic field is applied, we find a strong nonmonotonous magnetoresistance for temperatures  below $T_c$.
\end{abstract}


\maketitle

\section{Introduction}
\label{s1}

Superconductivity [\onlinecite{Kamerlingh-Onnes,BCS}] and Anderson localization
[\onlinecite{Anderson58}] are among most important and fundamental quantum phenomena in
condensed matter physics. These two phenomena are in a sense antagonists: in
the case of superconductivity the Cooper interaction creates a collective state
with vanishing resistivity, while the Anderson localization resulting from
disorder-induced quantum interference drives the system into a state with
zero conductivity. Therefore, when both interaction and disorder are present, a
competition between the superconductivity and localization naturally arises.
This competition is of particular interest in two-dimensional (2D) geometry,
where even a weak disorder makes the system an Anderson insulator. Thus, a 2D
system may be expected to undergo a direct quantum phase transition (QPT) between the
insulating and superconducting states, the superconductor-insulator transition
(SIT).

Experimentally SIT has been studied in a variety of 2D structures,
including amorphous Bi and Pb [\onlinecite{Haviland89,Parendo05}], MoC
[\onlinecite{Lee90}], MoGe [\onlinecite{Yazdani95}], Ta [\onlinecite{Qin06}], InO [\onlinecite{Exp-InO,Exp-InO-2}], NbN [\onlinecite{Exp-NbN}] and
TiN films [\onlinecite{Exp-TiN,BatVinBKT}], see also the reviews [\onlinecite{SITReview}]. In recent
years, there have been also a growing experimental activity on SIT in novel 2D
materials and nanostructures, such as LaAlO$_3$/SrTiO$_3$ interfaces
[\onlinecite{Caviglia2008,Ilani2014}], SrTiO$_3$ surfaces [\onlinecite{Kim2012,Iwasa2014}],
MoS$_2$ flakes [\onlinecite{Ye2012,Taniguchi2012}], FeSe thin films [\onlinecite{Exp-FeSe}], LaSrCuO surfaces
[\onlinecite{Bollinger2011}], and Li$_x$ZrNCl layered materials [\onlinecite{Exp-LiZrNCl}].
Characteristic for many of the novel structures is a strong screening of the
Coulomb interaction due to a large dielectric constant of the substrate (such
as SrTiO$_3$). In addition, strong spin-orbit coupling is present in many
of the novel materials (MoS$_2$, LaAlO$_3$/SrTiO$_3$, SrTiO$_3$).

To drive the system through SIT, one changes a parameter (film
thickness, gate voltage, doping) controlling the high-temperature sheet
resistivity. With lowering temperature, systems with lower resistivity become
superconducting (resistivity drops to zero), while those with higher
resistivity get insulating (resistivity becomes exponentially large).
The most salient observations common to the majority of the above experiments
are as follows:

\begin{enumerate}

\item[(i)] Most of the experiments are interpreted as supporting a direct
transition between the superconducting and insulating phases, although
some of them suggest a possibility of existence of an intermediate metallic
phase. The critical resistivity $R_c$ (the low-temperature limit of the
separatrix curve separating the temperature dependence of resistivity in the
insulating and superconducting phases) is of the order of the quantum resistance
$R_q = h/4e^2 \simeq 6.5\: {\rm k}\Omega$. However, the precise value of $R_c$
varies from one experiment to another, roughly in the range between $R_q/2$ and
$3 R_q$.

\item[(ii)] For those systems that are superconducting (at low temperature $T$ and magnetic
field $H$), a non-monotonous dependence of resistivity on $T$ and
$H$ is observed. In particular, a giant non-monotonous magnetoresistance
is found in such systems at very low temperatures,  $T \ll T_c$.

\item[(iii)] The temperature dependence of resistivity on the insulating side is very
fast (activation or even stronger).

\end{enumerate}

Theoretical investigation of the interplay of interaction and disorder in
systems with Cooper attraction has a long history. Soon after the development
of the microscopic theory of
superconductivity by Bardeen, Cooper, and Schrieffer (BCS) [\onlinecite{BCS}],
the question of influence of disorder on superconductivity
attracted a great deal of attention. It was
found [\onlinecite{AG59,Anderson59}] that the diffusive motion of electrons
does not affect essentially the temperature $T_c$ of superconducting
transition, i.e the mean free path does not enter the expression for
$T_c$. This statement is conventionally called ``Anderson theorem''.

Effects of disorder-induced Anderson localization
[\onlinecite{Anderson58}] on superconductivity were considered in
Refs.~[\onlinecite{BuSa,KK85}]. It was found
that, within the BCS approach, the superconductivity in a disordered
system persists up to the localization threshold and even in the
localized regime near the Anderson transition. Furthermore,
Refs.~[\onlinecite{BuSa,KK85}] came to the conclusion that the mean-field
transition temperature $T_c$ in these regimes remains unaffected by
disorder (i.e. the Anderson theorem holds). In a parallel line of research, it
was discovered [\onlinecite{Maekawa82,Anderson83,Muttalib}] that an interplay of
long-range ($1/r$) Coulomb interaction and disorder leads to
suppression of $T_c$. These ideas were put on the solid basis by
Finkelstein [\onlinecite{Fin}] who developed the nonlinear sigma model (NLSM)
renormalization-group (RG) formalism.

An alternative approach to the SIT known as ``bosonic mechanism'' was proposed
in Refs.~[\onlinecite{Fisher90}]. It takes into account the superconducting phase
fluctuations and discards completely all other degrees of freedom, in
particular, the localization effects. It was also proposed that an intermediate
``Bose metal'' phase may separate the superconductor and insulator
[\onlinecite{bose-metal}]. A relation between the bosonic and fermionic mechanisms  as
a well as a status of the Bose metal conjecture remain quite obscure.

Recently, Feigelman {\it et al.} [\onlinecite{Feigelman07,feigelman10}] found that the
eigenfunction multifractality near the localization threshold in three
dimensions strongly affects properties
of a superconductor. Their remarkable finding is that $T_c$ is
dramatically enhanced: its
dependence on the coupling constant is no longer exponential (as in the
conventional BCS solution) but rather of
a power-law type. This result was obtained on the basis of the BCS-type
self-consistency equation,
with Cooper attraction being the only interaction included.

In a preceding work by the present authors [\onlinecite{PRL2012}] the influence of
disorder-induced Anderson localization on the temperature of superconducting
transition $T_c$ was studied within the field-theoretical
framework. Electron-electron interaction in particle-hole and Cooper channels
was taken into account. The focus was put on the case of a weak short-range
interaction (which
is relevant to materials with large dielectric constant, as well to
cold atom systems). Two-dimensional systems in the weak localization and
antilocalization regime,
as well as systems near mobility edge were investigated.
A systematic analytical approach to
the problem was developed in the framework of the interacting NLSM and
its RG treatment. The approach took into account the
mutual renormalization of disorder and all interaction constants (that,
in particular, leads to mixing of different interaction channels).
This methodology allows us to explore both the cases of a long-range
(Coulomb) interaction previously studied by Finkelstein [\onlinecite{Fin}] and of a
weak short-range interaction within a unified formalism.
More specifically, in the case of short-range interactions a system of coupled RG equations for the
problem was derived in the lowest order in disorder and three interaction couplings (singlet, triplet, and
Cooper channels).

The analysis of RG equations for the weak short-range interaction showed the
behavior which is \textit{exactly opposite} to that predicted by
Ref.~[\onlinecite{Fin}] for Coulomb interaction. It was found that
the interplay of such interactions and Anderson localization leads to
\textit{strong enhancement} of superconductivity in a broad range of parameters in dirty 2D systems,
as well as in three dimensional (3D) systems near the Anderson transition (in contrast to the
\textit{suppression} in the Coulomb case).
In the latter case (vicinity of the Anderson transition), the microscopic
theory of Ref. [\onlinecite{PRL2012}] justified previous theoretical results obtained
from the self-consistency equation [\onlinecite{Feigelman07,feigelman10}].

This result of Ref. [\onlinecite{PRL2012}] is of fundamental importance and
represents an unexpected physics (enhancement of superconductivity by localization, which is
naively its exact antagonist). Indeed, remarkably, the localization
physics, responsible for the increase of resistivity and thus driving the system
towards an insulating state, favors at the same time the superconductivity.
The key condition is a suppression of the long-range component of the Coulomb
interaction (see also Ref. [\onlinecite{Kravtsov12}]). This opens a new way for searching novel materials exhibiting
high-temperature superconductivity: one needs the combination of a large
dielectric background constant and disorder in layered structures.

In this paper, we extend the formalism of Ref.~[\onlinecite{PRL2012}]
by deriving the RG equations to the lowest order in disorder but, formally, for arbitrary interaction
couplings.
We use this framework to explore systematically the interplay of
superconductivity, interaction, and localization
in 2D systems, with a focus on the SIT in thin films. More specifically:

\begin{enumerate}

\item[(i)]
We evaluate the  temperature dependence of the resistivity $\rho(T)$ for
given bare (high-temperature) couplings
down to the temperature $T_c$ at which the finite expectation value of the
superconducting order parameter emerges,
or else, down to the temperature where the system enters the
insulating regime.

\item[(ii)] We use the RG equations to determine the structure of the phase diagram.
In particular, we identify parameter regions where the superconductivity is
enhanced by localization. Our results also indicate that in some cases the
phase diagram may include a critical-metal phase.

\item[(iii)] We study the magnetoresistance near the SIT within two-step RG approach.
Since the magnetic field suppresses both superconductivity and localization, a
non-monotonous magnetoresistance arises, as observed experimentally.
Furthermore, this magnetoresistance becomes very strong at low temperatures,
again in agreement with experiments. Both orbital and Zeeman effects of
the magnetic field are incorporated in the unifying RG scheme.

\end{enumerate}
All the above analysis is performed for the cases of short-ranged and
long-ranged Coulomb interaction, both with and without spin-orbit interaction.

The structure of the article is as follows.
In Sec.~\ref{s2} we introduce the NLSM formalism.  The corresponding
RG equations (valid to the lowest order in disorder and for arbitrary interaction strength) are
presented in Sec.~\ref{sec:one-loop}. The RG equations are used in Sec.~\ref{s4} to analyze the phase diagram in zero magnetic field.
The temperature dependence of resistivity in zero magnetic field is discussed in Sec.~\ref{s5_0}.
In Sec. ~\ref{s5} this analysis is extended to calculate the magnetoresistance
in a transverse and in a parallel magnetic field.
Section~\ref{s6} contains a discussion of obtained results, their
implications, limitations, possible extensions, comparison with numerical and experimental results.
Finally, our results and conclusions are summarized in Sec.~\ref{s8}.
Several Appendices contain technical details of the derivation of RG equations and of their analysis.

\section{Formalism}
\label{s2}

\subsection{NLSM action}

The action of the  NLSM is given as a sum of the non-interacting part, $S_\sigma$,
and contributions arising from the interactions in the particle-hole singlet, $S_{\rm int}^{(\rho)}$,
particle-hole triplet, $S_{\rm int}^{(\sigma)}$, and particle-particle (Cooper), $S_{\rm int}^{(c)}$, channels (see Refs. [\onlinecite{Fin,KB}] for review):
\begin{gather}
S=S_\sigma + S_{\rm int}^{(\rho)}+S_{\rm int}^{(\sigma)}+S_{\rm int}^{(c)},
\label{eq:NLSM}
\end{gather}
where
\begin{align}
S_\sigma & = -\frac{g}{32} \int\!\! d\bm{r} \Tr (\nabla Q)^2 + 4\pi T Z_\omega \int\!\! d\bm{r} \Tr \eta  Q ,
\notag  \\
S_{\rm int}^{(\rho)}& =-\frac{\pi T}{4} \Gamma_s\! \sum_{\alpha,n} \sum_{r=0,3}
\int\!\! d\bm{r} \Tr \Bigl [I_n^\alpha t_{r0} Q\Bigr ] \Tr \Bigl [I_{-n}^\alpha t_{r0} Q\Bigr ] ,
\notag \\
S_{\rm int}^{(\sigma)}& =-\frac{\pi T}{4} \Gamma_t\! \sum_{\alpha,n} \sum_{r=0,3}
\sum_{j=1}^3
\int\!\! d\bm{r} \Tr \Bigl [I_n^\alpha \bm{t_{r}} Q\Bigr ]
 \Tr \Bigl [I_{-n}^\alpha \bm{t_{r}} Q\Bigr ] ,
\notag \\
S_{\rm int}^{(c)}& =-\frac{\pi T}{2} \Gamma_c\! \sum_{\alpha,n} \sum_{r=0,3}(-1)^r
\int\!\! d\bm{r} \Tr \Bigl [I_n^\alpha t_{r0} Q I_{n}^\alpha t_{r0} Q\Bigr ] .
\notag
\end{align}
Here $g$ is the total Drude conductivity
(in units $e^2/h$ and including spin), $\bm{t_{r}}=\{t_{r1},t_{r2},t_{r3}\}$, and we use the following matrices
\begin{gather}
\Lambda_{nm}^{\alpha\beta} = \sgn n \, \delta_{nm} \delta^{\alpha\beta}t_{00}, \notag \\
\eta_{nm}^{\alpha\beta}=n \, \delta_{nm}\delta^{\alpha\beta} t_{00}, \\
(I_k^\gamma)_{nm}^{\alpha\beta}=\delta_{n-m,k}\delta^{\alpha\beta}\delta^{\alpha\gamma} t_{00} , \notag
\end{gather}
with $\alpha,\beta = 1,\dots, N_r$ standing for replica indices and $n,m$ corresponding to the
Matsubara fermionic energies $\varepsilon_n = \pi T (2n+1)$. The sixteen matrices,
\begin{equation}
\label{trj}
t_{rj} = \tau_r\otimes s_j, \qquad r,j = 0,1,2,3  ,
\end{equation}
operate in the particle-hole (subscript $r$) and spin (subsrcipt $j$) spaces
with the corresponding Pauli matrices denoted by
\begin{gather}
\tau_1 = \begin{pmatrix}
0 & 1\\
1 & 0
\end{pmatrix}, \, \tau_2 = \begin{pmatrix}
0 & -i\\
i & 0
\end{pmatrix}, \, \tau_3 = \begin{pmatrix}
1 & 0\\
0 & -1
\end{pmatrix} , \label{eq:tau-def}\\
s_1 = \begin{pmatrix}
0 & 1\\
1 & 0
\end{pmatrix}, \, s_2 = \begin{pmatrix}
0 & -i\\
i & 0
\end{pmatrix}, \, s_3 = \begin{pmatrix}
1 & 0\\
0 & -1
\end{pmatrix} .
 \label{eq:spin-def}
\end{gather}
Matrices $\tau_0$ and $s_0$ stand for the $2\times 2$ unit matrices. The matrix field $Q(\bm{r})$ (as well as the trace $\Tr$) acts in the replica, Matsubara,
spin, and particle-hole spaces. It obeys the following constraints:
\begin{gather}
Q^2=1, \qquad \Tr Q = 0, \qquad Q^\dag = C^T Q^T C .
\end{gather}
 The charge conjugation matrix $C = i t_{12}$ satisfies the following relation $C^T = -C$.
Matrix $Q$ can be parameterized as $Q=T^{-1} \Lambda T$ where the matrices $T$ obey (symbol $*$ denotes the complex conjugation)
 \begin{equation}
 C T^* = T C,\qquad (T^{-1})^* C = C T^{-1} . \label{TC}
 \end{equation}

In order to avoid notational confusion, it is instructive to compare our notation with that
of the reviews~[\onlinecite{Fin}] and~[\onlinecite{KB}]. In both references, a  different definition of Pauli matrices in the particle-hole space has been used, namely, $i\tau_j$ instead of $\tau_j$ for $j=1,2,3$. In Ref.
~[\onlinecite{Fin}] Pauli matrices in the spin space coincide with our definition \eqref{eq:spin-def}. In Ref.~[\onlinecite{KB}] the spin-space Pauli matrices  $-i s_j$ (for $j=1,2,3$)  were used instead of our definition \eqref{eq:spin-def}. The interaction terms $S_{\rm int}^{(\rho)}$, $S_{\rm int}^{(\sigma)}$ and $S_{\rm int}^{(c)}$ coincide
with terms in Eqs. (3.9a), (3.9b), and (3.9b) of Ref.~[\onlinecite{Fin}] provided the following relations between the couplings $\Gamma_s$, $\Gamma_t$ and $\Gamma_c$ in $S_{\rm int}^{(\rho)}$, $S_{\rm int}^{(\sigma)}$ and $S_{\rm int}^{(c)}$ and $Z$, $\Gamma_2$ and $\Gamma_c$ in Ref.~[\onlinecite{Fin}] hold:
$\Gamma_s \equiv -(\pi\nu/4) Z$, $\Gamma_t \equiv (\pi\nu/4)\Gamma_2$, and $\Gamma_c \equiv (\pi\nu/4) \Gamma_c$. Here the thermodynamic density of states $\nu$ includes the spin-degeneracy factor. Note that Ref.~[\onlinecite{Fin}] focuses on the case of unscreened (long-ranged) Coulomb interaction. Hence the interaction
amplitude $\Gamma_s$ in the singlet particle-hole channel is expressed through the frequency renormalization factor $Z$ there. We consider both long-ranged (Coulomb) and short-ranged interactions. In the latter case the quantities $\Gamma_s$ and $Z_\omega$ are independent variables. The interaction terms $S_{\rm int}^{(\rho)}$, $S_{\rm int}^{(\sigma)}$ and $S_{\rm int}^{(c)}$ coincide
with the terms in Eqs. (3.92d), (3.92e), and (3.92f) of Ref.~[\onlinecite{KB}] provided
$\Gamma_s \equiv K^{(1)}$, $\Gamma_t \equiv K^{(2)}$, and  $\Gamma_c\equiv K^{(3)}/2$.
The parameters $g$ and $Z_\omega$ in $S_\sigma$ are related to the corresponding quantities $D, Z, \nu$ of Ref.~[\onlinecite{Fin}] as $g = 4\pi \nu D$ and $Z_\omega= (\pi \nu/4) Z$ and to the parameters $G$ and $H$ in Ref.~[\onlinecite{KB}] as $g=16/G$ and $Z_\omega=H/2$ .

\subsection{Interaction in the Cooper channel}

The Cooper-channel interaction term can be rewritten as
\begin{gather}
S_{\rm int}^{(c)} = -\frac{\pi T}{4}  \Gamma_c\! \sum_{\alpha,n} \sum_{r=1,2} \sum_{j=0}^{3} \int\!\! d\bm{r}\Tr \bigl [ t_{rj} L_n^\alpha Q \bigr ] \Tr \bigl [ t_{rj} L_n^\alpha Q \bigr ]  .
\end{gather}
Here the matrix $L_n^\alpha$ is defined as
\begin{equation}
(L_n^\alpha)^{\beta \gamma}_{km} = \delta_{k+m,n}\delta^{\alpha\beta}\delta^{\alpha\gamma} t_{00} .
\end{equation}
However, for $j=1,2,3$ we find
\begin{gather}
\Tr \bigl [ t_{rj} L_n^\alpha Q \bigr ] =
-\Tr \bigl [ C t^T_{rj} C L_n^\alpha Q \bigr ]
=-\Tr \bigl [ t_{rj} L_n^\alpha Q \bigr ]=0
.
\end{gather}
Therefore, the term $S_{\rm int}^{(c)} $ describing the interaction in the Cooper channel is fully determined by the Cooper-singlet channel:
\begin{gather}
S_{\rm int}^{(c)} = -\frac{\pi T}{4}  \Gamma_c \sum_{\alpha,n} \sum_{r=1,2}  \int d\bm{r} \Tr \bigl [ t_{r0} L_n^\alpha Q \bigr ] \Tr \bigl [ t_{r0} L_n^\alpha Q \bigr ] . \label{Oc}
\end{gather}

\subsection{Relation with the BCS hamiltonian \label{InitialBCS}}

In general, bare values of the interaction parameters $\Gamma_s, \Gamma_t$ and $\Gamma_c$
can be estimated for a given electron-electron interaction $U(\bm{r}-\bm{r^\prime})$
in a microscopic hamiltonian. It is convenient to introduce the dimensionless
 parameters $\gamma_{s,t,c} = \Gamma_{s,t,c}/Z_\omega$. Then their bare values  can be written as
\begin{equation}
 \gamma_{s0}=-\frac{F_s}{1+F_s}, \quad
\gamma_{t0} =-\frac{F_t}{1+F_t}, \quad \gamma_{c0} = - F_c ,
\end{equation}
where $F_s = \nu U(q) + F_t$,
\begin{align}
F_t &= -\frac{\nu}{2} \Bigl \langle U_{\rm scr}(2k_F\sin(\theta/2))\Bigr  \rangle_{FS}, \notag
\\
F_c & = \frac{F_t}{2} -\frac{\nu}{4} \Bigl \langle U_{\rm scr}(2k_F\cos(\theta/2)) \Bigr \rangle_{FS} .
\end{align}
Here $U_{\rm scr}(q)$ stands for the statically screened interaction and $\langle\dots\rangle_{FS}$
denotes averaging over the Fermi surface. In the BCS case (for example, for a weak short-range attraction mediated by phonons),
the interaction can be written as $U(\bm{r}) = - (\lambda/\nu) \delta(\bm{r})$ where $0< \lambda \ll 1 $.
Neglecting screening in this case we find
\begin{equation}
F_s \approx - \lambda/2, \qquad F_t \approx  \lambda/2,\qquad F_c \approx \lambda/2 .
\end{equation}
Thus, for the BCS case (i.e. when neither screened nor unscreened Coulomb repulsion is taken into account),
we get the following interaction parameters at the ultraviolet scale (which is given by Debye frequency $\omega_D$ in the case of phonon-induced superconductivity):
\begin{equation}
-\gamma_{s0} \approx \gamma_{t0} \approx \gamma_{c0} \approx  -\lambda/2 .
\label{eq:rge1}
\end{equation}
If disorder is strong, $\omega_D \tau \ll 1$, the relations \eqref{eq:rge1} determine initial values
of the interaction parameters for the action \eqref{eq:NLSM}. In what follows, we will refer to the line determined by relations $-\gamma_s=\gamma_t=\gamma_c$ as the ``BCS line''.
When disorder is weak, $\omega_D \tau \gg 1$, the relations \eqref{eq:rge1} hold at the scale corresponding
to the Debye frequency $\omega_D$. Then the Cooper interaction constant is renormalized at ballistic
scales (between $\omega_D$ and $1/\tau$) such that
\begin{gather}
-\gamma_{s0}=\gamma_{t0}= -\lambda/2, \notag \\
 \gamma_{c0} = -\frac{\lambda/2}{1-(\lambda/2)\ln\omega_D\tau}=\frac{1}{\ln T_c^{BCS}\tau} .
\end{gather}
where  $T_c^{BCS} =\omega_D \exp(-2/\lambda)$.

\subsection{$\mathcal{F}$ algebra and $\mathcal{F}$ invariance}

The NLSM action \eqref{eq:NLSM} involves the matrices which are formally defined
in the infinite Matsubara frequency space.
To perform calculations with these matrices, it is convenient to introduce an ultraviolet
cutoff  $N_M^\prime$ for the Matsubara frequencies.
In addition, it is useful to introduce another cutoff $N_M < N_M^\prime$
indicating the size of a non-trivial part of the $Q$ matrix (beyond which the $Q$ matrix equals $\Lambda$).
At the end of calculations both cutoffs should be sent to infinity.

Global rotations of the $Q$ matrix with any matrix of the type $\exp (i \hat \chi)$, where
$\hat\chi=\sum_{\alpha,n} \chi_n^\alpha I^\alpha_n t_{00}$, play
an important role [\onlinecite{Baranov1999,Kamenev1999}].
In the limit $N_M, N_M^\prime \to \infty$ and $N_M/N_M^\prime \to 0$,
the set of rules known as $\mathcal{F}$ algebra [\onlinecite{Baranov1999}]
allows one to establish the following relations (for $r=0,3$ and $j=0,1,2,3$):
\begin{align}
\Tr I^\alpha_n t_{rj}e^{i \hat\chi} Q e^{-i \hat\chi} & = \Tr I^\alpha_n t_{rj} e^{i \chi_0} Q e^{-i\chi_0}\notag
 \\ & + 8 i n \chi_{-n}^\alpha \delta_{r0}\delta_{j0}, \notag \\
\Tr \eta e^{i \hat\chi} Q e^{-i \hat\chi} & = \Tr \eta Q  + \sum_{\alpha, n} i n \chi_{n}^\alpha
\Tr I^\alpha_n t_{00} Q
 \notag \\ &
- 4  \sum_{\alpha, n} n^2 \chi_{n}^\alpha
\chi_{-n}^\alpha , \notag \\
\Tr \Bigl [I_n^\alpha t_{r0} e^{i \hat\chi} Q e^{-i \hat\chi} \Bigr ]^2 & =
\Tr \Bigl [I_n^\alpha t_{r0} Q \Bigr ]^2 .
\label{eq:Falg}
\end{align}
Using Eqs. \eqref{eq:Falg}, one can check that, provided $\Gamma_s=-Z_\omega$, the action  \eqref{eq:NLSM} is invariant under global rotations of the matrix $Q$ with the matrix $\exp (i \hat \chi)$ (so called $\mathcal{F}$ invariance). The constraint $\Gamma_s=-Z_\omega$ corresponds to the case of Coulomb interaction [\onlinecite{Fin}].
Since the relation $\Gamma_s=-Z_\omega$ is dictated by the symmetry of the action \eqref{eq:NLSM} it should remain fulfilled under the RG flow.

\section{One-loop renormalization-group equations}
\label{sec:one-loop}

\subsection{Preserved spin-rotational symmetry}
\label{subsec:preserved}

To derive RG equations in the one-loop approximation (i.e., to the lowest order in disorder strength), we employ the background-field method and apply it to renormalization of the NLSM action \eqref{eq:NLSM}. Details of the derivation can be found in Appendix \ref{App:Der:BFM}. In $d=2$ dimensions the one-loop RG equations read [$t=2/(\pi g)$]:
\begin{align}
\frac{d t}{dy} & = t^2 \Bigl [ 1 + f(\gamma_s)+3 f(\gamma_t)- \gamma_c \Bigr ] , \label{eq:rg:final:t}\\
\frac{d\gamma_s}{dy}  & = - \frac{t}{2} (1+\gamma_s)\bigl ( \gamma_s+3\gamma_t+2\gamma_c+4\gamma_c^2\bigr ), \label{eq:rg:final:gs} \\
\frac{d\gamma_t}{dy}  & = - \frac{t}{2} (1+\gamma_t) \Bigl [\gamma_s-\gamma_t-2\gamma_c\bigl (1+2\gamma_t-2
\gamma_c\bigr ) \Bigr ], \label{eq:rg:final:gt} \\
\frac{d\gamma_c}{dy} & = - 2\gamma_c^2 - \frac{t}{2} \Bigl [ (1+\gamma_c)(\gamma_s-3\gamma_t) - 2\gamma_c^2+4\gamma_c^3\notag \\
 & \hspace{2cm} + 6\gamma_c \Bigl (\gamma_t-\ln(1+\gamma_t)\Bigr )\Bigr ] ,
\label{eq:rg:final:gc}
\\
\frac{d\ln Z_\omega}{dy} & = \frac{t}{2} \Bigl (\gamma_s+3\gamma_t+2\gamma_c +4\gamma_c^2\Bigr ) ,\label{eq:rg:final:z}
\end{align}
where $y=\ln(L/l)$ ($l$ denotes the mean free path) and $f(x) = 1-(1+1/x)\ln(1+x)$. These RG equations describe
the evolution of the system with spin-rotational and time-reversal symmetries upon changing the characteristic
length scale $L$.
We stress that RG equations \eqref{eq:rg:final:t} - \eqref{eq:rg:final:z} satisfy the particle number conservation since $d (Z_\omega +\Gamma_s)/dy=0$. Further, it is worth emphasising that the right-hand-sides of the equations  are nonsingular in the limit of Coulomb interaction, $\gamma_s=-1$.

The ultraviolet value of the NLSM coupling $t$ that describes the disorder strength
is given by the dimensionless Drude resistivity.
The renormalization of $t$ at larger scales involves the contributions to the resistivity induced by
interference effects and by virtual (elastic) processes due to interactions in particle-hole singlet ($\gamma_s$)
and triplet ($\gamma_t$), as well as in Cooper channel ($\gamma_c$).

We emphasize that  Eqs. \eqref{eq:rg:final:t} - \eqref{eq:rg:final:z} are obtained in the lowest order in $t$
but they are formally exact in interactions $\gamma_{s,t,c}$.
It is worth noting that the Cooper-interaction coupling $\gamma_c$ enters all the RG equations only in a polynomial way.
Interestingly, the contribution of Cooper channel to the renormalization of $t$ is fully described by the linear term
only, thus rendering Eq. \eqref{eq:rg:final:t} for arbitrary $\gamma_c$
the same as in the weak-coupling limit [\onlinecite{FinSC}], $|\gamma_c|\ll 1$.

The first term in Eq. \eqref{eq:rg:final:gc} describes the standard BCS instability; in accordance with the ``Anderson theorem'' this term is not affected by disorder. Moreover, the ``Anderson theorem'' manifests itself in Eq. \eqref{eq:rg:final:gc} through the absence of the terms
$t \mathcal{O}(\gamma_c)$ on the right hand side. To the lowest order in interaction couplings, the effect of disorder on the renormalization of $\gamma_c$ is solely due to the presence of the interaction in the particle-hole channels.

Somewhat counter-intuitively, Eq. \eqref{eq:rg:final:t} suggests an insulating behavior (an increase of the resistivity with
increasing $L$) for $\gamma_c\to -\infty$. We note, however, that the (dimensionless) physical resistivity $\rho$ is not exactly
equal to the NLSM coupling $t$ because of the inelastic contribution to the conductivity governed by superconducting
fluctuations, see Sec.~\ref{s5_0} below for details.
Near the superconducting instability (for large $|\gamma_c|\gg 1$), this antilocalizing inelastic contribution to the conductivity becomes large.

Furthermore, towards the superconducting  instability, $\gamma_c\to -\infty$, the disorder-induced
renormalization of $\gamma_c$ in Eq. \eqref{eq:rg:final:gc} is dominated by the term $-2 t \gamma_c^3$ which tends to
impede a development of the superconducting  instability. Thus, if Eqs. \eqref{eq:rg:final:t} - \eqref{eq:rg:final:z} would constitute the ultimate truth, the superconducting instability would not, strictly speaking, develop. An explanation for this apparent paradox is as follows. It turns out that the one-loop RG equations become insufficient in a  vicinity of the superconducting instability, namely, on scales
 larger than $L_X$ where $|\gamma_c|$ reaches a value $\sim 1/t \gg 1$. In other words,  the weak-disorder condition of validity of the one-loop RG, $t\ll 1$, should in fact be supplemented by the condition  $t|\gamma_c|\ll 1$.

The emergence of the latter condition (and thus of the scale $L_X$) becomes evident from a comparison of the terms of the zeroth and the first order in $t$  in Eq. \eqref{eq:rg:final:gc}. This scale $L_X$ arises also in the calculation of the conductivity (see Sec. \ref{s5_0}): at this
 scale the inelastic contribution to the conductivity reaches in magnitude the elastic one.
 We expect that in the vicinity of the superconducting instability higher-loop terms
of the type $t(t\gamma_c)^{k}$ in the beta-function for $t$ and $\gamma_c^2(t\gamma_c)^{k}$ in the equation governing renormalization of $\gamma_c$ should emerge. Upon resummation, they are expected to restore the divergence of $\gamma_c$ at a scale $L_c$ slightly larger than $L_X$.  At the same time, since the second-loop ($k=2$) terms  are similar to those describing
mesoscopic fluctuations of the superconducting order parameter [\onlinecite{SF,LDOSSC}], we expect for $|\gamma_c|t>1$ (i.e., for temperatures slightly above the transition) strong spatial fluctuations of observables (in particular, of the local tunneling density of states [\onlinecite{TDOS,LDOSSC}], as observed in experiments, see, e.g., Ref.~[\onlinecite{Exp-InO-2}]).

To the lowest order in $\gamma_c$, Eqs. \eqref{eq:rg:final:t} -  \eqref{eq:rg:final:z} coincide with results obtained by Finkelstein long ago~[\onlinecite{FinSC}]. Recently, one-loop RG equations beyond the lowest order in interactions were reported in Ref.~[\onlinecite{DellAnna}] for the case of preserved spin-rotational and time-reversal symmetries. It should be stressed, however, that our RG equations \eqref{eq:rg:final:t} -  \eqref{eq:rg:final:z} differ from those
of Ref.~[\onlinecite{DellAnna}]. It is instructive to highlight the difference.
First of all, the right hand side of a RG equation for $\gamma_s$ in Ref.~[\onlinecite{DellAnna}]
[see Eq. (A12) there] contains a term proportional to $t \gamma_c^2$ rather than to $t (1+\gamma_s) \gamma_c^2$ as in our Eq. \eqref{eq:rg:final:gs}.
Since the quantity $Z_\omega+\Gamma_s=Z_\omega(1+\gamma_s)$ should have no renormalization by virtue of the particle number conservation, this would imply the presence of a term proportional to $t \gamma_c^2/(1+\gamma_s)$ in the RG equation for $Z_\omega$.
Being divergent for the case of Coulomb interaction, $\gamma_s=-1$, such a term would, however, violate the $\mathcal{F}$-invariance of the NLSM action \eqref{eq:NLSM} and is thus not allowed. Second, the RG equation for $\gamma_t$ reported in Ref. [\onlinecite{DellAnna}] does not contain the term proportional to $t \gamma_c^2$,
 in contrast to our Eq. \eqref{eq:rg:final:gt}. Finally, the RG equation for $\gamma_c$ reported in Ref. [\onlinecite{DellAnna}] contains an additional term proportional to $t \gamma_c \ln(1+\gamma_s)$ as compared to our Eq. \eqref{eq:rg:final:gc}. We note that a similar term was reported by Belitz and Kirkpatrick in Ref. [\onlinecite{KB}] [see Eq. (6.8g) there]. In our opinion, such terms, divergent for the case of Coulomb interaction, $\gamma_s=-1$, cannot appear in the course of renormalization of $\mathcal{F}$-invariant operators, including $S_c$. In Ref. [\onlinecite{FinPhysica}], the appearance of a term proportional to $t \gamma_c \ln(1+\gamma_s)$ in the RG equation for $\gamma_c$ of Ref. [\onlinecite{KB}] was attributed to an improper treatment of the gauge invariance. In our background-field RG calculations, terms proportional to  $\ln(1+\gamma_s)$ do appear in the course of renormalization of $\Gamma_c$ at intermediate steps but cancel each other in the final results, in agreement with the $\mathcal{F}$-invariance, see Appendix \ref{App:Der:BFM}.

\subsection{General case}

The RG equations \eqref{eq:rg:final:t} - \eqref{eq:rg:final:z} have been derived for the case of preserved spin-rotational symmetry. We are now going to generalize them to systems with spin-rotational symmetry broken (partly or fully) due to spin-orbit coupling and/or spin-orbit impurity scattering. Both these symmetry-breaking mechanisms induce finite relaxation rates ($1/\tau_s^x$, $1/\tau_s^y$, $1/\tau_s^z$) for corresponding components of the electron spin.  The relaxation rates determine the mass of the corresponding triplet modes (diffusons and cooperons). As an example, the mode corresponding to the spin component $S_x$ acquires a mass proportional to $1/\tau^y_{s}+1/\tau_s^z$. This mode thus become effectively frozen and drops out of RG equations at length scales $L \gg L^x_s \sim [1/(D\tau^y_{s})+1/(D\tau_s^z)]^{-1/2}$.

  In the presence of spin-orbit coupling, there is the spin relaxation due to D'yakonov-Perel' mechanism.
  The corresponding relaxation rates are given by $1/\tau^{x,y,z}_{s} \sim  \Delta_{\rm so}^2 \tau$, where $\Delta_{\rm so}$ denotes the spin-orbit splitting [\onlinecite{DP1971}]. Therefore, all triplet modes (both for diffusons and cooperons) are suppressed at the length scales $L \gg L_{\rm so}=v_F/\Delta_{\rm so}$, i.e., the number of triplet modes contributing to the RG equations is $n=0$. In the case of a 2D electron system with the spin-orbit impurity scattering but without spin-orbit coupling, the spin relaxation is anisotropic: $1/\tau_s^z=1/\tau_{\rm so}$, $1/\tau_s^{x,y}=0$, where  $1/\tau_{\rm so}$ denotes the skew scattering rate [\onlinecite{HLN1980}]. Thus, for $L \gg L_s=\sqrt{D \tau_{\rm so}}$ the triplet modes corresponding to the total spin component $S_z$ remain massless. Therefore, in this case $n=1$ triplet mode still contributes to the RG equations.

  If the spin-orbit coupling and spin-orbit scattering are both present,
then different regimes with $n=3$, $n=1$ and $n=0$ can be realized depending on the relations
between $L$, $L_{\rm so}$ and $L_s$. For all three cases the one-loop RG equations can be written as
\begin{align}
\frac{d t}{dy} & = t^2 \Bigl [ \frac{n-1}{2} + f(\gamma_s)+n f(\gamma_t)- \gamma_c\Bigr ] , \label{eq:rg:final:t:G}
\\
\frac{d\gamma_s}{dy}  & = - \frac{t}{2} (1+\gamma_s)\bigl ( \gamma_s+n \gamma_t+2\gamma_c+4\gamma_c^2\bigr ), \label{eq:rg:final:gs:G} \\
\frac{d\gamma_t}{dy}  & = - \frac{t}{2} (1+\gamma_t) \Bigl [ \gamma_s-(n-2)\gamma_t \notag \\
& \hspace{2cm} -2\gamma_c \bigl (1+2\gamma_t-2 \gamma_c \bigr ) \Bigr ], \label{eq:rg:final:gt:G}
\\
\frac{d\gamma_c}{dy} & =  - 2\gamma_c^2 - \frac{t}{2} \Bigl [ (1+\gamma_c)(\gamma_s- n\gamma_t) - 2\gamma_c^2+4\gamma_c^3 \notag \\
 & \hspace{2cm} + 2 n\gamma_c \Bigl (\gamma_t-\ln(1+\gamma_t)\Bigr )\Bigr ] ,
\label{eq:rg:final:gc:G}
\\
\frac{d\ln Z_\omega}{dy} & = \frac{t}{2} \Bigl (\gamma_s+n\gamma_t+2\gamma_c +4 \gamma_c^2\Bigr ) .\label{eq:rg:final:z:G}
\end{align}
In the case $n=0$, Eq. \eqref{eq:rg:final:gt:G} should be omitted. The RG equations \eqref{eq:rg:final:t:G} - \eqref{eq:rg:final:gc:G}
constitute one of the main results of the paper.
In the rest of the paper, we will analyze these equations to investigate
phase diagrams and observables for the cases of preserved and broken spin-rotational symmetry.

The system of RG equations \eqref{eq:rg:final:t:G} - \eqref{eq:rg:final:gc:G} has the fixed plane $\gamma_s=-1$ corresponding to the case of long-ranged Coulomb interaction. In fact, this statement is not restricted to the one-loop RG equations. The existence of such a fixed plane is a consequence of the particle-number conservation and of the $\mathcal{F}$-invariance of the NLSM action \eqref{eq:NLSM}. Due to the charge conservation, RG equations for $\gamma_{s}$ and $z$ are related to all orders in $t$:
\begin{equation}
\frac{d\gamma_s}{d y} = -(1+\gamma_s) \zeta_{z}, \qquad \frac{d Z_\omega}{dy} = Z_\omega \zeta_z .
\label{fp:stab:gen}
\end{equation}
The value $\zeta_z^*$ of the anomalous dimension $\zeta_z$ at a fixed point determines the dynamical critical exponent $z=d+\zeta_z^*$. The latter controls the temperature behavior of the specific heat, $c_v \sim T^{d/z}$ [\onlinecite{CC1986}].
Typically, one expects that $z\leqslant d$ ($\zeta_z^*<0$) which implies the
instability of the fixed point in the plane $\gamma_s=-1$  with respect to the increase of $\gamma_s$.

It is worth reminding the reader that RG equations \eqref{eq:rg:final:t:G} - \eqref{eq:rg:final:z:G} are of one-loop order with respect to diffusive modes (i.e., are derived by expansion of the right hand side to the lowest nontrivial order in $t$) but are exact in interaction.
Typically, one expects that one-loop RG equations are valid until entering the insulating (strong-disorder) phase, i.e. for $t\lesssim 1$. This requires a tacit assumption that in the expansion of the right hand side of RG equations in powers of $t$ all coefficients (which are functions of interaction amplitudes) are of the order of unity. In the case of superconducting instability, $\gamma_c$ diverges at some scale $L_c$, so that coefficients of the expansion in powers of $t$ become much larger than unity. As discussed in Sec.~\ref{subsec:preserved}, near the superconducting instability (i.e., at  $|\gamma_c|\gg 1$) the general condition of validity of the one-loop approximation $t\lesssim 1$ becomes more restrictive:  $t|\gamma_c|\lesssim 1$. Similarly, near the Stoner instability  (which corresponds to the divergence of  $\gamma_t$) the two-loop analysis [\onlinecite{KB1990,FinScience}] demonstrates that expansion in $t$ is justified for $t\lesssim 1/\gamma_t \ll 1$.

Up to now we have discussed the renormalization as a flow of couplings with the length scale. In practice, one usually has a sufficiently large system and the infrared cutoff is controlled not by the system size but rather by the temperature $T$.   In this situation,
the renormalization due to the contributions to RG equations \eqref{eq:rg:final:t:G} - \eqref{eq:rg:final:z:G} induced by interactions should be stopped at the length scale $L_T$ which is determined as follows [see Eq.  \eqref{app:prop:free}]:
\begin{equation}
T = \frac{1}{\tau} \left (\frac{l}{L_T}\right )^2 \frac{t_0 Z_{\omega 0}}{t(L_T) Z_\omega(L_T)} ,
\label{eq:TL_T}
\end{equation}
where $t_0=t(l)$ and $Z_{\omega0} = Z_\omega(l)$.
This transformation of temperature into the length scale [\onlinecite{footnote-T0}] allows us to
investigate the temperature dependence of observables. In particular, the electrical
resistivity in the absence of magnetic field is addressed in Sec. \ref{s5_0}.
The inclusion of magnetic field induces two additional length scales, $l_H$ and $l_Z$,
related to the orbital and Zeeman effect of magnetic field and leading to the
magnetoresistivity, Sec. \ref{s5}.

\section{Phase diagram at zero magnetic field}
\label{s4}

\subsection{Preserved spin rotational symmetry}

We start our analysis of RG equations \eqref{eq:rg:final:t:G} - \eqref{eq:rg:final:z:G} from the case in which spin rotational and time reversal symmetries are preserved, i.e., there are $n=3$ triplet modes. We note that in notations of Ref. [\onlinecite{KB}] this case is termed as G(LR) for Coulomb interaction and G(SR) for short-ranged interaction.

\subsubsection{Coulomb interaction}
\label{subsec:preserved-coulomb}

For the case of Coulomb interaction, $\gamma_s=-1$, which is the fixed plane of Eqs.~\eqref{eq:rg:final:t:G} - \eqref{eq:rg:final:gc:G}, the RG equations can be simplified as (we set $n=3$)
\begin{align}
\frac{d t}{dy} & = t^2 \Bigl [ 2+3 f(\gamma_t)- \gamma_c \Bigr ] , \label{eq:rg:final:t:Coulomb}\\
\frac{d\gamma_t}{dy}  & = \frac{t}{2} (1+\gamma_t) \Bigl (1+\gamma_t+2\gamma_c(1+2\gamma_t-2\gamma_c)\Bigr ) , \label{eq:rg:final:gt:Coulomb} \\
\frac{d\gamma_c}{dy} & =- 2\gamma_c^2 + \frac{t}{2} \Bigl [ (1+\gamma_c)(1+3\gamma_t) + 2\gamma_c^2(1-2\gamma_c)  \notag \\
& \hspace{2cm} - 6\gamma_c \Bigl (\gamma_t-\ln(1+\gamma_t)\Bigr )\Bigr ] .
\label{eq:rg:final:gc:Coulomb}
\end{align}
Let us now analyze fixed points of Eqs. \eqref{eq:rg:final:t:Coulomb} - \eqref{eq:rg:final:gc:Coulomb}.
It turns out that the structure of the set of fixed points and of the three-dimensional phase diagram is very rich.
Specifically:

\begin{itemize}

\item
There is a marginally unstable line of fixed points at $t=\gamma_c=0$ (with arbitrary $\gamma_t$). These fixed points describe a conventional {\it clean Fermi liquid} without Cooper-channel attraction.

\item
There is a line of fixed points at $t=0$ and $\gamma_c=-\infty$ (with arbitrary $\gamma_t$) corresponding to the {\it superconducting} (SC) phase.

\item
Further, Eqs. \eqref{eq:rg:final:t:Coulomb} - \eqref{eq:rg:final:gc:Coulomb} contain also the attractive line of fixed points at $\gamma_t=\infty$ and  $\gamma_c=1$. The divergence of $\gamma_t$ corresponds to a {\it ferromagnetic} instability.

\item
Formally, in Eqs. \eqref{eq:rg:final:t:Coulomb} - \eqref{eq:rg:final:gc:Coulomb}, there exists also a fixed point at $\gamma_t=-1$, $\gamma_c=0$ and $t=\infty$. While the range of $t\gtrsim 1$ is beyond the accuracy
of the one-loop RG, it is expected on general grounds that full RG equations should contain an attractive fixed point (or a family of fixed points) with $t=\infty$ describing the {\it insulating} phase.

\item
Within Eqs. \eqref{eq:rg:final:t:Coulomb} - \eqref{eq:rg:final:gc:Coulomb} there is a possibility at some length scale to enter the phase with $\gamma_t=-1$.  At this length scale there are finite values $\gamma_c<0$ and $t$. We note that $\gamma_t=-1$ corresponds to the infinitely strong attraction in the triplet particle-hole channel indicating a possibility of {\it exciton condensation}. Since the value $\gamma_t=-1$ is reached at a length scale close to $L_X$,  full RG equations are needed to study a competition of exciton condensation in the spin channel and superconductivity in the Cooper channel. We leave this as a prospect for future research and do not discuss a possibility of exciton condensation in the rest of the paper.

\item
 Going beyond the one-loop RG equations \eqref{eq:rg:final:t:Coulomb} - \eqref{eq:rg:final:gc:Coulomb}, we
 expect a fixed point  at $\gamma_c=-\infty$, $t\sim 1$, and a certain value of $\gamma_t$
  governing the {\it transition between the superconductor and insulator phases}. The corresponding phase boundary is a critical surface with a flow towards this SIT fixed point originating at  the trivial fixed point with $t=\gamma_c=\gamma_t=0$. We will discuss the SIT fixed point
 in more detail in Sec. \ref{s6} below.

\item
Similarly, we expect strong-coupling fixed points that control the ferromagnet-insulator and the ferromagnet-superconductor transitions. We will not discuss these fixed points in the present paper [\onlinecite{footnoteA}].

\end{itemize}

\begin{figure}[t]
\centerline{\includegraphics[width=7cm]{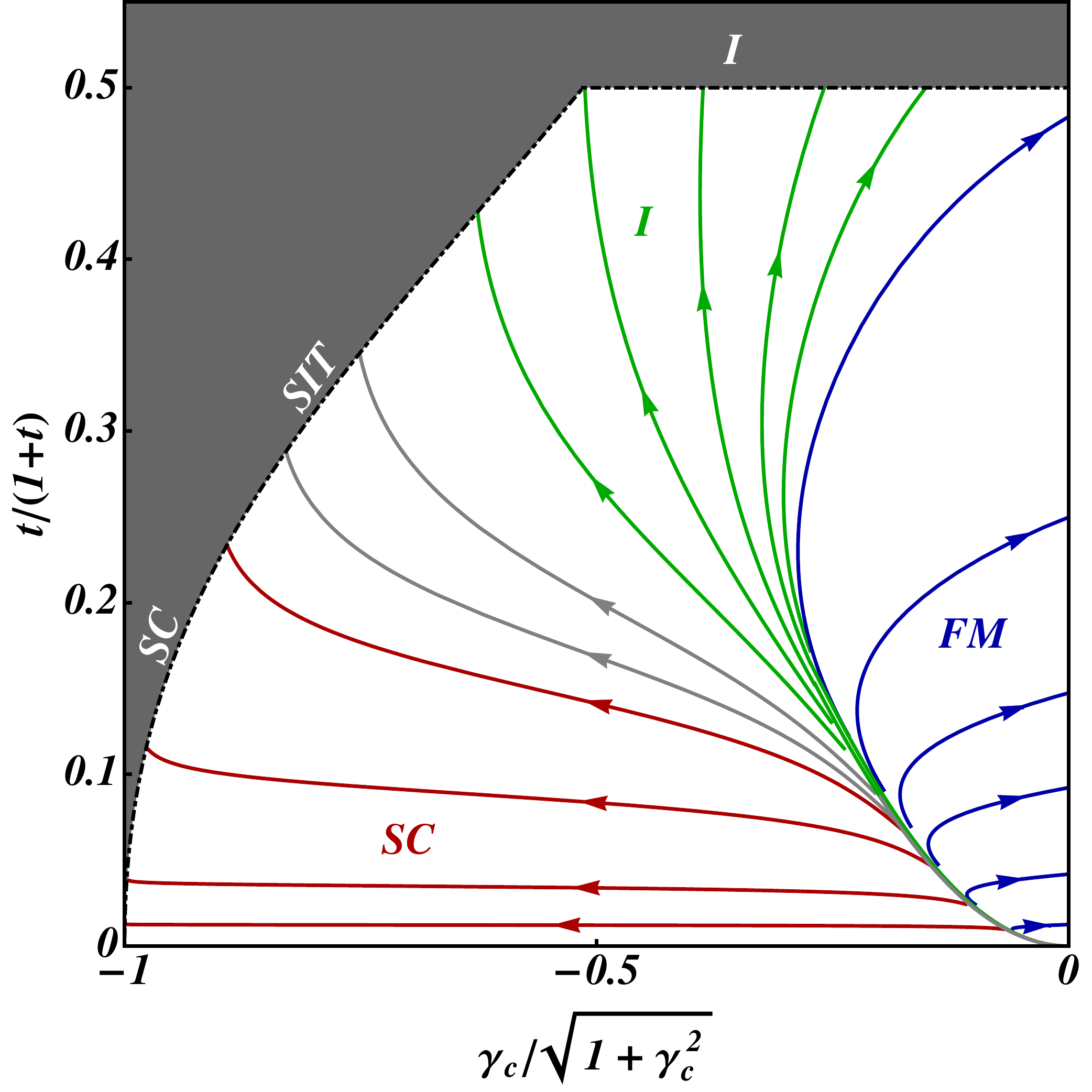}}
\caption{(Color online) The case of preserved spin rotational symmetry with Coulomb interaction, $\gamma_s=-1$:
the RG flow obtained from numerical solution of Eqs. \eqref{eq:rg:final:t:Coulomb} - \eqref{eq:rg:final:gc:Coulomb}.
The initial condition fixes
$\gamma_{t0}=0.2$. The arrows indicate the flow towards the infrared.
The gray region indicates the part of the phase diagram which is not accessible within one-loop RG equations.
The lines describing the flow to the superconducting (SC), insulating (I), and
ferromagnetic (F) phases are shown in red, green, and blue, correspondingly.
Gray flow lines correspond to the region of superconductor-insulator transition (SIT).}
\label{fig:RGO:PD00}
\end{figure}

\begin{figure}[t]
\centerline{\includegraphics[width=7.5cm]{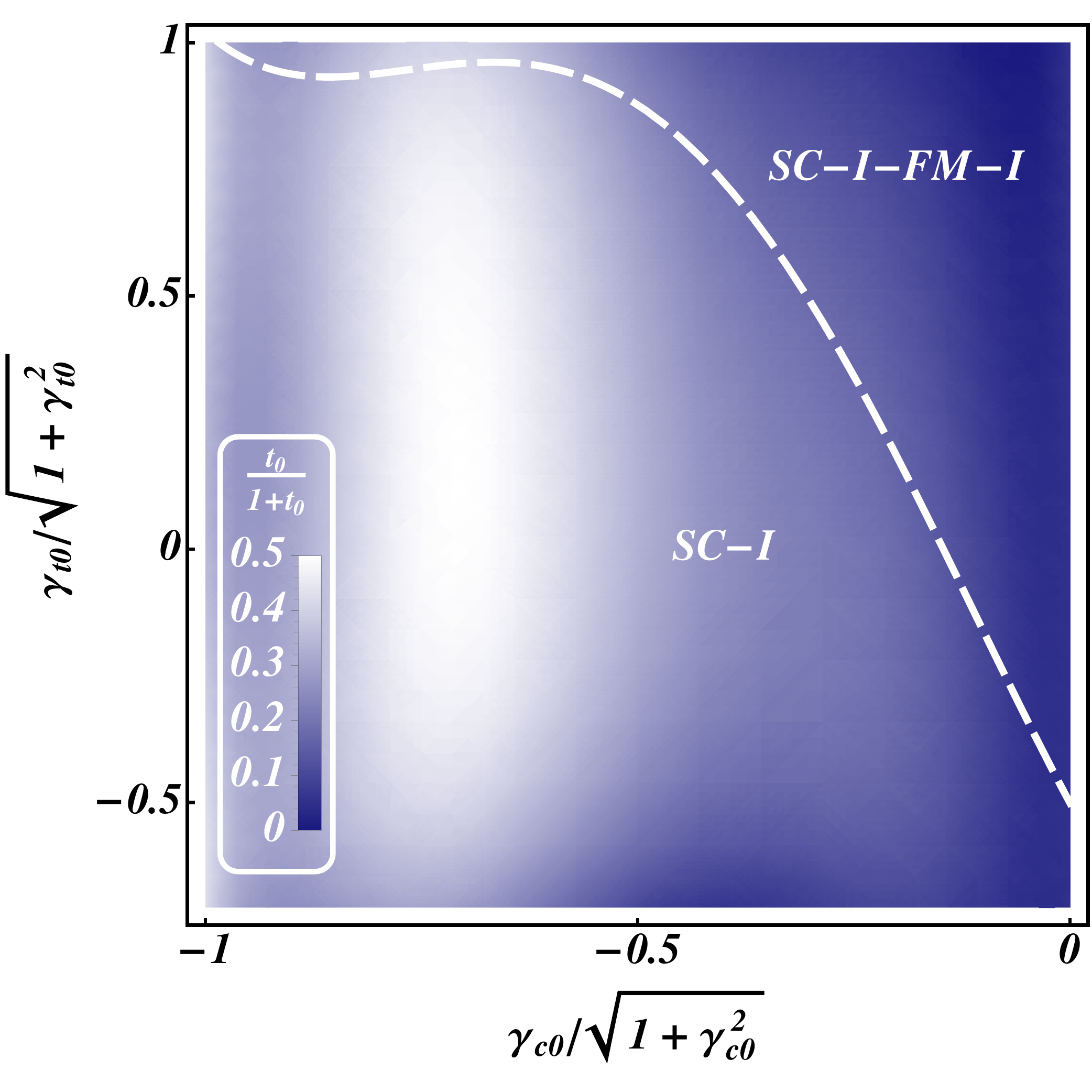}}
\caption{(Color online) The case of preserved spin rotational symmetry with Coulomb interaction, $\gamma_s=-1$: a projection of the phase diagram on the $\gamma_{t0} - \gamma_{c0}$ plane. The color indicates the value of the Drude resistivity $t_0$ at which the quantum phase transition from SC to I occurs. Above the dashed line, the FM phase appears in addition to SC and I phases. The figure is obtained from numerical solutions of RG Eqs. \eqref{eq:rg:final:t:Coulomb} - \eqref{eq:rg:final:gc:Coulomb}. }
\label{fig:RGO:PD}
\end{figure}

Let us now discuss properties of the emerging phases (see Figs. \ref{fig:RGO:PD00} and \ref{fig:RGO:PD}) and corresponding fixed points, in more detail.

{\it Superconducting phase.} We first note that within the RG equations \eqref{eq:rg:final:t:Coulomb} - \eqref{eq:rg:final:gc:Coulomb} the superconducting line of fixed points at $t=0$ and $\gamma_c=-\infty$ is unstable, which makes the superconducting phase formally unreachable. As we have already discussed, this indicates a failure of the one-loop  (lowest order in $t$) RG equations near the superconducting instability. In the absence of disorder (i.e., at $t=0$), Eq. \eqref{eq:rg:final:gc:Coulomb} describes the usual BCS-type scenario. The Cooper-channel interaction $\gamma_c$ diverges at some finite length scale $L_c$ as $\gamma_c(L\to L_c) \sim - 1/(L_c-L)$. To estimate the length scale $L_c$ in the case of finite disorder, we shall use the scale $L_X$ defined by the condition $|\gamma_c(L_X)| = 1/t(L_X) \gg 1$. Assuming that the divergence of $\gamma_c$ is of the BCS type, we get an estimate $(L_c - L_X)/L_X \sim t(L_X) \ll 1$.  Thus, while the one-loop RG is not sufficient to follow the flow up to the singularity scale $L_c$, it works up to a scale $L_X$ which is only slightly smaller than $L_c$.


{\it Insulating phase and superconductor-insulator transition.}
On general grounds, we assume that once the RG flow reaches $t\sim 1$, the system is in the insulating phase, i.e., it flows into the insulating (I) fixed point with $t=\infty$. On the other hand, as discussed above, if $t$ remains small when $|\gamma_c|$ reaches a value $1/t$, the system flows into a superconducting fixed point. There should be thus a fixed point at $t \sim 1$ (i.e., with resistivity of order of quantum resistance $R_q$) and certain values of $\gamma_c$ and $\gamma_t$ that controls the quantum phase transition between superconductor and insulator, see Sec. \ref{s6} for a further discussion. At small values of $t$ and $\gamma_c<0$, $\gamma_t>0$, the separatrix surface between the two phases is parametrized by the following equation: $t = 4\gamma_c^2/(1+3\gamma_t)$.

{\it ``Ferromagnetic'' phase.}
For the attractive line of fixed points at $\gamma_t=\infty$ and  $\gamma_c=1$, the value $\gamma_c=1$
is fixed by a cancelation of terms in the right-hand side of Eq. \eqref{eq:rg:final:gc:Coulomb} which are proportional to $\gamma_t \gg 1$.
The divergence of $\gamma_t$ occurs at some finite length scale $L_{FM}$. Due to a delocalizing effect of the interaction (Altshuler-Aronov) contribution to renormalization of the resistance at large $\gamma_t$,  the fixed point value $t(L_{FM})$ remains finite and is non-universal (i.e., determined by the initial conditions). Therefore, Eqs. \eqref{eq:rg:final:t:Coulomb} - \eqref{eq:rg:final:gc:Coulomb} predict ferromagnetic metallic phase with a non-universal resistivity. Strictly speaking, one-loop equations are insufficient to describe accurately the regime $t \gamma_t \gtrsim 1$ (see Refs. [\onlinecite{KB1990,FinScience}]) but this is not expected to modify essentially the emergence of instability.

However, since the emergent fixed points are characterized by a finite value of dimensionless resistivity $t(L_{FM})$, the diffusive RG
continues at larger scales. Specifically, to describe properly the system at scales larger than $L_{FM}$, one needs to take into account breaking of spin rotational symmetry and derive a new set of RG equations. In this case all triplet diffusive modes in the particle-hole channel and  singlet and triplet modes in the Cooper channel are suppressed. One can thus assume that the system at $L>L_{FM}$ is described by RG equations\eqref{eq:rg:final:t:G} with $n=0$, $\gamma_c=0$ and $\gamma_s=-1$ that results in insulating behavior at large length scales. Moreover, due to enhanced spin fluctuations near the Stoner instability, the system at $L>L_{FM}$ can demonstrate a spin-glass behavior [\onlinecite{NAL}]. In what follows,  we shall term this phase ferromagnetic (FM) for simplicity.

\begin{figure}[t]
\centerline{\includegraphics[width=7cm]{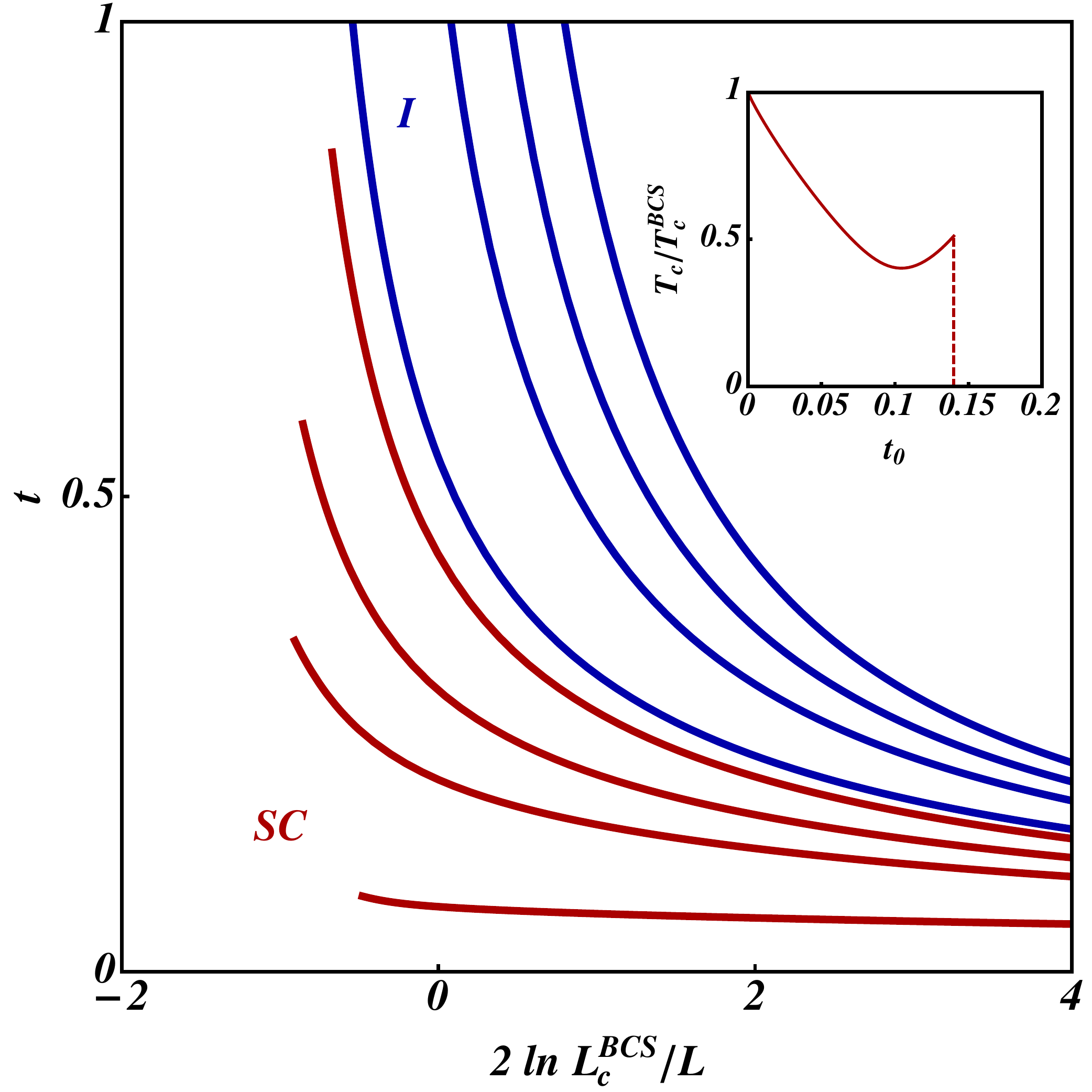}}
\caption{(Color online) The case of preserved spin rotational symmetry with Coulomb interaction, $\gamma_s=-1$:
dependence of $t$ (``renormalized Drude resistivity'') on the length scale across the quantum phase transition between superconducting (SC, red curves) and insulating (I, blue curves) phases.
The curves are obtained from numerical solutions of RG equations \eqref{eq:rg:final:t:Coulomb} - \eqref{eq:rg:final:gc:Coulomb} for $\gamma_{c0}=-0.25$, $\gamma_{t0}=0.01$ and $t_0=0.05,\ 0.1,\ 0.12,\ 0.14,\ 0.15,\ 0.18,\ 0.2,\ 0.22$ (from the bottom to the top).
}
\label{fig:RGO:PD:Res:Ad}
\end{figure}
\begin{figure}[t]
\centerline{\includegraphics[width=7cm]{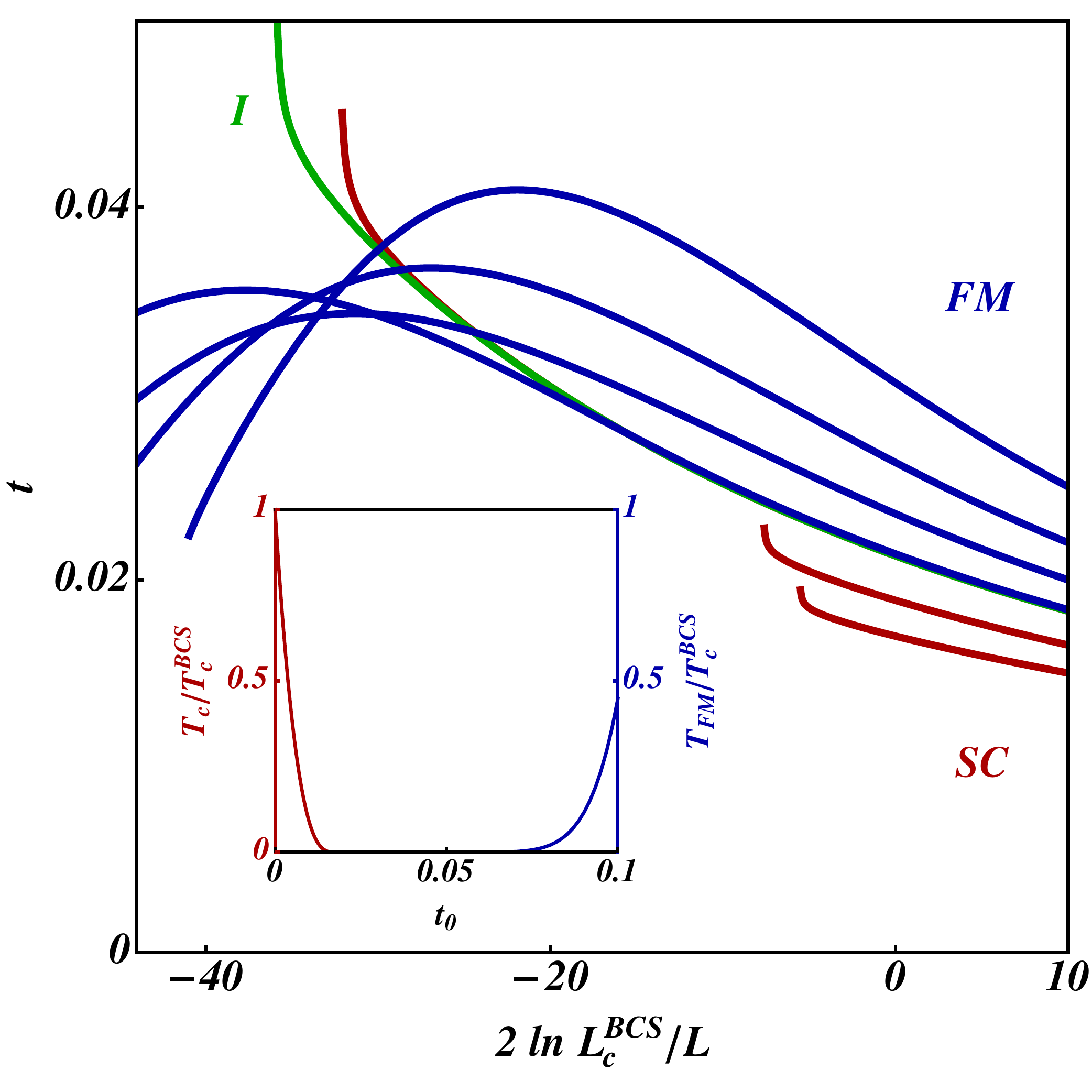}}
\caption{(Color online) The case of preserved spin rotational symmetry with Coulomb interaction, $\gamma_s=-1$:
dependence of $t$ on the length scale across the quantum phase transition between superconducting (SC, red curves), insulating (I, green curve),
 and ferromagnetic (FM, blue curves) phases.
The curves are obtained from numerical solutions of RG equations \eqref{eq:rg:final:t:Coulomb} - \eqref{eq:rg:final:gc:Coulomb} for $\gamma_{c0}=-0.1$, $\gamma_{t0}=0.4$ and $t_0=0.015,\ 0.0165,\ 0.18325,\ 0.18326,\ 0.184,\ 0.02,\ 0.022,\ 0.025$ (from the bottom to the top). For higher values of $t_0$,
another insulating phase (not shown on this scale) emerges.
Inset: dependence of $T_c/T_c^{BCS}$  and $T_{FM}/T_c^{BCS}$ on $t_0$.
}
\label{fig:RGO:PD:Res}
\end{figure}

{\it Overall RG flow and phase diagram.}
A part of the RG flow for Eqs. \eqref{eq:rg:final:t:Coulomb} - \eqref{eq:rg:final:gc:Coulomb} is shown in Fig.  \ref{fig:RGO:PD00}.
In general, a projection of the flow in a three-dimensional parameter space onto a 2D plane, as in Fig. \ref{fig:RGO:PD00} depends
on initial conditions for the couplings. For the plot shown in Fig.  \ref{fig:RGO:PD00}, we have assumed a realistic relation between the triplet (third axis) and Cooper amplitudes,
which has allowed us to avoid intersections in the projected flows. Furthermore, the RG flow is shown only in the region of validity of the one-loop approximation: $t\,\text{max}\{1,|\gamma_c|\}\lesssim 1$. The flows towards the superconducting, insulating, and ferromagnetic phases are plotted in red,
green, and blue, correspondingly. The grey part of the flow describes the vicinity of the SIT. One of the grey curves is the separatrix between the superconducting and insulator phases. However, the one-loop precision is insufficient to determine the separatrix in the region $t \, \text{max}\{1,|\gamma_c|\}\gtrsim 1$. At small values of $\gamma_c$ the separatrix is parametrized by $t=4\gamma_c^2/(1+3\gamma_t)$.

The phase diagram expected on the basis of the RG equations \eqref{eq:rg:final:t:Coulomb} - \eqref{eq:rg:final:gc:Coulomb} is shown in Fig. \ref{fig:RGO:PD} in the plane of bare interaction couplings $\gamma_{c0}$ and $\gamma_{t0}$. For $\gamma_{c0}<0$, the superconducting phase exists at small values of $t_0$. For given $\gamma_{c0}$ and $\gamma_{t0}$ the quantum phase transition from superconductor to insulator occurs with increase of $t_0$.
In addition, for a sufficiently large $\gamma_{t0}$ (above the dashed line) a ferromagnetic phase emerges. In this part of the $\gamma_{c0} - \gamma_{t0}$ plane, a sequence of transitions S -- I -- FM -- I takes place with increasing bare resistivity $t_0$.
For $\gamma_{c0}>0$, there is no superconducting phase; changing $t_0$ drives a transition from the ferromagnetic to the insulator phase.

The dependence of the NLSM coupling $t$ on the length scale $L$ across the quantum phase transition from the superconducting to insulating phase (in the part of the phase diagram in Fig. \ref{fig:RGO:PD} where FM phase does not occur) is shown in Fig.  \ref{fig:RGO:PD:Res:Ad}. This dependence dominates the corresponding evolution of the total electrical resistivity $\rho$ (apart from a narrow region close to the superconducting instability, where the inelastic contributions due to fluctuating Cooper pairs becomes dominant, see Sec. \ref{s5_0} for details).

In Fig.~\ref{fig:RGO:PD:Res} we choose the values of $\gamma_{c0}$ and $\gamma_{t0}$ such that the FM phase exists in addition to the SC and I ones. We thus show the length dependence of $t$ across the quantum phase transitions from SC to I and from I to FM phases. We note that within RG Eqs. \eqref{eq:rg:final:t:Coulomb} - \eqref{eq:rg:final:gc:Coulomb} the insulating phase (between SC and FM phases) exists in a very narrow interval of $t_0$, see Fig. \ref{fig:RGO:PD:Res}. As one can see, the scale $L_X$ (at which red curves in Fig. \ref{fig:RGO:PD:Res} are stopped), which yields approximately the superconducting coherence length,  is larger than the BCS coherence length $L_c^{BCS} = l \exp (-1/2\gamma_{c0})$. In the ferromagnetic phase, the corresponding length scale $L_{FM}$ (where blue curves end) is still larger than $L_X$.

At finite temperature, the interaction contributions to the RG equations \eqref{eq:rg:final:t:Coulomb} - \eqref{eq:rg:final:gc:Coulomb} are stopped at the length scale $L_T$. Neglecting the difference between $L_T$ and the temperature-induced dephasing length $L_\phi$ (which cuts off the localization corrections), we can stop the whole RG at $L_T$. Then the transition temperatures to superconducting ($T_c$) and ferromagnetic phases  ($T_{FM}$) is estimated as follows (see also a discussion in the end of Sec.~\ref{sec:one-loop}):
$T_c\approx (1/\tau) (l/L_X)^{2}$ and $T_{FM}  \approx (1/\tau) (l/L_{FM})^{2}$. A typical dependence of $T_c$ and $T_{FM}$ on $t_0$ is shown in the insets to  Figs. \ref{fig:RGO:PD:Res:Ad} and \ref{fig:RGO:PD:Res}. The effect of disorder on $T_c$ depends on the sign of the term in the square brackets in the right hand side of Eq. \eqref{eq:rg:final:gc:Coulomb}. It occurs that for $\gamma_c<0$ and $\gamma_t>-1$ this term is always  positive, except for a small region at small negative values of $\gamma_c$ and $-1<\gamma_t<-1/3$. Therefore, as was first found by Finkelstein [\onlinecite{Fin}], disorder in the presence of  Coulomb interaction suppresses the superconducting phase (i.e., lowers $T_c$). At the same time, disorder induces the ferromagnetic phase which exists in an intermediate range of disorder. This implies a nonmonotonous dependence of $T_{FM}$ on $t_0$.

We note that $T_c$ evaluated from the RG equations \eqref{eq:rg:final:t:Coulomb} - \eqref{eq:rg:final:gc:Coulomb} is in fact somewhat larger than the true superconducting (Berezinskii-Kosterlitz-Thouless) transition temperature $T_{BKT}$ due to the presence of phase fluctuations of the order parameter at temperatures below $T_c$, see Sec. \ref{s5_0} for more detail. The relative difference between $T_c$ and $T_{BKT}$ is, however, small for weak disorder, and thus does not essentially affect a much stronger variation of $T_c$ with disorder explored in this paper.

\subsubsection{Short-ranged interaction}

In the case of short-ranged interaction, RG equations \eqref{eq:rg:final:t:G} - \eqref{eq:rg:final:gc:G} with  $n=3$ read
\begin{align}
\frac{d t}{dy} & = t^2 \Bigl [1 + f(\gamma_s)+3 f(\gamma_t)- \gamma_c\Bigr ] , \label{eq:rg:final:t:SR}\\
\frac{d\gamma_s}{dy}  & = - \frac{t}{2} (1+\gamma_s)\bigl ( \gamma_s+3 \gamma_t+2\gamma_c+4\gamma_c^2\bigr ), \label{eq:rg:final:gs:SR} \\
\frac{d\gamma_t}{dy}  & = \frac{t}{2} (1+\gamma_t) \Bigl [-\gamma_s+\gamma_t +2\gamma_c\bigl (1+2\gamma_t-2 \gamma_c \bigr ) \Bigr ], \label{eq:rg:final:gt:SR} \\
\frac{d\gamma_c}{dy} & =  - 2\gamma_c^2 + \frac{t}{2} \Bigl [ (1+\gamma_c)(-\gamma_s+3\gamma_t) +2\gamma_c^2(1-2\gamma_c) \notag \\
 & \hspace{1.5cm} - 6\gamma_c \Bigl (\gamma_t-\ln(1+\gamma_t)\Bigr )\Bigr ] .
\label{eq:rg:final:gc:SR}
\end{align}
Contrary to the Coulomb-interaction case (where we had $\gamma_s=-1$), the singlet particle-hole amplitude $\gamma_s$ is not fixed now, so that the RG flow occurs in the four-dimensional parameter space. However, the structure of the set of attractive fixed points (quantum phases) and of fixed points describing quantum phase transitions between them remains qualitatively the same as in the Coulomb case.
Specifically, the fixed points of the RG flow for the short-ranged interaction are as follows [\onlinecite{footnoteA}]:

\begin{itemize}

\item
There is a surface of clean-Fermi-liquid fixed points at $t=\gamma_c=0$ (with arbitrary $\gamma_t$ and $\gamma_s$).

\item
The fixed-point surface at $t=0$ and $\gamma_c=-\infty$ corresponds to the superconducting phase.

\item
The line of fixed points with  $\gamma_s=-1$, $\gamma_t=\infty$, $\gamma_c=1$, and arbitrary $t$, is attractive in the $\gamma_s$ direction.
Therefore, the RG equations \eqref{eq:rg:final:t:SR}-\eqref{eq:rg:final:gc:SR} lead to the same ferromagnetic phase that exists in the case of Coulomb interaction.

\item Exactly as in the Coulomb case, there should be a fixed point (or a family of fixed points) with $t=\infty$ describing the insulating phase.

\item For the same token as in the Coulomb case, a SIT fixed point with $t\sim 1$ should separate the superconducting and insulating phases.


\end{itemize}

\begin{figure}[t]
\centerline{\includegraphics[width=7.5cm]{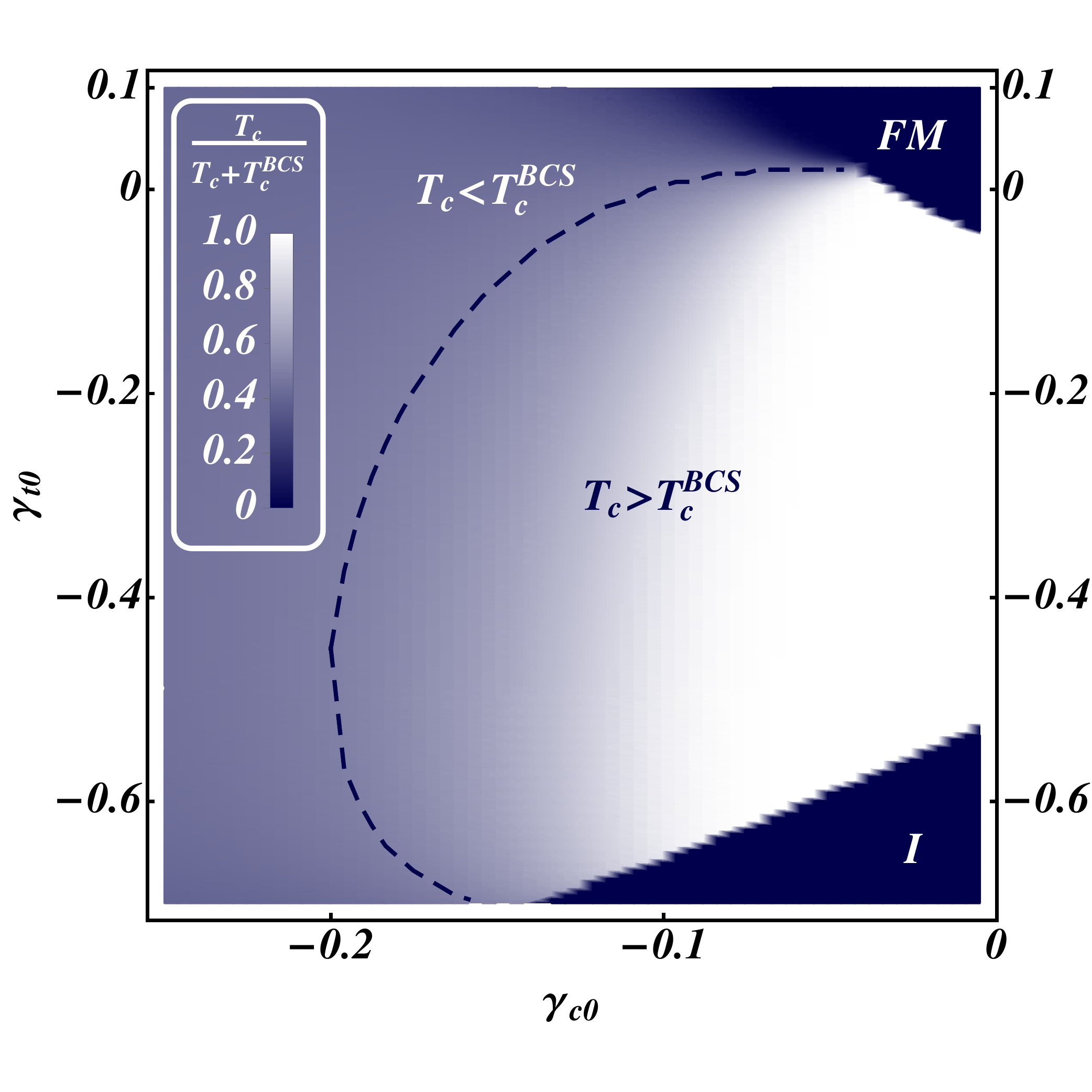}}
\caption{(Color online) The case of preserved spin rotational symmetry with short-ranged interaction, $\gamma_{s0}=-0.05$: a projection of the phase diagram on the $\gamma_{t0} - \gamma_{c0}$ plane.
The color indicates the ratio of $T_c/(T_c+T_c^{BCS})$ for $t_0=0.06$.
The dashed curve separates the regions with $T_c<T_c^{BCS}$ and with $T_c>T_c^{BCS}$. The figure is obtained from numerical solutions of RG Eqs. \eqref{eq:rg:final:t:SR} - \eqref{eq:rg:final:gc:SR}.}
\label{fig:RGO:PD2}
\end{figure}

\begin{figure}[t]
\centerline{\includegraphics[width=7cm]{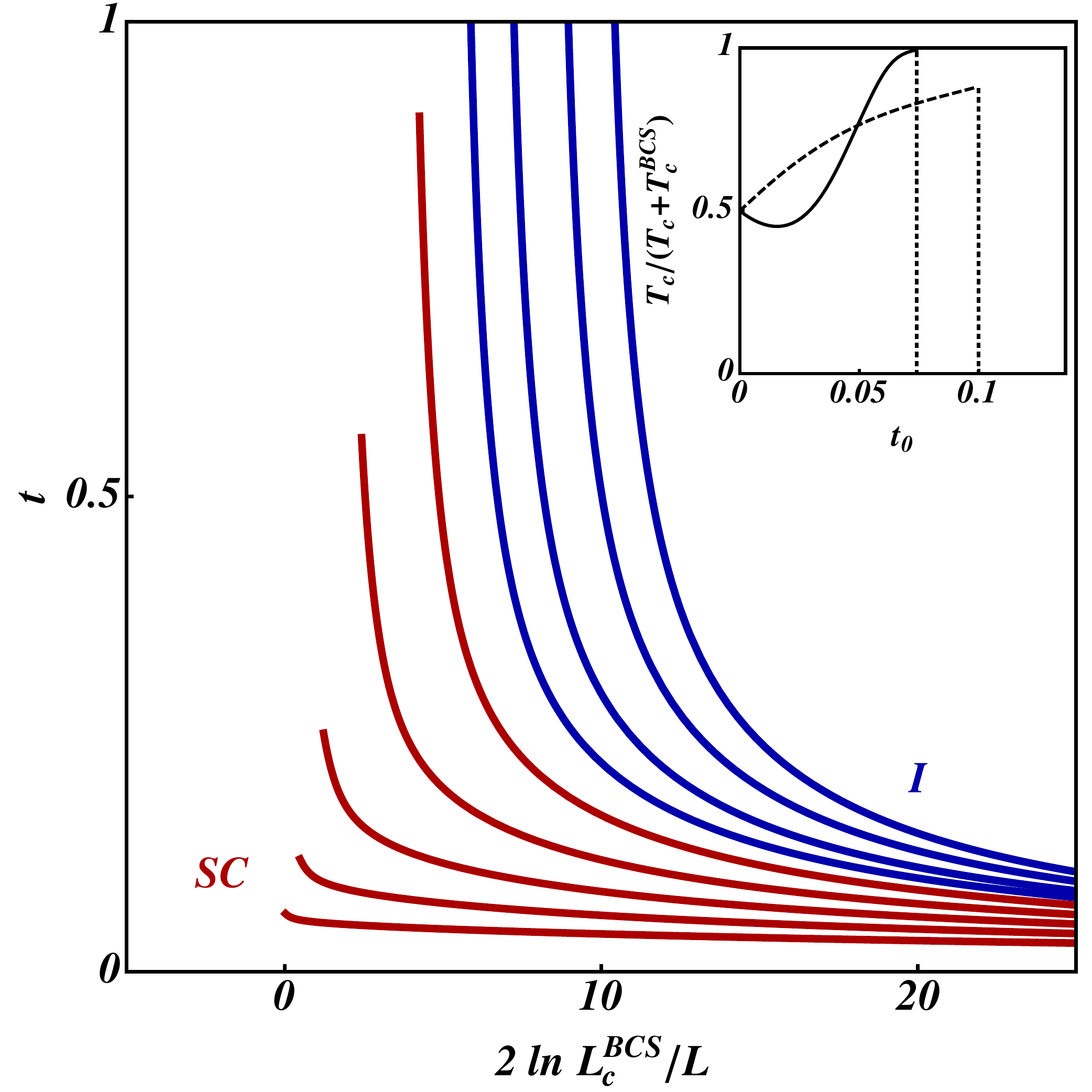}}
\caption{(Color online) The case of preserved spin rotational symmetry with short-ranged interaction, $\gamma_{s0}=-0.05$: dependence of $t$ on the length scale across the quantum phase transition between superconducting (SC, red curves) and insulating (I, blue curves) phases. The curves are obtained from the numerical solution of RG equations \eqref{eq:rg:final:t:SR} - \eqref{eq:rg:final:gc:SR} for $\gamma_{c0}=-0.04$, $\gamma_{t0}=0.005$, and $t_0=0.01,\ 0.03,\ 0.05,\ 0.06,\ 0.07,\ 0.078,\ 0.085,\ 0.095,\ 0.105$ (from bottom to top). For higher values of $t_0$,
an insulating phase (not shown on this scale) emerges.
The dependences of $T_c/(T_c+T_c^{BCS})$ on $t_0$ for $\gamma_{c0}=-0.04$, $\gamma_{t0}=0.005$ and $\gamma_{s0}=-0.05$ (solid curve) and for $\gamma_{c0}=-\gamma_{s0}=\gamma_{t0}=-0.1$ (dashed curve)
are shown in the inset. }
\label{fig:RGO:PD2:SIT}
\end{figure}

The phase diagram for a given $\gamma_{s0}>-1$ is similar to that  for the case of Coulomb interaction, $\gamma_{s0}=-1$ (shown in Fig. \ref{fig:RGO:PD}). With increase of $\gamma_{s0}$, the destruction of the superconducting phase gets shifted towards larger values of $t_0$. The crucial difference between the cases of short-ranged and Coulomb interactions is the existence of large region of the phase diagram with $L_X<L_c^{BCS}$ (and thus $T_c > T_c^{BCS}$). In the case of a bare repulsion in the particle-hole channel, $\gamma_{s0}<0$ and $\gamma_{t0}>0$, the superconducting transition temperature is typically lower than the clean BCS result, $T_c<T_c^{BCS}$ (see Fig. \ref{fig:RGO:PD2}). However, the situation changes if the bare interaction in the triplet particle-hole channel is attractive, $\gamma_{t0}<0$. As illustrated in Fig. \ref{fig:RGO:PD2}, a significant part of the phase diagram is occupied by superconductor with $T_c>T_c^{BCS}$.
It should be emphasized that the superconducting phase with enhanced $T_c$ exists also for $\gamma_{t0}>0$. However, it occurs only in a small region of $\gamma_{t0}, |\gamma_{c0}|, |\gamma_{s0}| \ll 1$ (see Fig. \ref{fig:RGO:PD2}).
Typical RG evolution of the resistance $t$ in this region of initial values of interactions is
shown in Fig. \ref{fig:RGO:PD2:SIT}. Being initially suppressed by disorder,  $T_c$ can be significantly (several orders of magnitude) enhanced with respect to $T_c^{BCS}$ near the superconductor-insulator quantum phase transition, as illustrated in the inset to Fig. \ref{fig:RGO:PD2:SIT}. This is in agreement with the conclusion of our work [\onlinecite{PRL2012}] where RG equations \eqref{eq:rg:final:t}-\eqref{eq:rg:final:gc} with the right-hand sides expanded to the lowest nontrivial order in $\gamma_s$, $\gamma_t$ and $\gamma_c$ were analyzed.

The mechanism of enhancement of the transition temperature is as follows. For small initial values of interaction parameters $|\gamma_{s0}|, |\gamma_{t0}|, |\gamma_{c0}| \ll t_0 \ll 1$, the renormalization of the Cooper interaction amplitude occurs in two distinct steps. At the first step of the RG flow, the interaction is renormalized
due to the presence of disorder (the terms proportional to $t$), while at the second step the standard BCS-type renormalization  (the term $-2 \gamma_c^2$) takes place.
At the first step of renormalization,
we can linearize the RG equations \eqref{eq:rg:final:t:SR}-\eqref{eq:rg:final:gc:SR} in interaction parameters and neglect the term $-2\gamma_c^2$. Then, in the course of RG flow, the interaction amplitudes approach the BCS line $\gamma_s=\gamma_t=-\gamma_c$, converting the repulsion in singlet and triplet particle-hole channels into attraction. This is the consequence  of the (weak) multifractality of the noninteracting fixed point. At some scale $L_1$ such that $\ln L_1/l = 1/t_0-1/t$ the interaction couplings become of the order of the
resistance: $|\gamma_{s,t,c}|\sim t$. Provided $|\gamma_{s0}|, |\gamma_{t0}|, |\gamma_{c0}| \gg t_0^2$, the resistance at this scale $t(L_1) \sim t_0^2/\max\{|\gamma_{s0}|, |\gamma_{t0}|,|\gamma_{c0}|\} \ll 1$ and all interaction parameters are still much smaller than unity.
After the length scale $L_1$ the second step of RG flow starts, where in Eq.~\eqref{eq:rg:final:gc} one can neglect terms proportional to $t$ compared to the disorder-independent
term $-2\gamma_c^2$. Thus, the Cooper interaction $\gamma_c$ flows according to the standard BCS RG equation for a clean system with the initial value $\gamma_c(L_1) \sim t(L_1)$ rather than $\gamma_{c0}$.
Hence, we find the following rough estimate for the transition temperature: $T_c \sim (1/\tau) (l/L_1)^{2} \exp(-1/|\gamma_c(L_1)|) \sim (1/\tau) \exp (-2/t_0) \gg T_c^{BCS}$ (see Appendix \ref{AppTcEnh} for details).

In short, the role of the  non-interacting disorder-induced multifractality is to enhance the interaction in the
Cooper channel such that it becomes comparable to the resistance. After
that, the divergence in the Cooper channel is driven by the standard mechanism  (the same as in a
clean system).
The enhancement of $T_c$ occurs in an intermediate range of disorder ($t_0$ between $|\gamma_{i0}|$ and $|\gamma_{i0}|^{1/2}$).
For a weaker disorder $t_0 \lesssim \max\{|\gamma_{c0}|, |\gamma_{t0}|,|\gamma_{s0}|\}$, one can find a suppression of the transition temperature instead of enhancement, see solid curve in the inset to Fig. \ref{fig:RGO:PD2:SIT}.

\begin{figure}[t]
\centerline{\includegraphics[width=7cm]{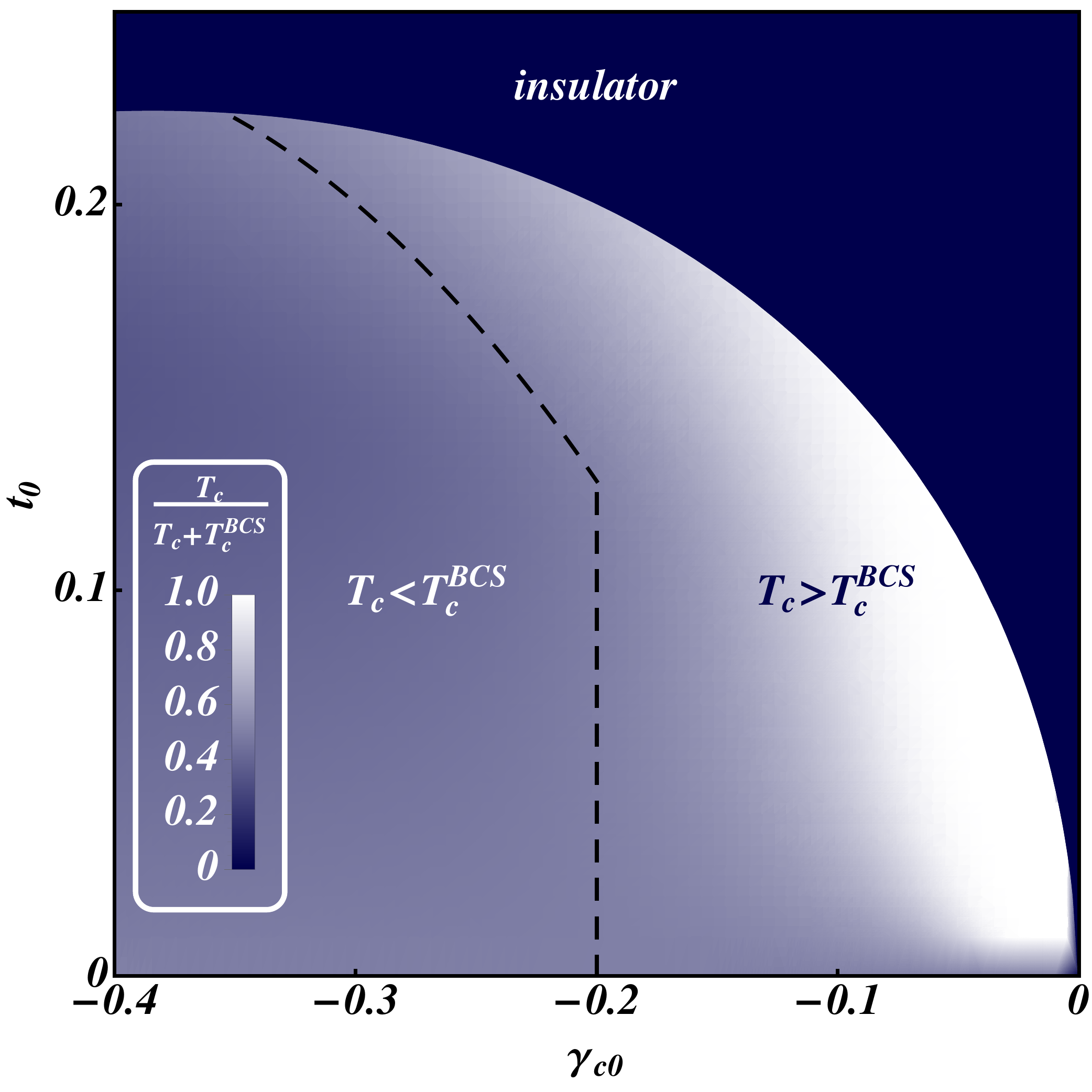}}
\caption{(Color online) The case of preserved spin rotational symmetry with short-ranged interaction on the BCS line: the color density plot for the ratio $T_c/(T_c+T_c^{BCS})$ in the $t_0 - \gamma_{c0}$ plane. The dashed black lines separate the regions with $T_c>T_c^{BCS}$ and $T_c<T_c^{BCS}$.}
\label{fig:RGO:PD2:SIT:BCSLine}
\end{figure}

If the disorder scattering rate $1/\tau$ exceeds the Debye frequency $\omega_D$, the starting point of the RG flow will be likely located not far from the  BCS line, $\gamma_{s0}=-\gamma_{t0}=-\gamma_{c0}$ (see Sec. \ref{InitialBCS}). For such initial conditions, the dependence of $T_c/(T_c+T_c^{BSC})$ on $t_0$ is shown in the inset to Fig. \ref{fig:RGO:PD2:SIT} by dashed curve. It is worth stressing that in the case of initial interaction parameters on the BCS line there is no initial decrease of transition temperature with increase of $t_0$. This is because the second term in the right hand side of Eq. \eqref{eq:rg:final:gc:SR} is negative on the BCS line for $0>\gamma_c> -0.41$.
The dependence of $T_c/T_c^{BSC}$ on $\gamma_{c0}$ and $t_0$ on the BCS line is shown in Fig. \ref{fig:RGO:PD2:SIT:BCSLine}.

\begin{figure}[t]
\centerline{\includegraphics[width=7cm]{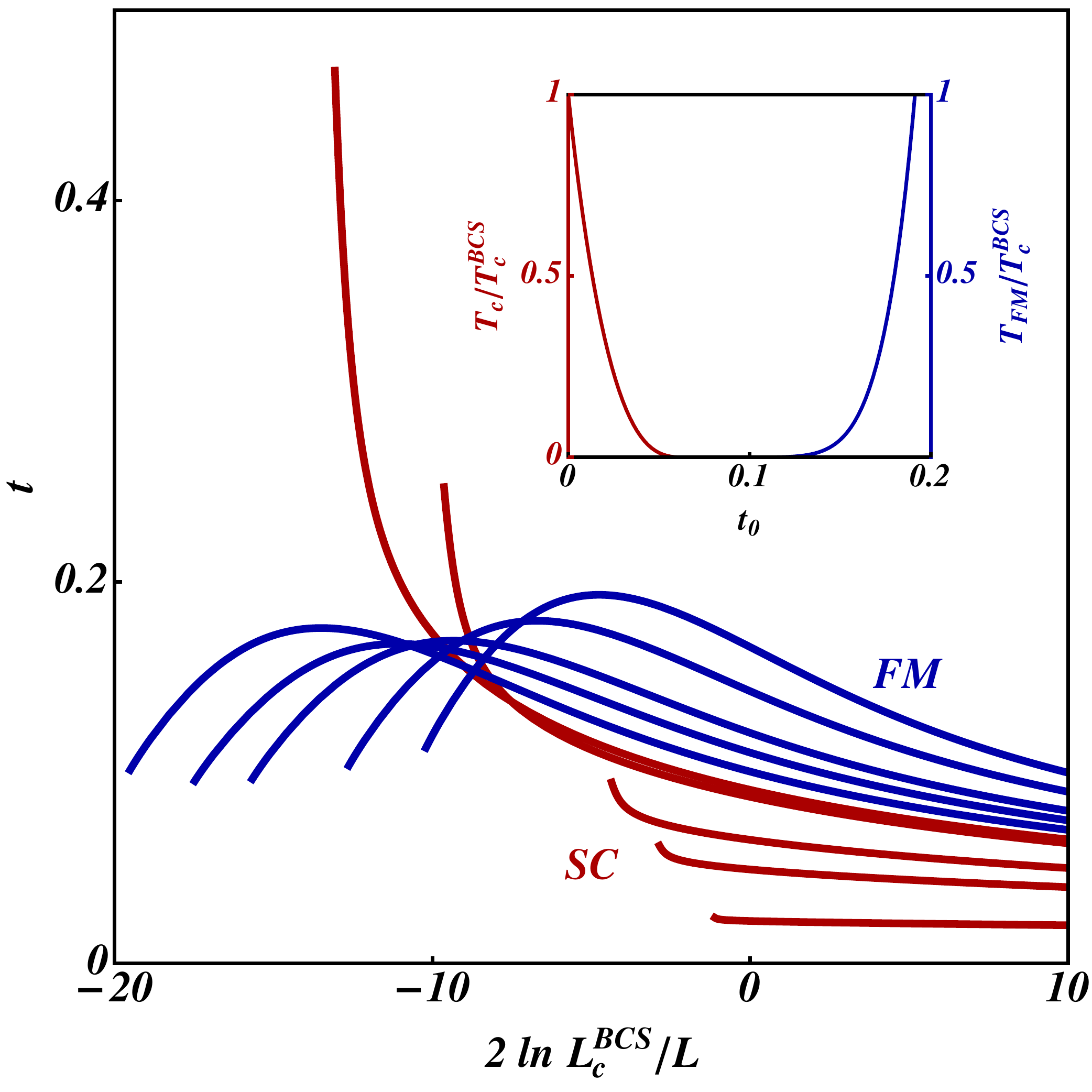}}
\caption{(Color online) The case of preserved spin rotational symmetry with the short-ranged interaction, $\gamma_{s0}=-0.05$: dependence of $t$ on the length scale across the quantum phase transition between superconducting (SC, red curves) and ferromagnetic (FM, blue curves) phases. The curves are obtained from the numerical solution of RG equations \eqref{eq:rg:final:t:SR} - \eqref{eq:rg:final:gc:SR} for $\gamma_{c0}=-0.1$, $\gamma_{t0}=0.2$, and $t_0=0.02,\ 0.04,\ 0.05,\ 0.063,\ 0.065,\ 0.07,\ 0.075,\ 0.08,\ 0.09,\ 0.1$ (from bottom to top). The dependences of $T_c/T_c^{BCS}$ and $T_{FM}/T_c^{BCS}$  on $t_0$ are shown in the inset.}
\label{fig:RGO:PD2:SIT:a}
\end{figure}

Let us now turn to the region in the phase diagram, Fig. \ref{fig:RGO:PD2}, where the ferromagnetic (FM) phase emerges.
Figure \ref{fig:RGO:PD2:SIT:a} shows typical dependences of resistance $t$ on the length scale $L$ across the transition from SC to FM phases with increasing $t_0$.
The inset presents dependences of both critical temperatures ($T_c$ and $T_{\rm FM}$) on $t_0$.

\subsection{Broken spin rotational symmetry}

In the presence of spin-orbit coupling and/or spin-orbit scattering the spin-rotational symmetry is broken.
At length scales $L\gg \max\{L_{\rm so},L_s\}$ the spin-rotational symmetry is completely broken and all triplet modes are suppressed such that $n=0$. In Ref. [\onlinecite{KB}] this case is referred to as SO(LR) for Coulomb interaction and SO(SR) for short-ranged interaction [\onlinecite{footnote-TITS}].

\subsubsection{Coulomb interaction}

For $n=0$ and for the case of Coulomb interaction, $\gamma_{s}=-1$, the one-loop RG equations, Eqs. \eqref{eq:rg:final:t:G} - \eqref{eq:rg:final:gc:G}, take the form
\begin{align}
\frac{d t}{dy} & = t^2 \left( \frac{1}{2} - \gamma_c \right) , \label{eq:rg:final:t:SO:Coulomb}\\
\frac{d\gamma_c}{dy} & =- 2\gamma_c^2 + \frac{t}{2} \Bigl [ 1+\gamma_c + 2\gamma_c^2(1-2\gamma_c)  \Bigr ] .
\label{eq:rg:final:gc:SO:Coulomb}
\end{align}
The first, perturbative study of the effect of interaction on conductivity of a disordered system in the presence of spin-orbit scattering has been performed in Ref. [\onlinecite{F1982}].
 To the lowest order in $\gamma_c$ Eqs. \eqref{eq:rg:final:t:SO:Coulomb} - \eqref{eq:rg:final:gc:SO:Coulomb} coincide with the one-loop RG equations derived in Refs. [\onlinecite{CCFS1984,MF1986,Fin1987}].

Since the spin-orbit interaction kills the contribution of the triplet channel, while
the particle-hole singlet amplitude remains fixed, $\gamma_s=-1$, the RG flow now occurs in a 2D parameter space, $t$ and $\gamma_c$.
The structure of phase diagram is governed by the following fixed points:
\begin{itemize}

\item
Equations \eqref{eq:rg:final:t:SO:Coulomb} and \eqref{eq:rg:final:gc:SO:Coulomb} possess a clean-Fermi-liquid fixed point at $t=\gamma_c=0$ which is marginally unstable.

\item
There is the fixed point at $t=0$ and $\gamma_c=-\infty$ corresponding to superconducting (SC) phase.

\item As in all other symmetry classes, there should be the  insulating (I) phase with $t=\infty$. It is, however, not reachable within the one-loop RG equations.

\item
There is stable non-trivial fixed point at $\gamma_c^*=1/2$ and $t^*=2/3$ describing the critical metallic (CM) phase.
This fixed point appears at the borderline of applicability of one-loop RG equations, $t^*\sim 1$, so that we do not have a rigorous argument in favour of existence of the CM phase. We find, however, very plausible that the attractive character of this
fixed point is not destroyed by going beyond one loop.
The emergence of this fixed point can be traced back (i) to the competition of weak antilocalization (enhanced by delocalizing effect of repulsive Cooper-channel interaction) with the localizing Coulomb repulsion in Eq.~\eqref{eq:rg:final:t:SO:Coulomb}, and (ii) to the competition between Cooper instability and
the disorder-induced suppression of the interaction matrix element in Eq.~\eqref{eq:rg:final:gc:SO:Coulomb}.
The CM phase (if indeed exists) should be separated from the I phase by a CM--I quantum phase transition fixed point which is, however, located well beyond the limit of our one-loop RG.

\item
As in other symmetry classes, we expect existence of a fixed point at $\gamma_c=-\infty$ and $t\sim 1$ (region marked by `SIT')  such that the transition between superconductor and insulator occurs through the separatrix connecting
this fixed point and the trivial fixed point at $t=\gamma_c=0$. At small values of $t$ and $\gamma_c<0$ the separatrix is parametrized by the following equation, $t = 4\gamma_c^2$ [\onlinecite{Fin1987}].

\end{itemize}

\begin{figure}[t]
\centerline{\includegraphics[width=7cm]{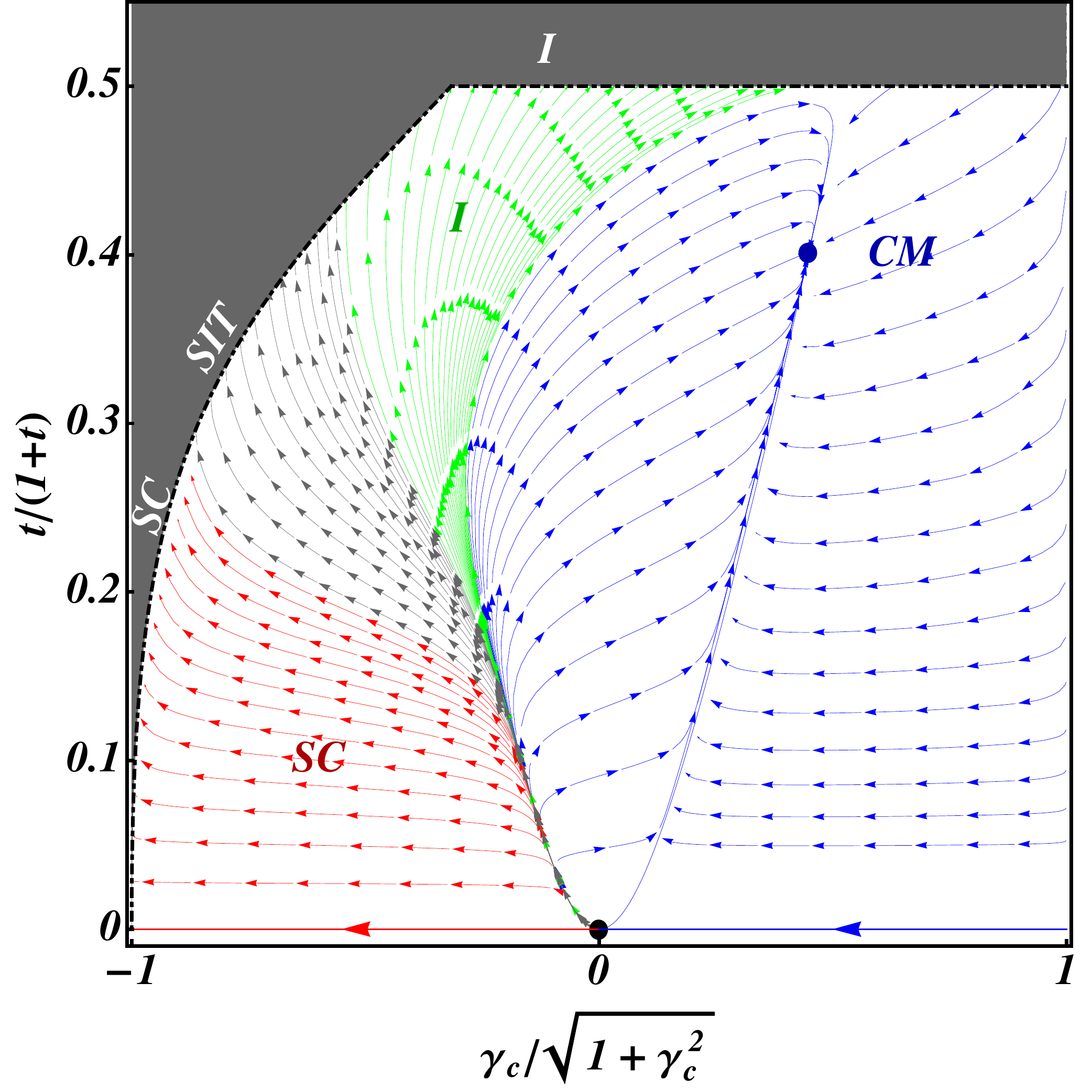}}
\caption{(Color online) The case of broken spin rotational symmetry with Coulomb interaction, $\gamma_{s}=-1$: the RG flow obtained from numerical solution of Eqs. \eqref{eq:rg:final:t:SO:Coulomb} and \eqref{eq:rg:final:gc:SO:Coulomb}. The arrows indicate the flow towards the infrared. The gray region indicates the part of the phase diagram which is not accessible within one-loop RG equations. The lines describing the flow to the superconducting (SC), insulating (I), and critical-metal (CM) phases are shown in red, green, and blue, correspondingly. Gray flow lines corresponds to the region of superconductor-insulator transition (SIT).
}
\label{fig:RGSO+Coulomb}
\end{figure}

The RG flow (and the corresponding phase diagram) for equations \eqref{eq:rg:final:t:SO:Coulomb} - \eqref{eq:rg:final:gc:SO:Coulomb} is shown in Fig. \ref{fig:RGSO+Coulomb}.
As in the case of preserved spin rotational symmetry, we stop the RG flow when either $|\gamma_c|$ reaches the value $1/t \gg 1$ at a certain scale $L_X$  (superconducting phase, red RG flow lines) or the resistance $t$ reaches the value unity (insulating phase, green flow lines). In addition, we have now the critical metal phase {(blue flow lines)}.

\begin{figure}[t]
\centerline{\includegraphics[width=7cm]{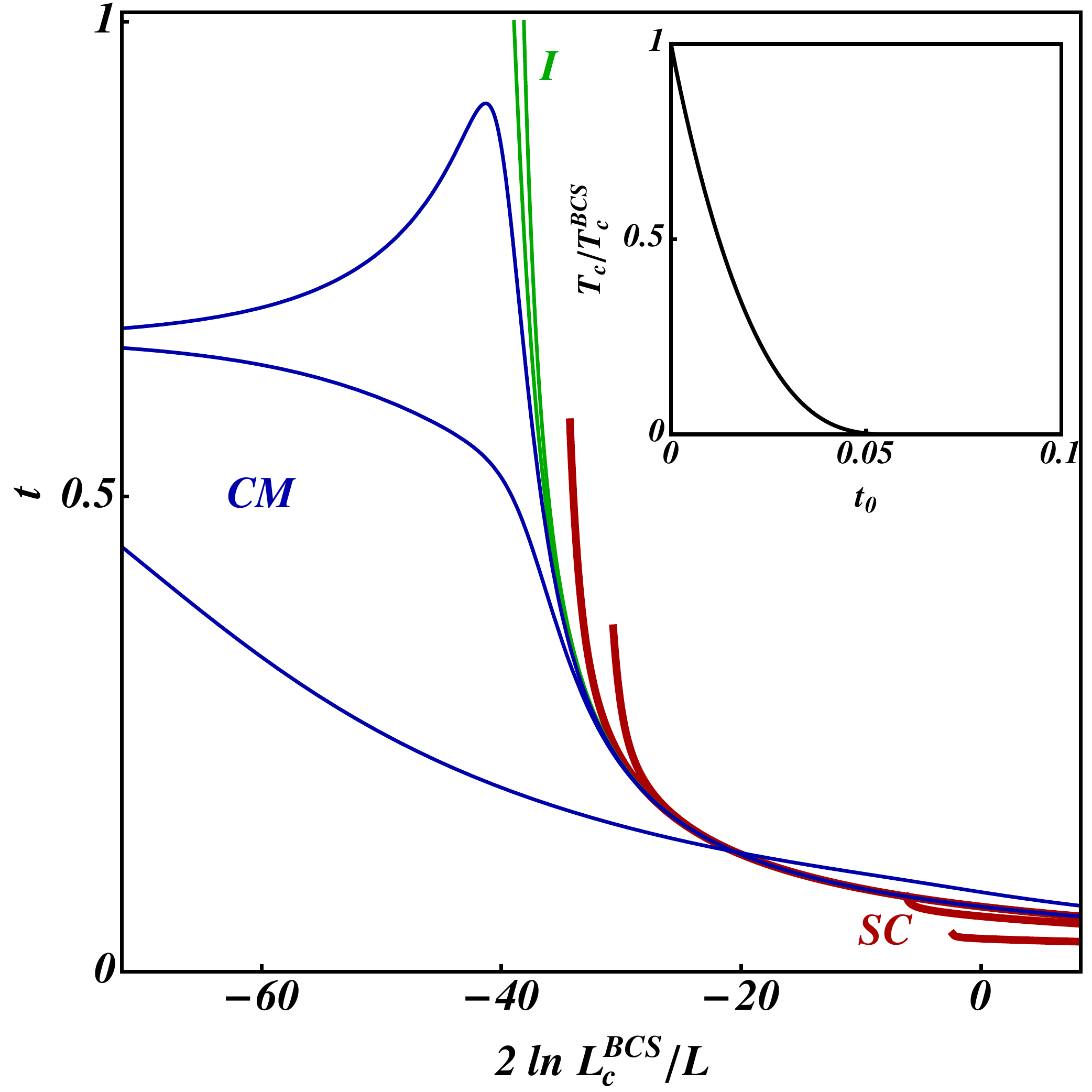}}
\caption{(Color online) The case of broken spin rotational symmetry with Coulomb interaction, $\gamma_{s}=-1$: dependence of $t$ on the length scale across the transitions between superconducting (SC, red curves), insulating (I, green curves), and critical-metal (CM, blue curves) phases. The curves are obtained from numerical solution of Eqs. \eqref{eq:rg:final:t:SO:Coulomb} and \eqref{eq:rg:final:gc:SO:Coulomb} for $t_0 = 0.03,\ 0.05,\ 0.058039,\ 0.0580414,\ 0.0580416,\  0.0580418$, $0.0580419,\ 0.0580421,\ 0.07$ (from bottom to top). With further increasing the Drude resistivity $t_0$, the system enters again the insulating phase (not shown here), see Fig.~\ref{fig:RGSO+Coulomb}. The inset: dependence of $T_c/T_c^{BCS}$ on $t_0$. The value of the Cooper-channel attraction is $\gamma_{c0}=-0.12$.
}
\label{fig:RGSO+Coulomb+Res}
\end{figure}

The dependence of $t$ on the length scale across the consecutive SC-I-CM transitions is shown in Fig. \ref{fig:RGSO+Coulomb+Res}. As one can see, there is a very narrow interval of $t_0$  values in which the insulating phase separating the SC and CM phases exists. We mention that at not too large length scales (or, equivalently, at not too low temperatures) the resistance curves for SC, I, and CM phases cross each other.
As expected, the Coulomb interaction suppresses the superconductivity, so that  $L_X > L_c^{BCS}$, and consequently, $T_c < T_c^{BCS}$ (as shown in the inset of Fig. \ref{fig:RGSO+Coulomb+Res}).

From Eq. \eqref{eq:rg:final:z:G} with $n=0$ we find the following one-loop RG result for $Z_\omega$ in the case of Coulomb interaction:
\begin{equation}
\label{zeta-crit-metal}
 \frac{d\ln Z_\omega}{dy} = \frac{t}{2} \Bigl (\gamma_s+2\gamma_c +4 \gamma_c^2\Bigr ) .
\end{equation}
As explained below Eq. \eqref{fp:stab:gen}, the fixed-point value of the $\zeta$ function  $d\ln Z_\omega / dy$
determines the dynamical exponent controlling the temperature dependence of the specific heat. The one-loop result Eq.~\eqref{zeta-crit-metal} yields for the critical metal  $c_v \sim T^{2/z}$ with $z=7/3$. Since $z>2$, the CM phase is stable with respect to the deviations of $\gamma_s$ from $\gamma_s=-1$.

\subsubsection{Short-ranged interaction}

In the case of short-ranged interaction RG equations \eqref{eq:rg:final:t:G} - \eqref{eq:rg:final:gc:G}  with $n=0$ take the form
\begin{align}
\frac{d t}{dy} & = t^2 \Bigl [-\frac{1}{2} + f(\gamma_s) - \gamma_c\Bigr ] , \label{eq:rg:final:t:SO:SR}\\
\frac{d\gamma_s}{dy}  & = - \frac{t}{2} (1+\gamma_s)\bigl ( \gamma_s+2\gamma_c+4\gamma_c^2\bigr ), \label{eq:rg:final:gs:SO:SR} \\
\frac{d\gamma_c}{dy} & =  - 2\gamma_c^2 + \frac{t}{2} \Bigl [ -(1+\gamma_c)\gamma_s+2\gamma_c^2(1-2\gamma_c)\Bigr ] .
\label{eq:rg:final:gc:SO:SR}
\end{align}
These RG equations are richer than Eqs. \eqref{eq:rg:final:t:SR}-\eqref{eq:rg:final:gc:SR} and describe the RG flow in
a three-dimensional space of $t$, $\gamma_s$, and $\gamma_c$:
\begin{itemize}

\item
There is a line of  clean-Fermi-liquid fixed points at $t=\gamma_c=0$ and arbitrarily $\gamma_s$. The peculiarity of the present symmetry class  is that the clean non-interacting fixed point $t=\gamma_c=\gamma_s=0$ is attractive. It corresponds to a {\it supermetal} (SM) phase.

\item
The line of fixed points at $t=0$ and $\gamma_c=-\infty$ corresponds to the superconducting phase.

\item As in all other symmetry classes, there is the  insulating (I) phase with $t=\infty$ not reachable within the one-loop RG equations.

\item
In addition to the critical metal phase at $\gamma_c^* = 1/2$, $\gamma_s^*=-1$, and $t^*=2/3$ there is a second
fully attractive fixed point at intermediate resistivity (i.e. on the border of applicability of one-loop RG equations): $\gamma_c^{**}=-1/2$, $\gamma_{s}^{**}=0$, and $t^{**}=1$. So, one-loop RG equations suggest a possibility of two different CM phases.

\item
As in other symmetry classes, we expect existence of a fixed point $t\sim 1$  controlling the transition between superconductor and insulator phases. Further intermediate-coupling ($t\sim 1$) fixed points control other emerging quantum phase transitions (SM-I, SC-SM, SM-CM, and CM-I).

\end{itemize}

\begin{figure}[t]
\centerline{\includegraphics[width=7.5cm]{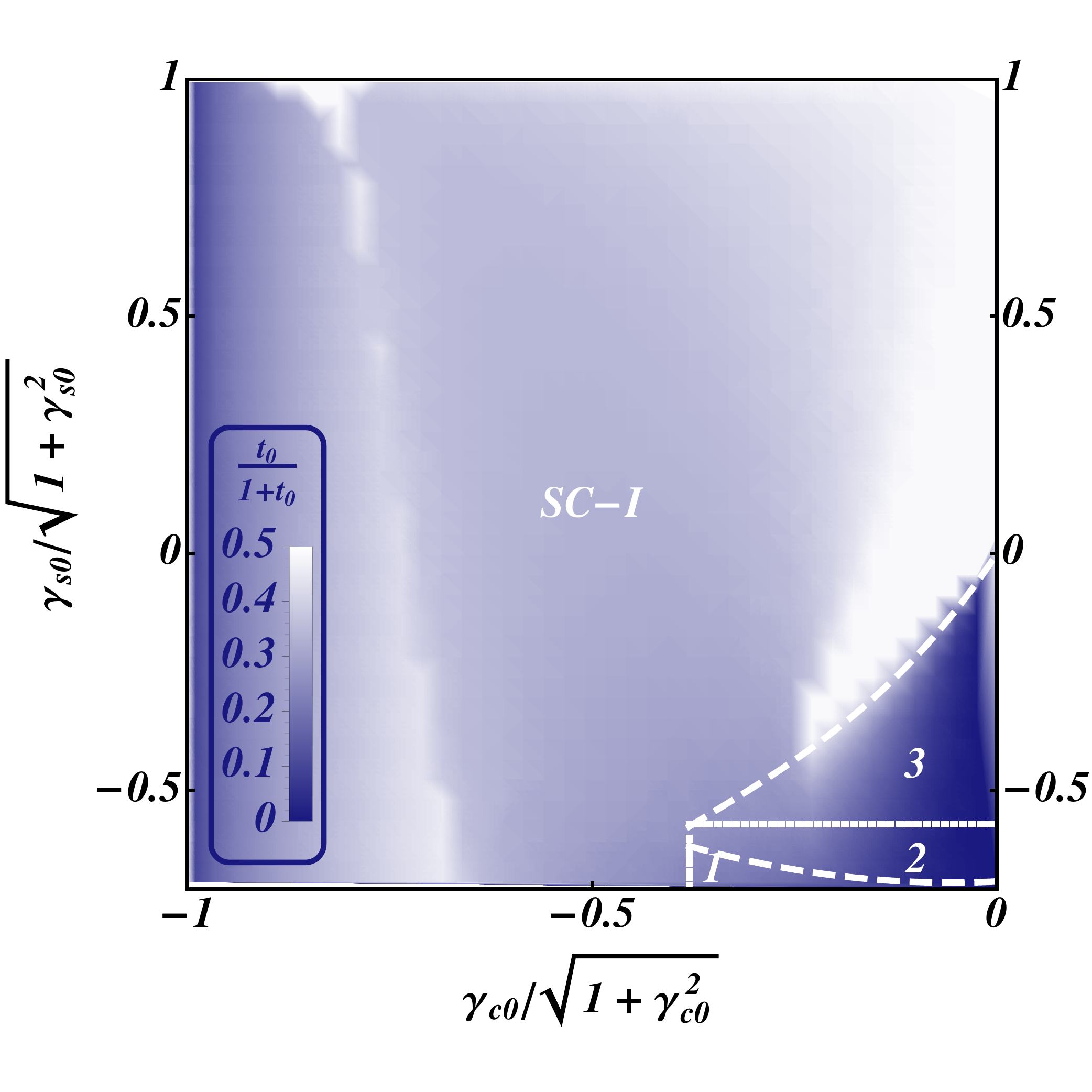}}
\caption{(Color online) The case of broken spin rotational symmetry with short-ranged interaction: a projection of the phase diagram on the plane $\gamma_{c0} - \gamma_{s0}$. The color indicates the value of $t_0$ at which the QPT from SC to SM or I occurs. In regions 1, 2 and 3 the increase of $t_0$ drives a sequence of QPTs: in 1 -- SC-SM-I, in 2 -- SC-SM-CM-I, and in 3 -- SC-I-SM-I.
The figure is obtained from numerical solutions of RG equations  \eqref{eq:rg:final:t:SO:SR} - \eqref{eq:rg:final:gc:SO:SR}.}
\label{fig:RGSO+SRI}
\end{figure}

The phase diagram expected on the basis of RG equations \eqref{eq:rg:final:t:SO:SR} - \eqref{eq:rg:final:gc:SO:SR}
is shown in Fig. \ref{fig:RGSO+SRI}. For $\gamma_{c0}<0$ the superconducting phase exists at small values of $t_0$. With increase of $t_0$ the QPT to insulator or supermetal occurs for given values of $\gamma_{c0}$ and $\gamma_{s0}$.  With increase of $\gamma_{s0}$ the superconducting phase proliferates. In the most part of the phase diagram the transition between superconductor and insulator occurs at $t_0\sim 1$.  Interestingly, there is a region of the phase diagram in which a sequence of quantum phase transitions, SC-SM-I, SC-SM-CM-I or SC-I-SM-I, occurs as $t_0$ grows (see Fig. \ref{fig:RGSO+SRI}). The dashed curve separating the region with multiple quantum phase transitions is parametrized by the condition $\gamma_{s0}=2\gamma_{c0}$, see Appendix \ref{AppTcEnh}.

\begin{figure}[t]
\centerline{\includegraphics[width=7cm]{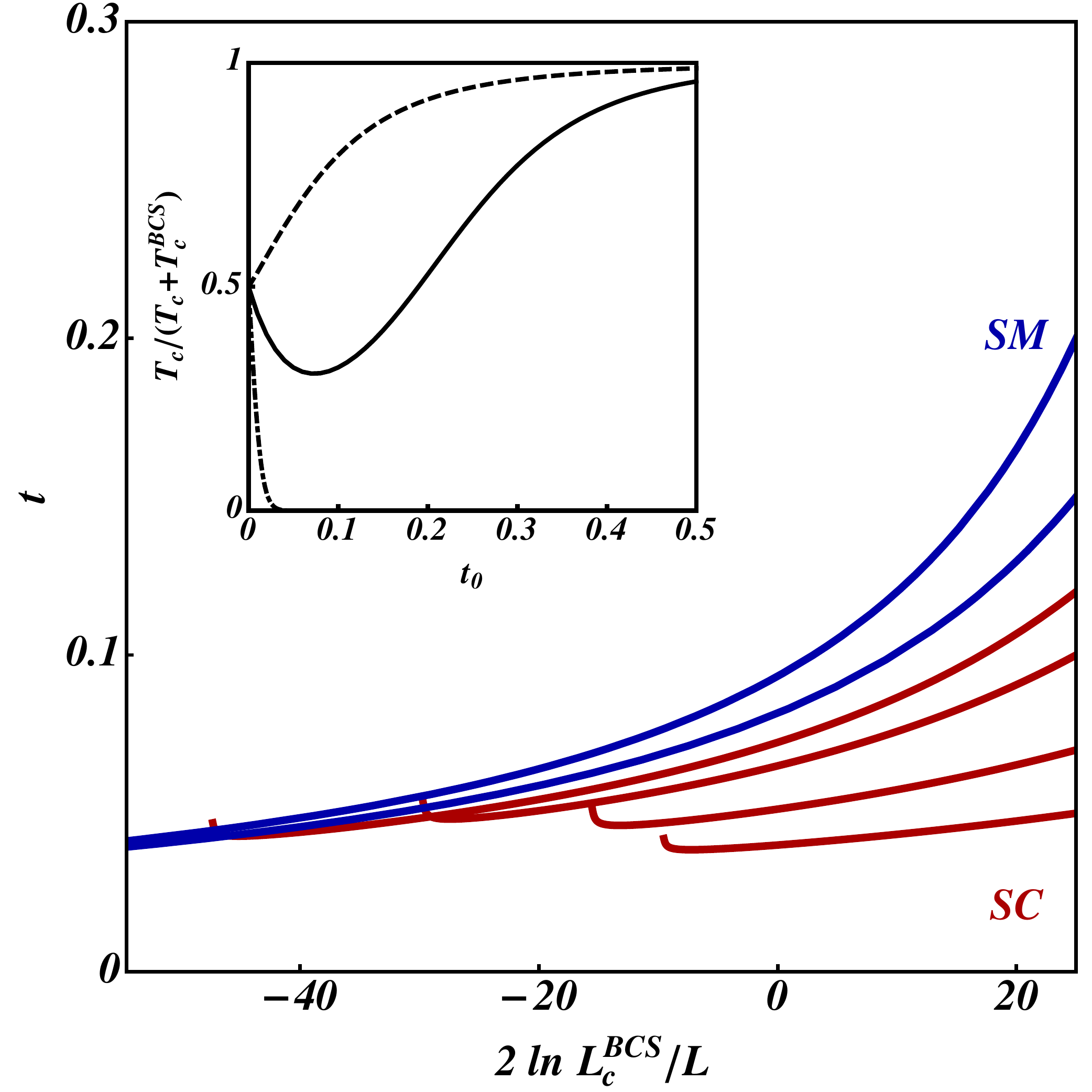}}
\caption{(Color online) The case of broken spin rotational symmetry with short-ranged interaction:
dependence of $t$ on the length scale across the QPT between superconducting and supermetallic phases. The curves are obtained from numerical solutions of RG equations  \eqref{eq:rg:final:t:SO:SR} - \eqref{eq:rg:final:gc:SO:SR} for $\gamma_{c0}=-0.04$, $\gamma_{s0}=-0.1$, and $t_0=0.05,\ 0.07,\ 0.1,\ 0.12,\ 0.15,\ 0.2$ (from bottom to top). Inset: Dependences of $T_c/(T_c+T_c^{BCS})$ on $t_0$ for $\gamma_{c0}=-0.04$ and $\gamma_{s0}=-0.005$ (solid curve), for $\gamma_{c0}=-\gamma_{s0}=-0.05$ (dashed curve) and for $\gamma_{c0}=-0.04$ and $\gamma_{s0}=-0.1$ (dot-dashed curve) are shown in the inset.
}
\label{fig:RGSO+SRI+R}
\end{figure}

\begin{figure}[t]
\centerline{\includegraphics[width=7.5cm]{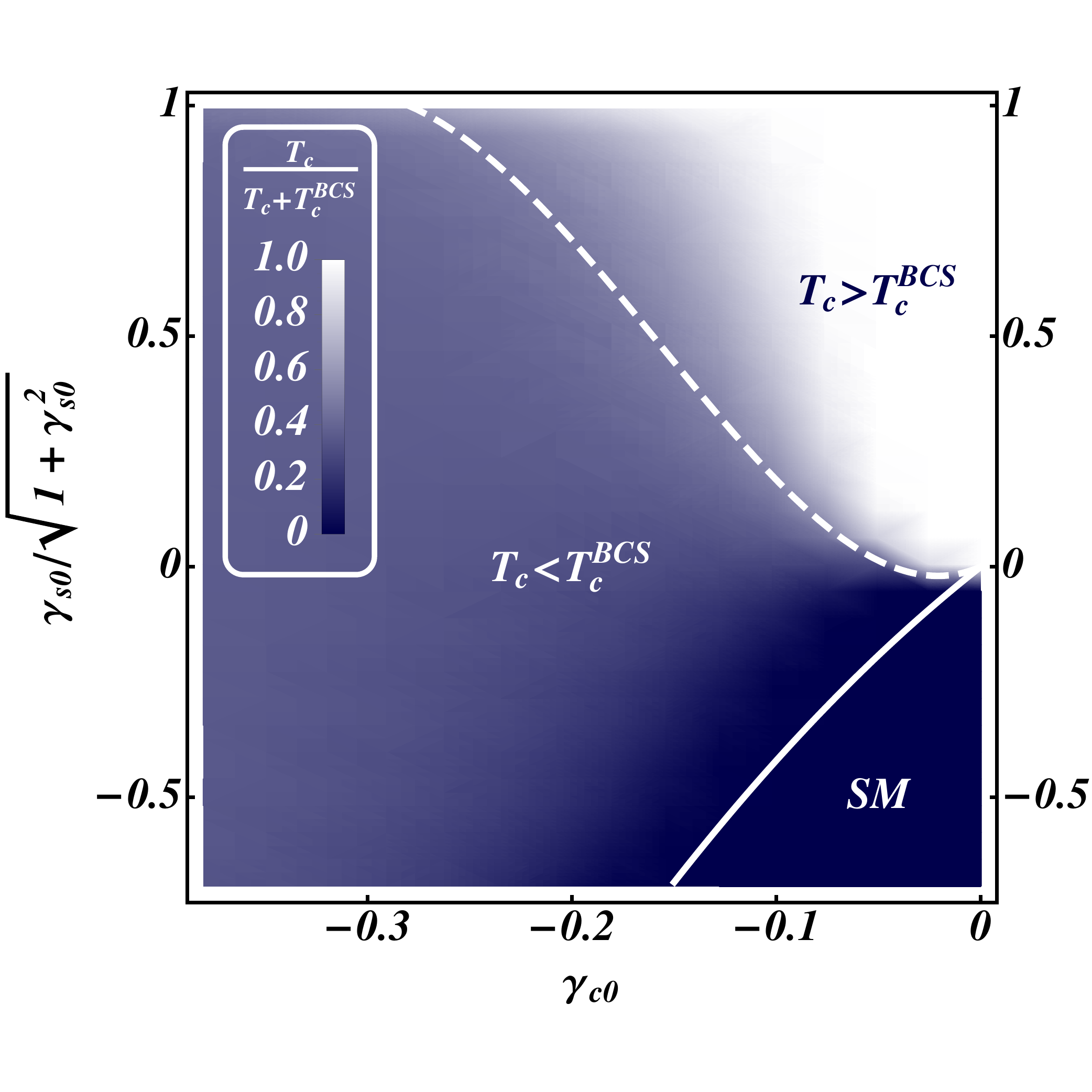}}
\caption{(Color online) The case of broken spin rotational symmetry with short-ranged interaction: the color plot for the ratio of $T_c/(T_c+T_c^{BCS})$ for $t_0=0.11$. The dashed curve separates the regions with $T_c<T_c^{BCS}$ and with $T_c>T_c^{BCS}$. The solid line indicate the boundary between superconductor and supermetal. The figure is obtained from numerical solutions of RG Eqs. \eqref{eq:rg:final:t:SO:SR} - \eqref{eq:rg:final:gc:SO:SR}.}
\label{fig:RGSO+SRI+Tc}
\end{figure}

A typical dependence of  $t$ on the length scale $L$ across the SC-SM transition governed by increase of disorder ($t_0$) is illustrated in Fig. \ref{fig:RGSO+SRI+R} for some initial values of $\gamma_{s0}$ and $\gamma_{c0}$. As shown in the inset, in this case (dot-dashed line) the disorder suppresses the transition temperature. However, with increase of $\gamma_{s0}$ one finds a nonmonotonous dependence of $T_c$ on $t_0$: at weak disorder $T_c$ is reduced in comparison with $T_c^{BCS}$ whereas at intermediate disorder $T_c$ is larger than $T_c^{BCS}$. There is the region in the phase diagram with small values of $\gamma_{c0}<0$ and $\gamma_{s0}<0$ in which $T_c>T_c^{BCS}$ (or more precisely, $L_X < L_c^{BCS}$).  The enhancement of $T_c$ in a certain range of bare couplings
is in agreement with conclusions of our work [\onlinecite{PRL2012}] where renormalization group equations \eqref{eq:rg:final:t:SO:SR} - \eqref{eq:rg:final:gc:SO:SR} with the right hand sides expanded to the lowest nontrivial order in $\gamma_s$ and $\gamma_c$ were analyzed. In fact, for attraction in the particle-hole channel, $\gamma_{s0}>0$, a significant part of the phase diagram is occupied by the superconducting phase with $T_c > T_c^{BCS}$, see Fig. \ref{fig:RGSO+SRI+Tc}.

The mechanism of enhancement of the transition temperature is similar to one for the case of preserved spin rotational symmetry. For small initial values of $\gamma_{c0}$ and $\gamma_{s0}$ the renormalization of the Cooper interaction amplitude occurs in two distinct steps. At the first step of RG flow, the interaction is renormalized due to the presence of disorder (the terms proportional to $t$), while at the second step the standard BCS-type renormalization  (the term $2 \gamma_c^2$) takes place. In the case $|\gamma_{s0}|, |\gamma_{c0}| \ll t_0 \ll 1$ at the first step of renormalization
we can linearize the RG equations \eqref{eq:rg:final:t:SO:SR}-\eqref{eq:rg:final:gc:SO:SR} in interaction parameters and neglect the term $-2\gamma_c^2$. Then, for $\gamma_{c0}<\gamma_{s0}/2$ in the course of RG flow, the interaction amplitudes approach the BCS line $\gamma_s=-\gamma_c$, thus converting the repulsion in the singlet particle-hole channel into attraction. This is the consequence  of the (weak) multifractality of the noninteracting fixed point.

At some scale $L_1$ such that $\ln L_1/l = 2/t-2/t_0$ the interaction couplings become of the order of the resistance: $|\gamma_{s,c}|\sim t$. The resistance at this scale is $t(L_1) \sim [t_0 (\gamma_{s0}-2\gamma_{c0})]^{1/2} \ll t_0$. We note that if the length scale $L_X$ is reached before $L_1$ which is typical for $t_0 \lesssim \max\{|\gamma_{c0}|,|\gamma_{s0}|\}$ one can find suppression of the transition temperature instead of enhancement (see solid curve in the inset to Fig. \ref{fig:RGSO+SRI+R}). After the length scale $L_1$ the second step of RG flow starts, where in Eq.~\eqref{eq:rg:final:gc:SO:SR} one can neglect terms proportional to $t$ compared to the disorder-independent term $-2\gamma_c^2$. Thus, the Cooper interaction $\gamma_c$ flows according to the standard BCS RG equation for a clean system with the initial value $\gamma_c(L_1) \sim t(L_1)$ rather than $\gamma_{c0}$.
Hence, we find the following rough estimate for the transition temperature: $\ln 1/(T_c\tau) \sim 2\ln L_1/l + 1/|\gamma_c(L_1)| \sim [t_0 (\gamma_{s0}-2\gamma_{c0})]^{1/2}  \gg \ln 1/(T_c^{BCS}\tau)$, see Appendix \ref{AppTcEnh} for details.

In full analogy with the case of preserved spin-rotation invariance,
if the disorder scattering rate $1/\tau$ exceeds the Debye frequency $\omega_D$, the starting point of the RG flow will be likely located not far from the  BCS line, $\gamma_{s0}=-\gamma_{t0}=-\gamma_{c0}$ (see Sec. \ref{InitialBCS}). For such initial conditions, the dependence of $T_c/(T_c+T_c^{BSC})$ on $t_0$ is shown in the inset to Fig. \ref{fig:RGSO+SRI+R} by dashed curve. As in the case of preserved spin invariance, for bare interaction parameters on the BCS line, there is no initial decrease of transition temperature with increase of $t_0$.
For a given $\gamma_{c0}$, a deviation from the BCS line in the initial conditions  towards larger (smaller) values of $\gamma_{s0}$ increases (decreases) the relative enhancement of the transition temperature, $T_c/T_c^{BSC}$. We emphasize that inspite of the antilocalization at the noninteracting fixed point the multifractality enhances the interaction in the Cooper channel.

\section{Resistance in zero magnetic field}
\label{s5_0}

Within the NLSM approach, physical observables
can be written as correlation functions of the matrix field $Q$. In particular, the conductivity obtained by evaluating a linear response to an electromagnetic field in the framework of the NLSM theory with the action \eqref{eq:NLSM}  can be expressed in the following way:
\begin{align}
\sigma(i\omega_n) = & -\frac{g}{16 n} \Bigl \langle \Tr [J_n^\alpha,Q(\bm{r})] [J_{-n}^\alpha,Q(\bm{r})] \Bigr \rangle \notag \\
& +\frac{g^2}{128 n} \int d\bm{r}^\prime \Bigl \langle \Tr J_n^\alpha Q(\bm{r}) \nabla Q(\bm{r})
\notag \\
& \hspace{1.5cm}\times
\Tr J_{-n}^\alpha Q(\bm{r}^\prime) \nabla Q(\bm{r}^\prime) \Bigr \rangle  .
\label{eq:PO:g}
\end{align}
Here $\omega_n=2\pi Tn$ is a Matsubara frequency,
expectation values are defined  with respect to the action \eqref{eq:NLSM}, and
\begin{equation}
J_n^\alpha = \frac{t_{30}-t_{00}}{2} I_n^\alpha + \frac{t_{30}+t_{00}}{2} I_{-n}^\alpha .
\end{equation}
As usual, the static conductivity $\sigma$ can be obtained after the analytic continuation of Eq. \eqref{eq:PO:g} to real frequencies: $i\omega_n \to \omega+i0^+$ and, then, taking the limit $\omega\to 0$.
At the classical level, $Q=\Lambda$, one finds $\sigma = g$.

\begin{figure*}[t]
\centerline{(a) \includegraphics[width=5cm]{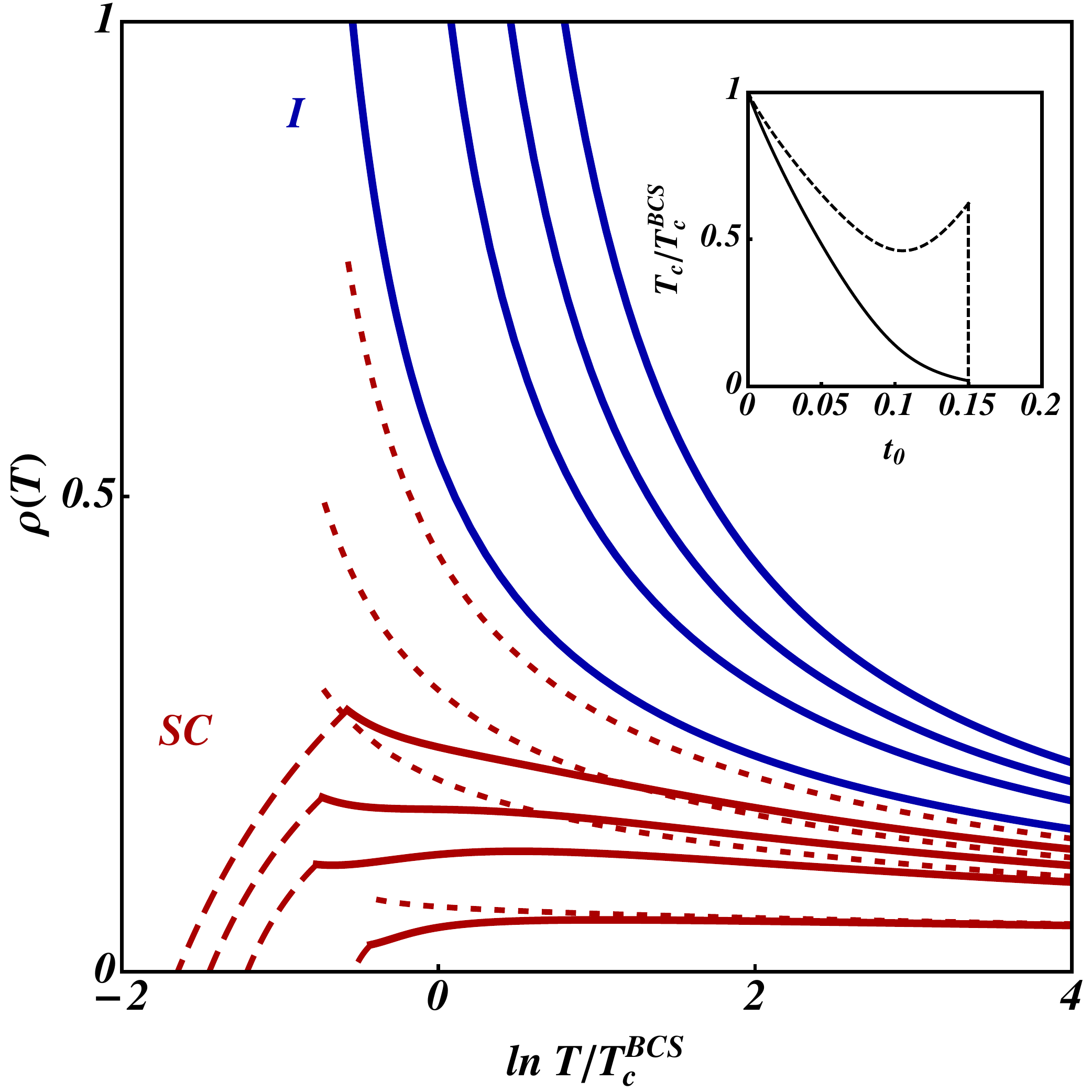} \qquad (b)\includegraphics[width=5cm]{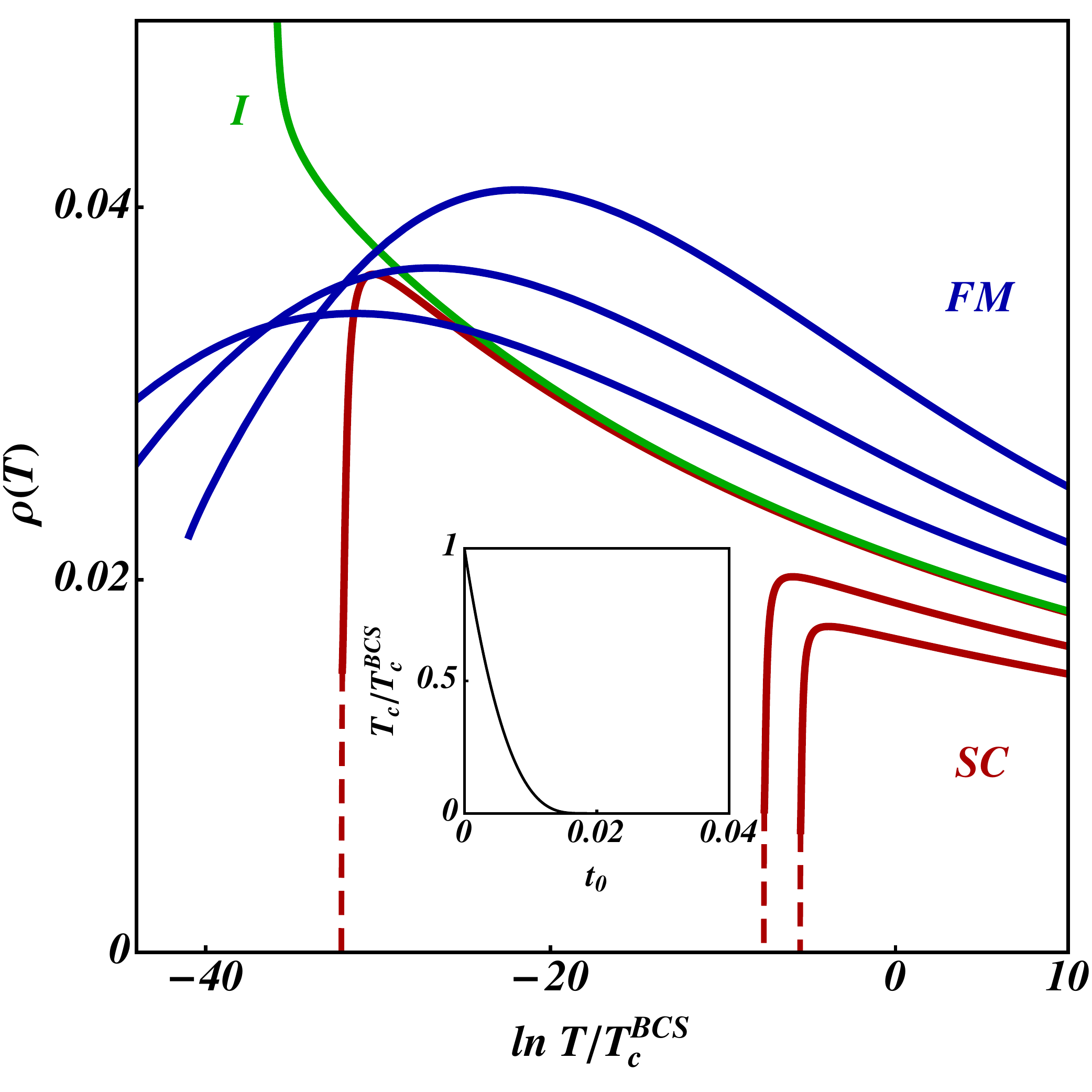} \qquad (c) \includegraphics[width=5cm]{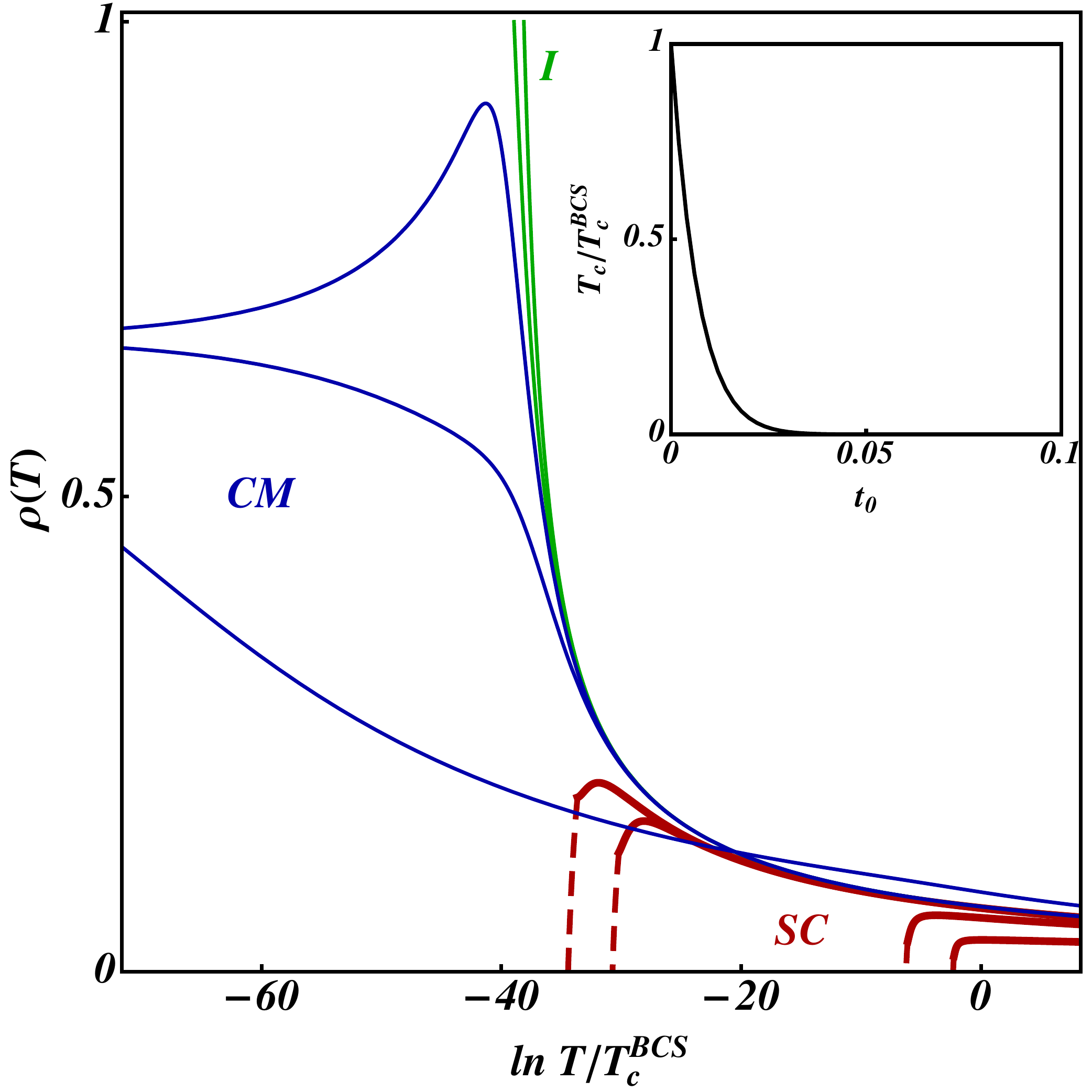}}
\centerline{(d) \includegraphics[width=5cm]{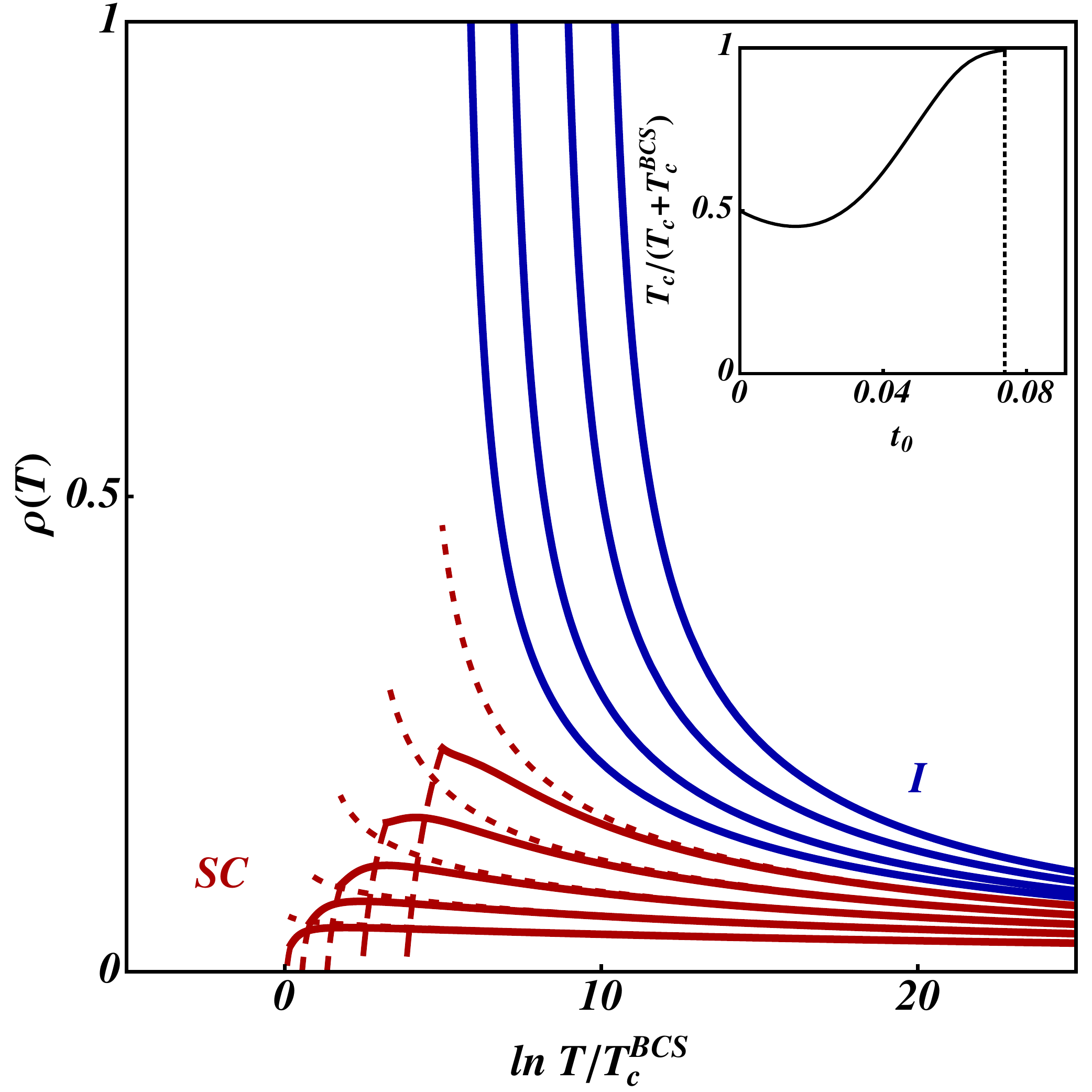}\qquad (e) \includegraphics[width=5cm]{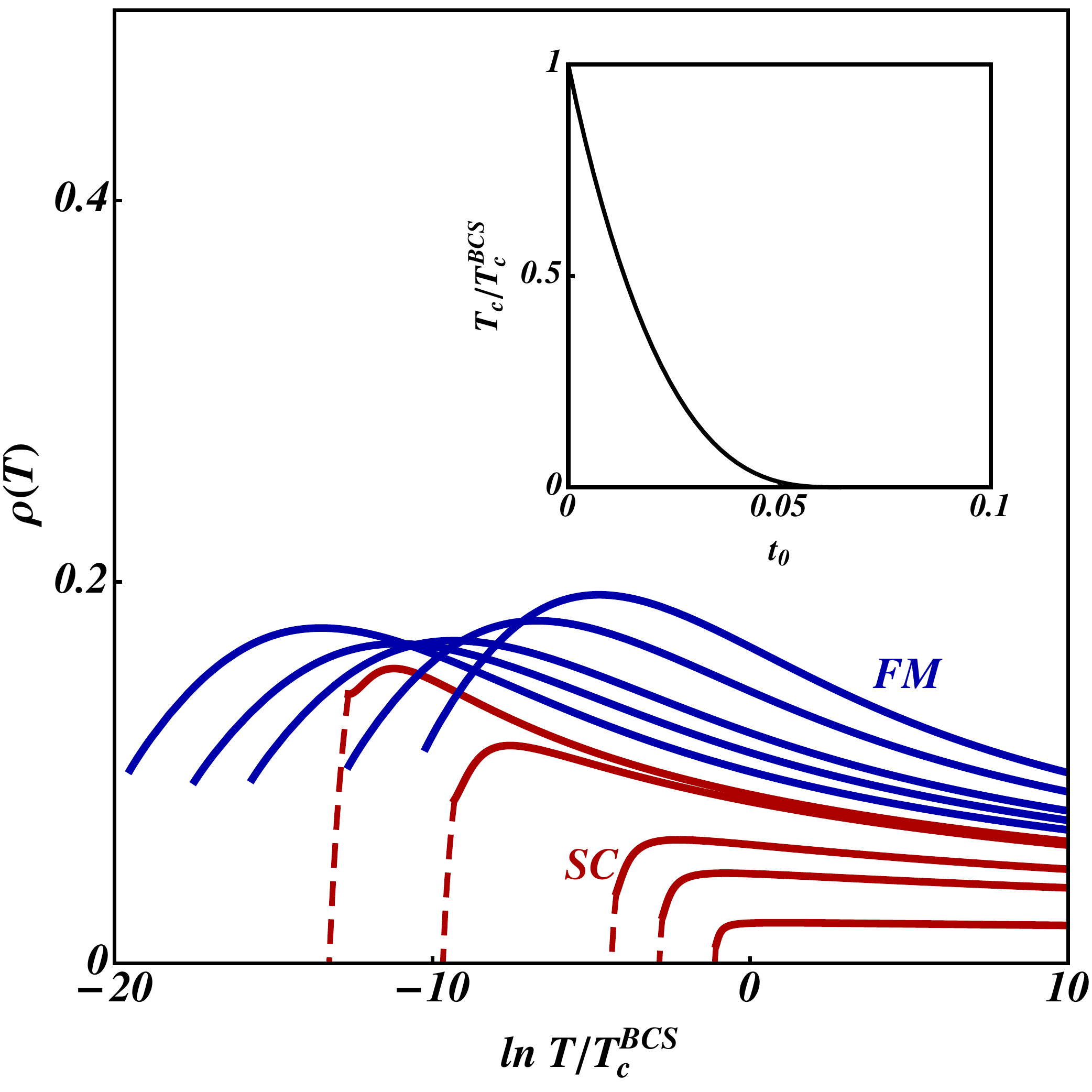} \qquad (f) \includegraphics[width=5cm]{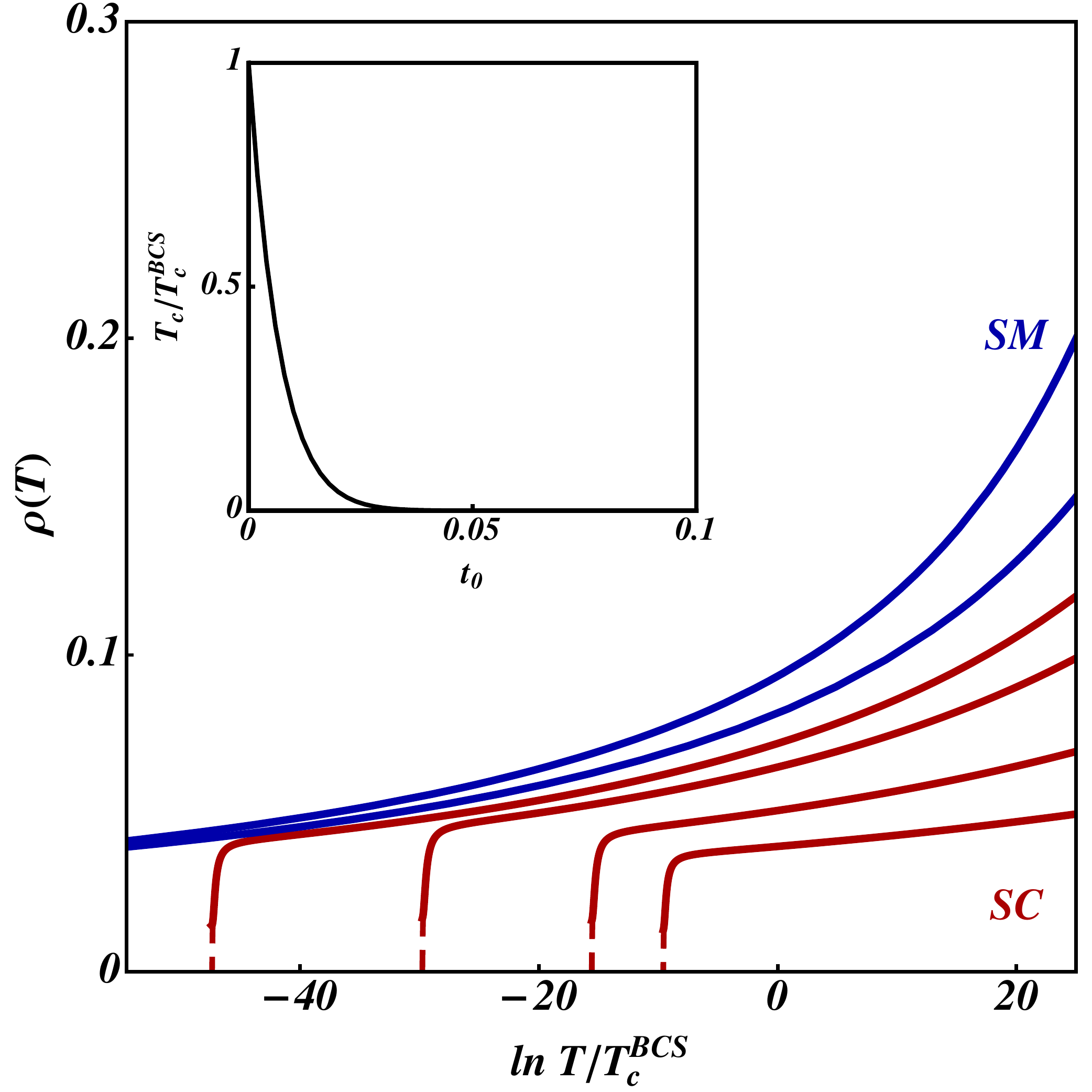}}
\caption{(Color online) Dependence of the resistance $\rho$ on temperature $T$ and $T_c$ on $t_0$ for the cases of Coulomb [(a), (b) and (c)] and short-ranged [(d), (e) and (f)] interaction. Parameters are the same as in Figs. \ref{fig:RGO:PD:Res:Ad}, \ref{fig:RGO:PD:Res}, \ref{fig:RGO:PD2:SIT}, \ref{fig:RGSO+Coulomb+Res}, \ref{fig:RGO:PD2:SIT:a} and \ref{fig:RGSO+SRI+R}, respectively. Figures (a), (b), (d) and (e) correspond to the case of preserved spin rotational symmetry, (c) and (f) --  broken spin rotational symmetry calculated from Eq. \eqref{eq:sigma:1}. The dashed parts of the curves correspond to the ``fluctuation'' region $T<T_X$. Dotted red curves in panels (a) and (d) show temperature dependences of $t$ for $T>T_X$.
The dependence of $T_c$ on $t_0$ is shown in the insets (solid curve). The dependence of $T_X$ on $t_0$ is shown by dashed line in the inset to figure (a). For all other insets $T_X$ coincides with $T_c$ within our accuracy.
}
\label{fig:RG:Res:vs:Cond}
\end{figure*}

The RG equations derived in this work describe  renormalization of
the couplings in the NLSM action with the running spatial scale $L$.
As such, these equations yield  physical observables (e.g., the resistivity)
of a finite size sample at $T=0$. At finite temperature $T$, conductivity can be evaluated in two steps. At the first step, the action \eqref{eq:NLSM} is renormalized from the energy scale $1/\tau$ down to $T$. Thus the bare parameters in the action \eqref{eq:NLSM} are substituted by the parameters at the length scale $L_T$: $g \to 2/[\pi t(L_T)]$, $\gamma_{s,t,c} \to \gamma_{s,t,c}(L_T)$, and $Z_\omega \to Z_\omega(L_T)$. Their dependence on $L_T$ is governed by RG equations \eqref{eq:rg:final:t:G} - \eqref{eq:rg:final:z:G}. At the second step, the Kubo formula \eqref{eq:PO:g} is evaluated under assumption that in the NLSM  action \eqref{eq:NLSM} $Q$ fields are restricted by the temperature in the ultraviolet. We assume $L_T$ to be such that $t(L_T) \ll 1$ and $|\gamma_c(L_T)|\gg 1$. Then, the conductivity can be written as
\begin{equation}
\sigma(T) \simeq \frac{2}{\pi t(L_T)} - \frac{\pi}{2} \gamma_c(L_T) \ln \frac{L_\phi}{L_T} .
\label{eq:sigma:1}
\end{equation}
This result illustrates the fact that at finite temperature there is always a difference  between the physical
resistance $\rho = 1/\sigma$ and the coupling parameter $t$ in the NLSM action.
The second term in Eq.~\eqref{eq:sigma:1} is the Maki-Thompson contribution [\onlinecite{M1968,T1970,LO2001,Levchenko2010}], which is the dominant inelastic contribution to the resistivity. (We neglect the smaller Aslamazov-Larkin contribution [\onlinecite{AL1968}].)

Comparing two terms in Eq.~\eqref{eq:sigma:1}, we find that the inelastic contribution becomes of the same order as the renormalized impurity-scattering conductivity (given by the NLSM coupling) at the scale $L_X$ determined by the condition  $t(L) |\gamma_c(L)| = 1$.  Remarkably, this is the same scale as is found from the condition of validity of the one-loop RG, see Sec.~\ref{sec:one-loop}. Thus, at scales shorter than $L_X$, or, equivalently, at temperatures larger than $T_X$, the temperature dependence of the physical resistivity is dominated by the
RG behavior of $t(L)$.  Upon approaching the transition temperature, the role of the inelastic contribution controlled by $\gamma_c(L)$
increases. In the narrow temperature interval $T_c < T < T_X$ the conductivity is dominated by the inelastic contribution. As we have already discussed in Sec.~\ref{s4}, the width of this interval is
\begin{equation}
(T_X-T_c)/T_c\sim t(T_X).
\label{eq:TX}
\end{equation}
At the transition point $T_c$ the running coupling $\gamma_c$ diverges, $|\gamma_c(T_c)|=\infty$ and Eq.~\eqref{eq:sigma:1} formally yields $\rho(T_c)=0$.

In fact, the behaviour of the conductivity in the fluctuation region, $|T-T_c|/T_c\sim t(T_X)$, is additionally affected by superconducting phase fluctuations. These fluctuations lead to the BKT character of the actual superconducting phase transition.
The corresponding shift of the transition temperature is, however, small (as has been already mentioned in the end of
Sec.~\ref{subsec:preserved-coulomb}) and is not important for our results.
More specifically, as was argued by Beasley, Mooij, and Orlando [\onlinecite{BKT}], the shift is determined by disorder
strength, $(T_c-T_\text{BKT})/T_c\sim t$, for $t\ll 1$, see also Refs. [\onlinecite{HalperinNelson79,BKT1}].
The behavior of the true $\rho(T)$ in the fluctuation (BKT) region will be addressed elsewhere [\onlinecite{future-Elio}].
Let us only mention here that the value of the NLSM coupling $t(T_X)$ at the entrance to the fluctuating region
determines [\onlinecite{future-Elio}] the stiffness of the phase fluctuations in the BKT region $T_\text{BKT}<T<T_X$
and the relevant value of $t$ in the shift of $T_\text{BKT}$ with respect to $T_c$. Because of the renormalization of $t$,
this value may strongly differ from the bare (high temperature) Drude value of the resistance $t_0$ (cf. Ref. [\onlinecite{BatVinBKT}]).

Ignoring the above-mentioned subtleties of the behavior of $\rho(T)$ in the narrow fluctuation region
$|T-T_c|/T_c\sim t(T_X)$ around the mean-field $T_c$, the behavior of the electrical resistance in the whole range of temperatures is well described by
$\rho(T)=1/\sigma(T)$ with $\sigma(T)$ given by Eq.~\eqref{eq:sigma:1}. To relate the length scale $L_c$ with the transition temperature $T_c$
one needs to know the relevant value $D_c$ of the diffusive coefficient $D\sim 1/(t Z_\omega)$. It is obtained by using the values for $t$ and $Z_\omega$ at the scale $L_X$ (which is the border of validity of the one-loop RG and simultaneously is the beginning of the fluctuation region). The transition temperature is given by $T_c \sim D_c L_c^{-2}$, see Eq. \eqref{eq:TL_T}.

We present the dependence of $\rho(T)$ obtained in accordance with Eq. \eqref{eq:sigma:1} (for simplicity, we dropped the logarithmic factor in the inelastic term; this does not qualitatively affect the plots) for the cases of preserved and broken spin rotational symmetry and Coulomb and short-ranged interactions in Fig. \ref{fig:RG:Res:vs:Cond}.
We mention that due to Maki-Thompson correction the resistance drops very fast (dashed curves in Fig. \ref{fig:RG:Res:vs:Cond}) since $|\gamma_c(L_X)| \gg 1$ and, consequently, $(T_c - T_X)/T_X \sim \rho(T_X) \ll 1$. Therefore, the temperature $T_X$, at which resistivity has the maximum, can be used as an estimate of the superconducting transition temperature. In order to plot $\rho(T)$ in the fluctuation region $T<T_X$ where the one-loop RG becomes insufficient, we evaluated $\gamma_c(T)$ near $T_c$ keeping only the BCS term $-2\gamma_c^2$ in the RG equation for $\gamma_c$.

We note that our approach for evaluating temperature dependence of resistivity is different form that of Refs. [\onlinecite{GVV2010,KSF2012}]. In these works the full set of the first-order perturbative quantum corrections to the conductivity was computed at finite temperature in the presence magnetic field.
However, such approach assumes that $T$ and $H$ dependent corrections to the bare Drude conductivity $1/t_0$ are small. In our approach we split effects leading to temperature variation of the conductivity into two parts: those related to virtual and real processes. Then virtual processes are taken into account within the RG formalism (they are included in the renormalisation of $t$). This allows us to consider also situations with strong renormalisation, $t(L_T) \gg t_0$, i.e.,  cases in which ``quantum corrections'' are large in comparison with $1/t_0$.

\section{Magnetoresistance}
\label{s5}

A transverse magnetic field introduces an additional length scale $l_H$ (magnetic length) into the problem. In what follows we assume that it is larger than the mean free path, $l<l_H$. Let us start from $T=0$. In case of weak magnetic field, $l_H \gg L_c$, the superconducting instability at $L=L_c$ remains unaffected. For strong magnetic fields, $l_H \ll L_c$, the superconducting phase is destroyed by magnetic field since the growth of $|\gamma_c|$ within RG equations is stopped at the length scale $l_H$. Then the critical magnetic field can be estimated as $l_{H_c}=L_c$.
This results in the standard relation, $H_c \sim T_c/D_c$.
At $H<H_c$ the physical resistance $\rho$ should vanish in the infinite system, $L=\infty$.
For systems of finite size $L>L_c$, the resistance is not zero and is determined by the nontrivial configurations of the
order parameter in the presence of magnetic field.

At finite temperature, the critical magnetic field is a function of temperature. However, at $T\ll T_c$ this effect is small and can be neglected.
In order to evaluate the magnetoresistance at $T\ll T_c$ and for $H>H_c$ we use a two-step RG procedure. At the first step of RG the interaction in the Cooper channel grows towards instability. The first step ends at the length scale $l_H$ with some values $t(l_H)$ and $\gamma_{s,t,c}(l_H)$.
At $L>l_H$ the cooperon modes become ineffective and the second step of the RG procedure starts.
For length scales $l_H<L<L_T$  RG equations do not contain cooperon contributions:
\begin{align}
\frac{d t}{dy} & = t^2 \Bigl [ f(\gamma_s)+n f(\gamma_t)\Bigr ] , \label{eq:rg:final:t:G:Hperp}\\
\frac{d\gamma_s}{dy}  & = - \frac{t}{2} (1+\gamma_s)\bigl ( \gamma_s+n \gamma_t\bigr ), \label{eq:rg:final:gs:G:Hperp} \\
\frac{d\gamma_t}{dy}  & = - \frac{t}{2} (1+\gamma_t) \Bigl [ \gamma_s-(n-2)\gamma_t \Bigr ], \label{eq:rg:final:gt:G:Hperp} \\
\frac{d\ln Z_\omega}{dy} & = \frac{t}{2} \Bigl (\gamma_s+n\gamma_t\Bigr ) .\label{eq:rg:final:z:G:Hperp}
\end{align}
Here $y= \ln L/l_H$ and the initial values of couplings in Eqs. \eqref{eq:rg:final:t:G:Hperp} - \eqref{eq:rg:final:z:G:Hperp} are given by $t(l_H)$ and $\gamma_{s,t,c}(l_H)$. Similarly to the case of zero magnetic field, in order to find the physical resistance $\rho(T,H)$  one needs to take into account non-RG corrections to conductivity due to superconducting fluctuations. In particular, for $|\gamma_c(l_H)|\gg 1$ the leading-oder correction (which is the dominant one for magnetic fields not too close to the critical field $H_c$) [\onlinecite{GL2001}] yields:
\begin{equation}
\sigma(T,H) = \frac{2}{\pi t(L_T)} - \frac{4}{3\pi} \ln |\gamma_c(l_H)| .
\label{eq:mag:1}
\end{equation}
It should be emphasized that $t(L_T)$ in Eq.~\eqref{eq:mag:1} in fact depends  on the magnetic field via the two-step RG procedure.
At $H=H_c$ the physical resistivity $\rho(T,H)$ should vanish. This happens due to higher-order Cooper-channel corrections of non-RG type in Eq. \eqref{eq:mag:1} which become important in the fluctuation region (i.e., near $H_c$) and make $\rho(T,H_c)=0$ in spite of finite value $t(L_T)$.

Let us mention that, due to renormalization on ballistic scales, $L<l$, [\onlinecite{ZNA1}], one can also expect an effect of magnetic field on the initial values of parameters ($t_0$, $\gamma_{c0}$, $\gamma_{t0}$ and $\gamma_{s0}$) for RG equations \eqref{eq:rg:final:t:G} - \eqref{eq:rg:final:z:G}. This leads to a shift of the length scale at which $\gamma_c$ diverges. As a consequence, an additional dependence of transition temperature on $H$ appears. We do not take this effect into account.

\begin{figure}[t]
\centerline{(a) \includegraphics[width=3.5cm]{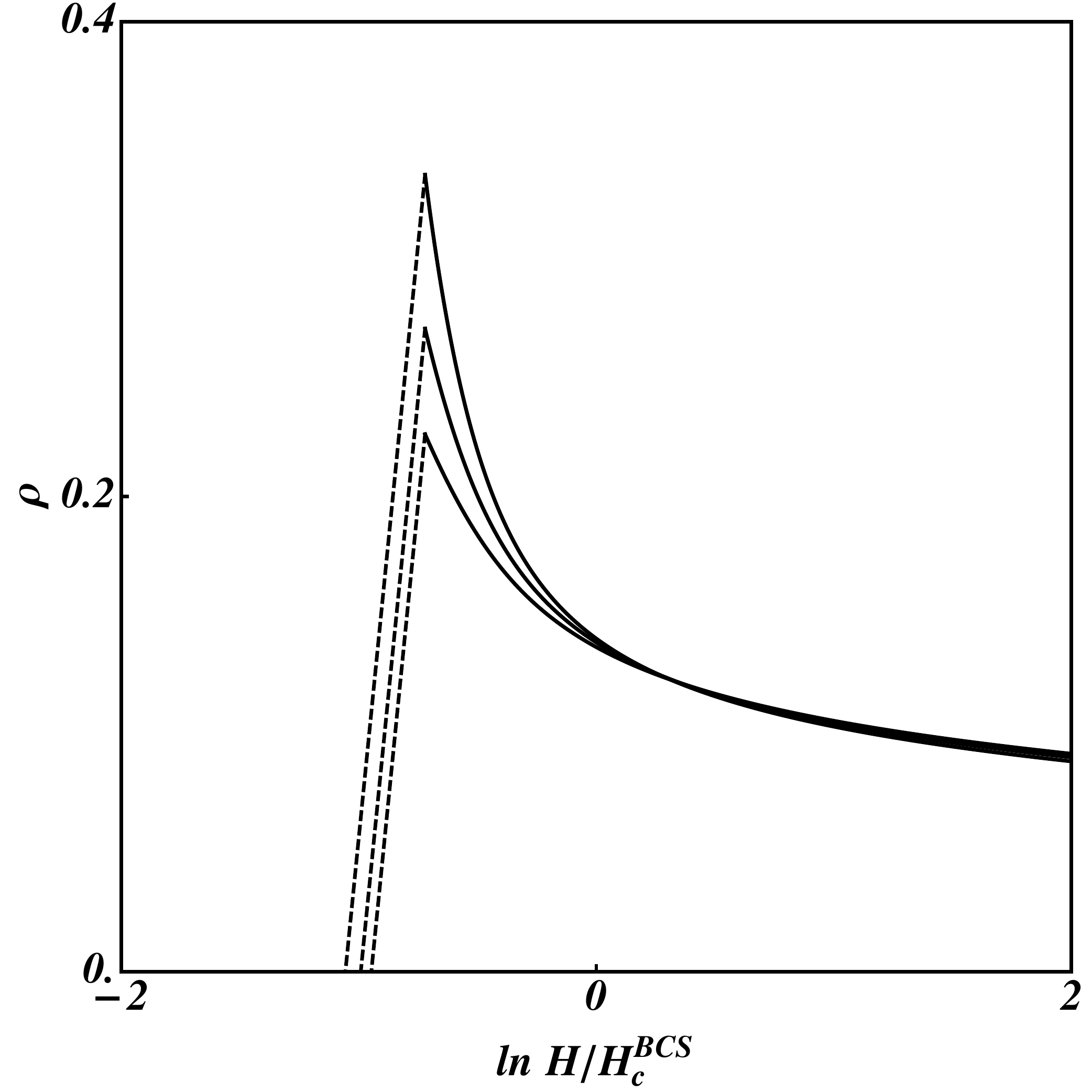}\quad (b) \includegraphics[width=3.5cm]{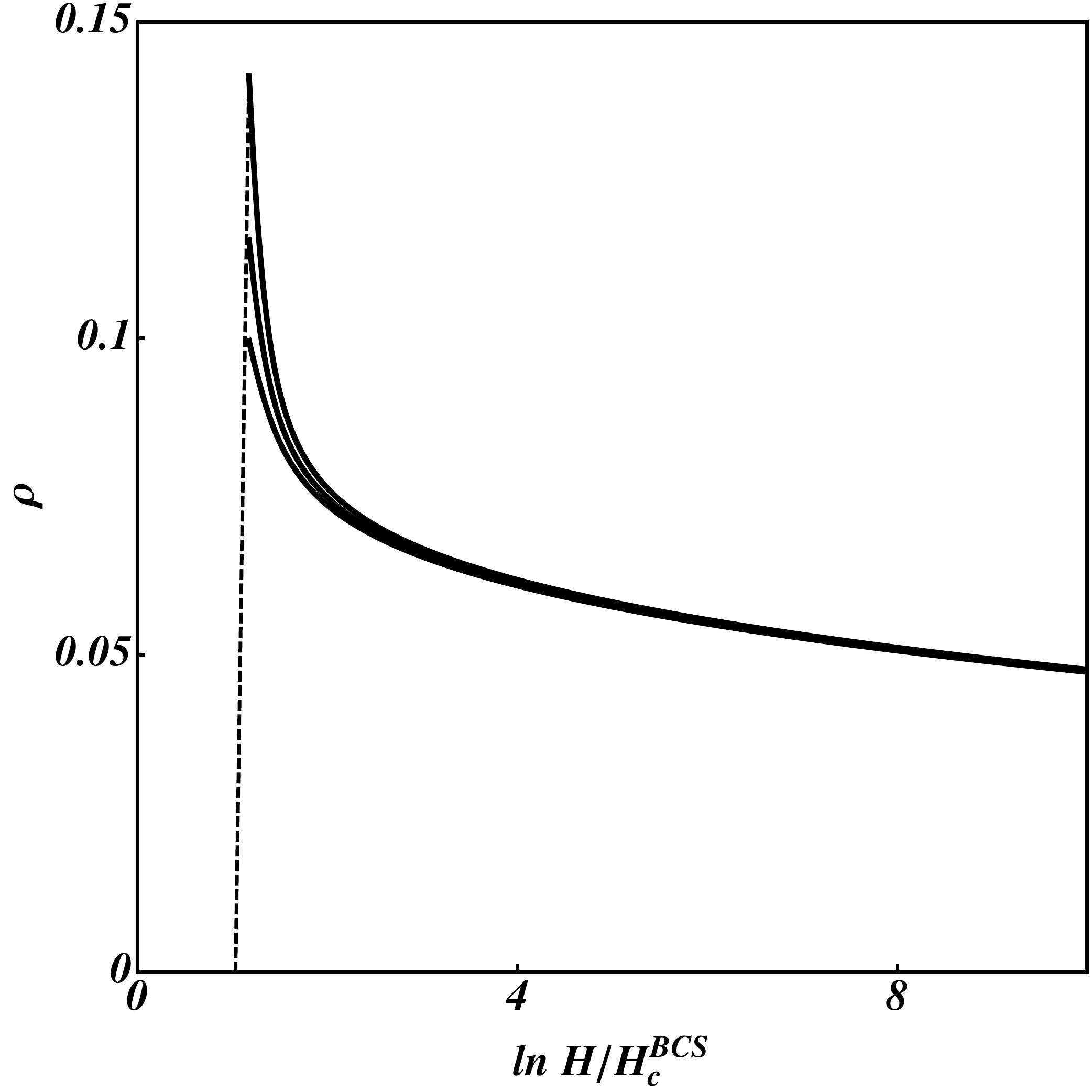}}
\centerline{(c) \includegraphics[width=3.5cm]{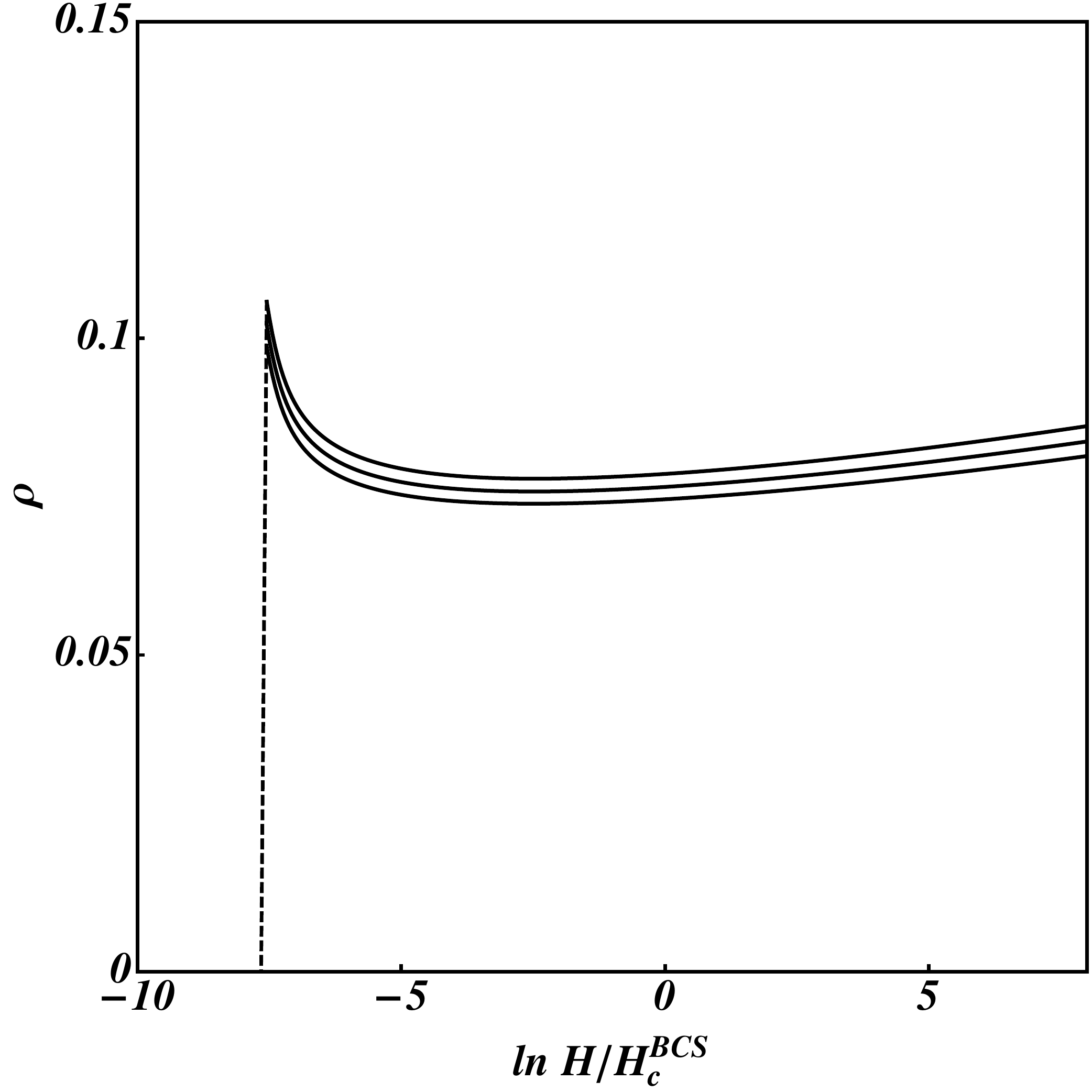}\quad (d) \includegraphics[width=3.5cm]{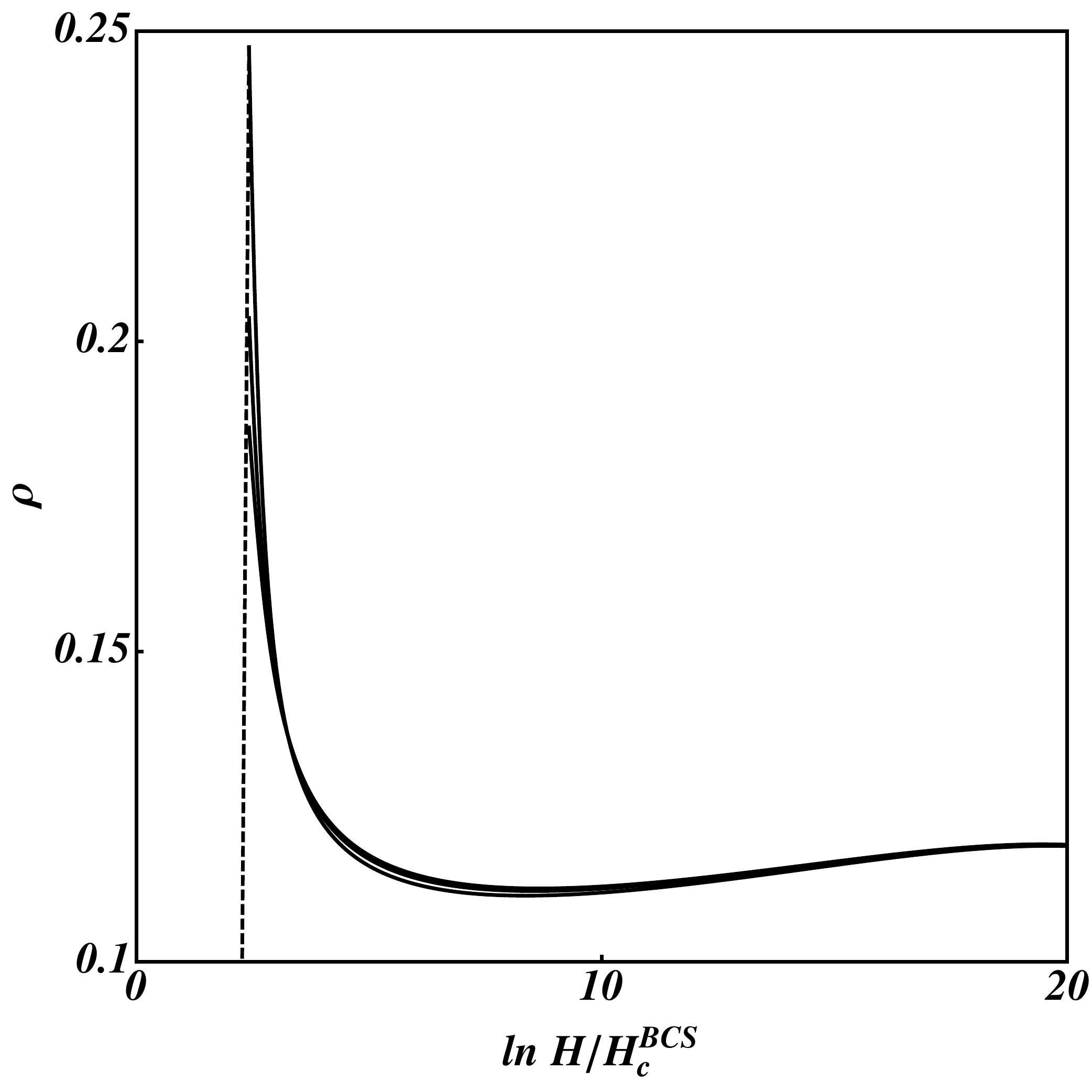}}
\caption{Dependence of the resistance $\rho$ on perpendicular magnetic field $H$ for the cases of preserved [(a) and (b)] and broken [(c) and (d)] spin rotational symmetries. Figures (a) and (c) correspond to the case of Coulomb interaction, (c) and (d) --  short-ranged interaction. Solid curves are obtained in from Eq. \eqref{eq:mag:1}. The dashed lines indicate the drop of magnetoresistance to zero at $H=H_c$. The parameters used are as follows
(a) $\gamma_s=-1$, $\gamma_{c0}=-0.45$, $\gamma_t=1$, $t_0=0.1$, $T=T_c/2,\ T_c/4,\ T_c/8$,
(b) $\gamma_{s0}=0.1$, $\gamma_{c0}=-0.1$, $\gamma_{t0}=-0.1$, $t_0=0.05$, $T=T_c/2,\ T_c/4,\ T_c/16$
(c) $\gamma_s=-1$, $\gamma_{c0}=-0.12$, $t_0=0.053$, $T=T_c/2,\ T_c/4,\ T_c/8$, and
(d) $\gamma_{s0}=0.05$, $\gamma_{c0}=-0.05$, $t_0=0.2$, $T=T_c/2,\ T_c/16,\ T_c/256$.
}
\label{fig:RG:Res:vs:Cond:H}
\end{figure}

The dependence of the resistivity on perpendicular magnetic field $H$ at different temperatures below $T_c$ is illustrated in  Fig. \ref{fig:RG:Res:vs:Cond:H}. The $H$ dependence of resistivity at fixed $T < T_c$ is qualitatively similar to the $\rho(T)$ dependence at zero magnetic field. For all four symmetry classes the magnetoresistance shows a maximum that grows with decreasing temperature.
It is worth mentioning that the maximum should become arbitrarily high as $T\to 0$, yielding a giant magnetoresistance in agreement with experimental observations. We do not plot these curves here since  we cannot controllably evaluate the resistance for $t\agt 1$ within the one-loop RG.

In addition to $l_H$, the perpendicular magnetic field induces another length scale $l_Z$ related to the Zeeman splitting. Usually one expects that $l_H \ll l_Z$ since the latter can be estimated as
$l_Z \sim l_H/ \sqrt{(1+\gamma_t(l_Z)) t(l_Z) g_L }$. Here $g_L$ stands for the Land\'e g-factor. Due to Zeeman splitting the magnetic field suppresses the triplet diffuson modes with $S_z=\pm 1$ and singlet and triplet cooperon modes with $S_z=0$ at length scales $L>l_Z$.

In the case of fully broken spin-rotational symmetry there is no triplet modes, and thus the orbital and Zeeman effects of magnetic field on RG equations are the same. Therefore, there is in fact only one length scale associated with magnetic field $l_{HZ}=\min\{l_H,l_Z\}$. A similar conclusion holds for the case of partially broken spin-rotational symmetry, $n=1$. Therefore, in the absence of spin-rotational symmetry we do not need to consider the effect of Zeeman splitting separately.

\begin{figure}[t]
\centerline{(a) \includegraphics[width=3.5cm]{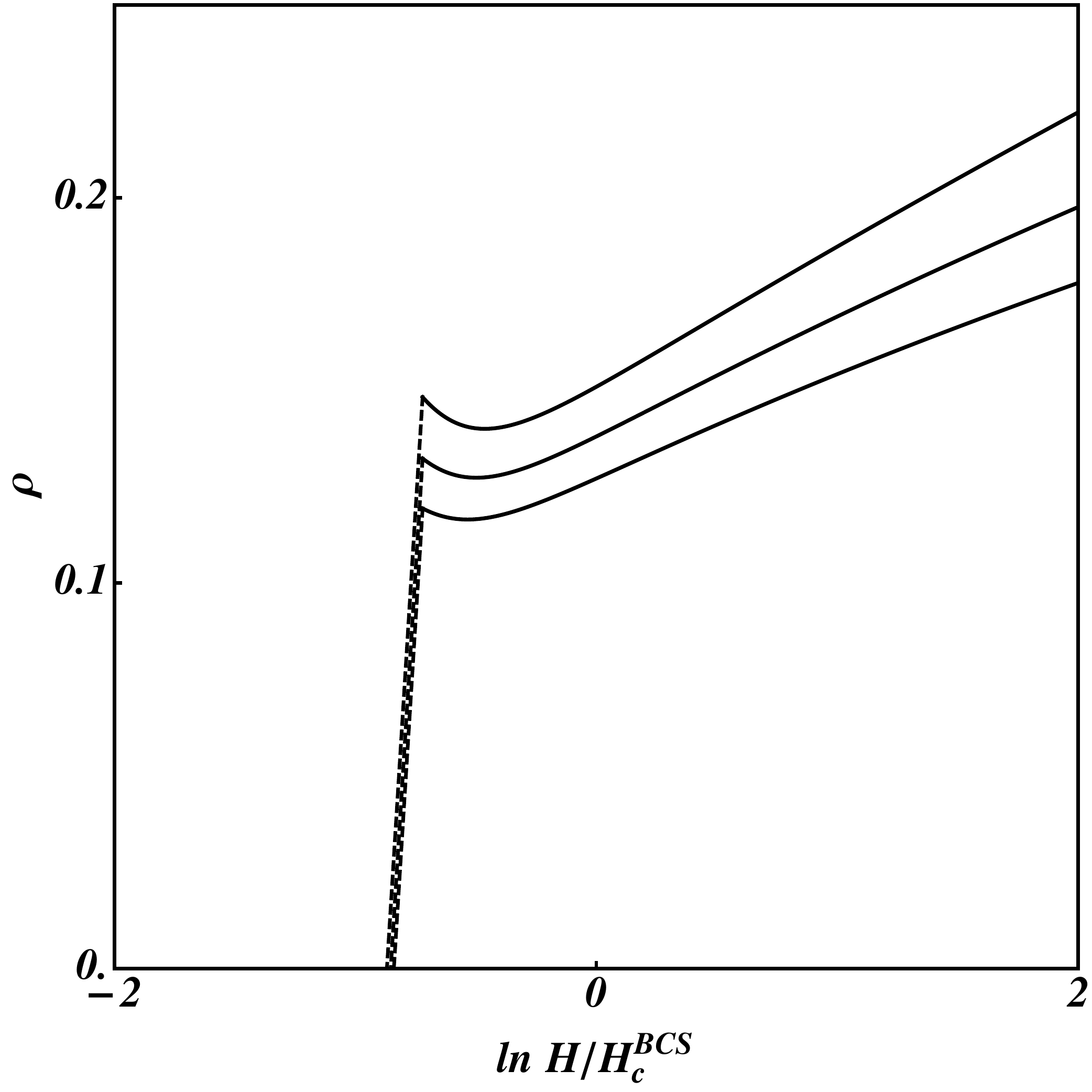}\quad (b) \includegraphics[width=3.5cm]{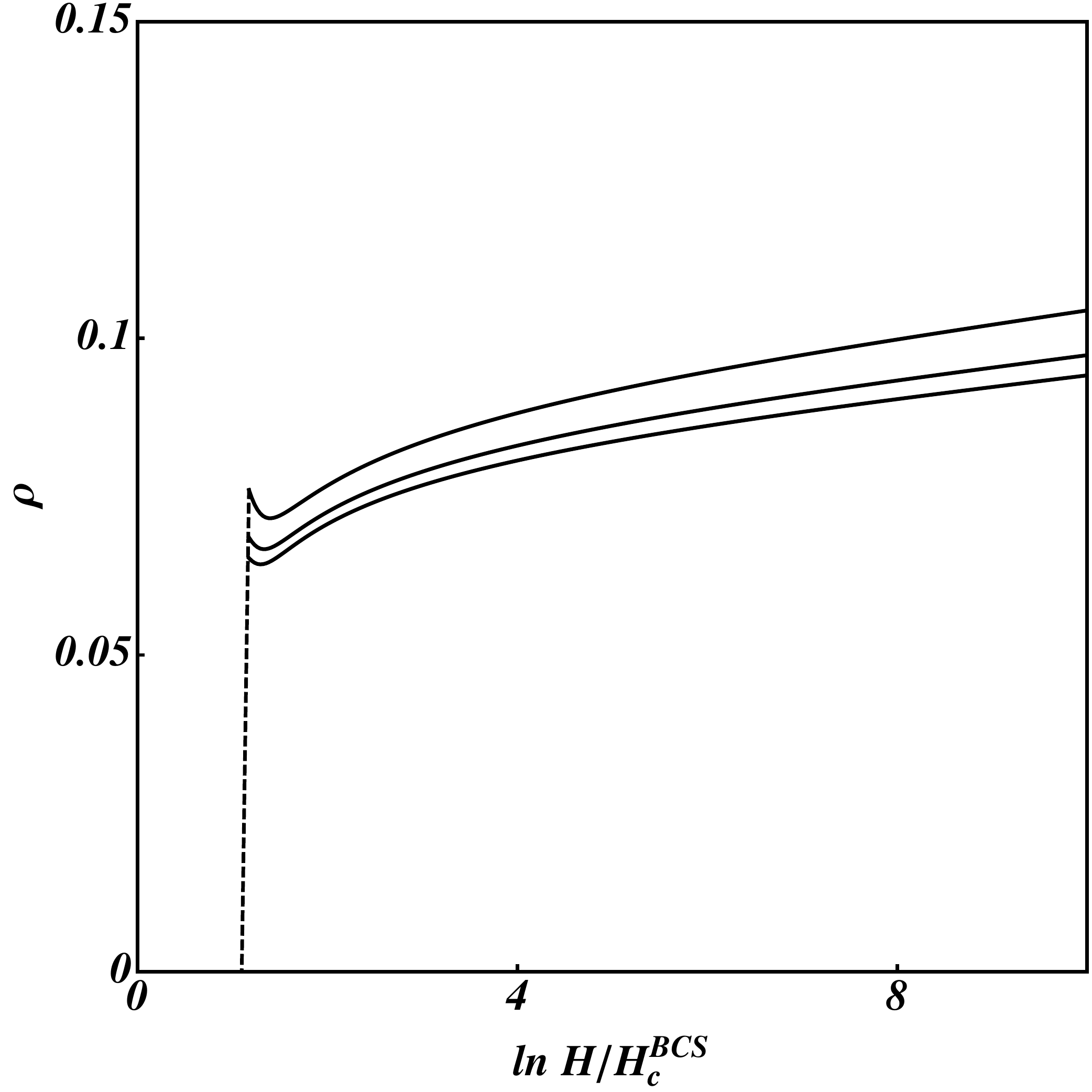}}
\caption{Dependence of the resistance $\rho$ on parallel magnetic field $H$ for the case of preserved spin rotational symmetry for the case of Coulomb (a) and short-ranged (b) interaction. Solid curves are obtained from Eq. \eqref{eq:mag:2}. The dashed lines indicate the drop of magnetoresistance to zero at $H=H_{cZ}$. The parameters used are as follows
(a) $\gamma_s=-1$, $\gamma_{c0}=-0.45$, $\gamma_t=1$, $t_0=0.1$, $T=T_c/2,\ T_c/4,\ T_c/8$,
(b) $\gamma_{s0}=0.1$, $\gamma_{c0}=-0.1$, $\gamma_{t0}=-0.1$, $t_0=0.05$, $T=T_c/2,\ T_c/4,\ T_c/16$
}
\label{fig:RG:Res:vs:Cond:HZ}
\end{figure}

For the case of preserved spin-rotational symmetry, the orbital and Zeeman effects are different.
For $l_H \ll l_Z$, the RG equations \eqref{eq:rg:final:t:G:Hperp} - \eqref{eq:rg:final:z:G:Hperp} should be modified at length scales $L>l_Z$ since two out of three triplet diffusive modes becomes massive and do not lead to  infrared divergences:
\begin{align}
\frac{d t}{dy} & = t^2 \Bigl [ f(\gamma_s)+ f(\gamma_t)\Bigr ] , \label{eq:rg:final:t:G:Hpar}\\
\frac{d\gamma_s}{dy}  & = - \frac{t}{2} (1+\gamma_s)\bigl ( \gamma_s+\gamma_t\bigr ), \label{eq:rg:final:gs:G:Hpar} \\
\frac{d\gamma_t}{dy}  & = - \frac{t}{2} (1+\gamma_t) \Bigl [ \gamma_s+\gamma_t \Bigr ], \label{eq:rg:final:gt:G:Hpar} \\
\frac{d\ln Z_\omega}{dy} & = \frac{t}{2} \Bigl (\gamma_s+\gamma_t\Bigr ) .\label{eq:rg:final:z:G:Hpar}
\end{align}
Here $y= \ln L/l_Z$ and the initial values of couplings in Eqs. \eqref{eq:rg:final:t:G:Hpar} - \eqref{eq:rg:final:z:G:Hpar} are given by $t(l_Z)$ and $\gamma_{s,t,c}(l_Z)$. Therefore, in this case, a three-step RG scenario is realized.

In the opposite case, $l_H \gg l_Z$, the system at length scales $l<L<l_Z$ is described by RG equations \eqref{eq:rg:final:t:G} - \eqref{eq:rg:final:z:G} (with $n=3$). Then for $l_Z<L<l_H$ RG equations transforms into
Eqs. \eqref{eq:rg:final:t:G:Hpar} - \eqref{eq:rg:final:z:G:Hpar} with $f(\gamma_s)$ substituted by $1+f(\gamma_s)$ (weak-localization correction remains intact in the presence of Zeeman splitting only). For larger length scales, $L>l_H$,  the system is governed by RG Eqs. \eqref{eq:rg:final:t:G:Hpar} - \eqref{eq:rg:final:z:G:Hpar}. The suppression of weak-localization correction does not change qualitative behavior of the RG flow.

In the case of a parallel magnetic field, the scale $l_H$ does not appear since the orbital effect of magnetic field can be neglected provided $l_Z \ll L_T\ll l_H^2/d$, where $d$ is the typical width of the film. We thus obtain a two-step RG scenario in which RG equations are modified at the length scale $l_Z$. As before, in order to find the physical resistance $\rho(T,H)$  one needs to take into account non-RG corrections to conductivity due to superconducting fluctuations. For $|\gamma_c(l_H)|\gg 1$ we get [\onlinecite{KLC2012}]
\begin{equation}
\sigma(T,H) = \frac{2}{\pi t(L_T)} +  \frac{4}{\pi} \ln |\gamma_c(l_Z)| .
\label{eq:mag:2}
\end{equation}

The parallel-field magnetoresistance for the case of preserved spin-rotational symmetry is illustrated in  Fig. \ref{fig:RG:Res:vs:Cond:HZ}
for several values of temperature below $T_c$.
The dependence of resistivity on parallel magnetic field at fixed temperate below $T_c$ is essentially different from the $\rho(H)$ dependence
in the case of transverse field. First, the maximum at an intermediate field is much less pronounced in the case of parallel field. Second, the parallel-field resistivity increases with $H$ in strong fields, contrary to the case of transverse magnetic field.

\section{Discussion}
\label{s6}

In this section, we discuss our results and their implications (in particular, for the phase diagrams of SITs),
relation to previous works, as well as limitations and possible generalizations of the RG scheme used.

\subsection{Relevant superconducting systems}

\subsubsection{Symmetry of the order parameter}

First of all, we note that our theory is derived for conventional (BCS) $s$-wave superconductors, where
the effect of $s$-wave non-magnetic impurities on the superconducting gap and $T_c$ is absent at the semiclassical level (``Anderson theorem'').
Our theory can be generalized to describe the multiband case with $s$-wave (or $s_{\pm}$-wave) pairing.
On the other hand, in unconventional ($p$-wave or $d$-wave) superconductors, impurities do suppress the superconductivity. As a result, the diffusion regime does not develop
there: either the pairing is so strong that the superconductivity is established already on the ballistic scales, or disorder
kills the superconductivity. Therefore, in such systems the enhancement of
superconductivity by localization (which occurs on the diffusive scales in $s$-wave
superconductors) is impossible.

However, in such superconductors a secondary superconducting transition due to
the pairing of Dirac quasiparticles is possible (which may change the true gap symmetry as, e.g., $d \to d+is$, thus opening the superconducting gap
at the nodal points of the spectrum), see e.g. Ref. [\onlinecite{Sachdev}] for review. This transition can be described by the RG equations generalized for
the novel symmetry classes (see Refs.~[\onlinecite{dwave1,DellAnna}]).
Furthermore, the peculiar form of the Fermi surface near half filling (nesting)
may lead to additional emergent symmetries specific to this problem [\onlinecite{dwave2}].
In particular, various novel interaction couplings would be possible by the enhanced
symmetry. Importantly, the Coulomb interaction between the
quasiparticles in this system is screened by the $d$-wave condensate. Thus, one
can expect a disorder-induced enhancement of the critical temperature for the secondary superconducting transition.

\subsubsection{Macroscopic homogeneity vs granularity}

In this paper we assume that the system is macroscopically homogeneous and
do not discuss granulated superconductors characterized by weak (Josephson) tunneling
between macroscopic superconducting islands. In granular systems, additional energy scales appear such
as Josephson and charging energies. We expect, however, that the peculiarities of inhomogeneous
superconductors, while leading to emergence of intermediate crossover regimes, do not
affect the universality of the (zero-$T$) SIT governed by the symmetries of the system.
The situation resembles the problem of Anderson metal-insulator
transition which is believed to be universal independently of whether
the microscopic disorder model is ``homogeneous'' (e.g. white-noise disorder) or
inhomogeneous (tunnel-coupled grains).

At the same time, the finite-$T$ behavior of
the resistivity in granular systems will be influenced by the presence of additional energy scales and thus differ from that of a homogeneous system.
On the other hand, the behaviour close to the transition will be governed by a similar BKT physics both for granular and homogeneous systems,
see the discussion in Sec.\ref{s5_0}.

\subsection{Screening of long-ranged Coulomb interaction}

Above, we have considered separately the two models of electron-electron interaction: long-range Coulomb interaction
and short-ranged interaction. In the latter case, the superconductivity was shown to be enhanced by
Anderson localization in a wide parametric range. In realistic electronic systems, there are two mechanisms that can suppress the Coulomb interaction and make it effectively short-range in a certain interval of length scales: (i) large dielectric constant of the medium, and (ii)  screening by a nearby external metallic layer which results in a less singular
dipole-dipole type interaction at scales larger than the distance to the gate.

In the presence of a dielectric medium, the interaction constant
in the singlet channel acquires a momentum dependence:
\begin{equation}
\gamma_s(q) = \tilde{\gamma}_s - (1+\tilde{\gamma}_s)\frac{\varkappa}{\varkappa+q},
\quad
\varkappa = \frac{2\pi e^2}{\varepsilon} \frac{\partial n}{\partial \mu}.
\label{eq:gammasq}
\end{equation}
Here $\tilde{\gamma}_s$ is the irreducible short-ranged part of the singlet interaction amplitude,
$\varepsilon$ is the dielectric constant of the medium, $\varkappa$ is the inverse screening length,
and $\partial n/\partial \mu$ is the thermodynamic density of states (which is not renormalized by the interplay of disorder and interaction).

Usually, the condition $\varkappa^{-1}\ll l$  (the screening radius is smaller than the mean free path) is fulfilled
and for length scales $L\geqslant l$ one finds $\gamma_s=-1$
which is a hallmark of long-ranged Coulomb interaction.
However, for large dielectric constant the opposite relation, $l\ll \varkappa^{-1}$, is possible.
In this case, at length scales $l\leqslant L \leqslant \varkappa^{-1}$ the long-ranged Coulomb interaction
provides small contribution to $\gamma_s$ and is indistinguishable from the short-ranged interaction
within the RG. If the scale $L_c$
is smaller than $\varkappa^{-1}$, then the long-ranged Coulomb interaction does not affect
the transition temperature.
Therefore, for large dielectric constant such that $\varkappa l\ll 1$, the long-range Coulomb interaction does not
influence the superconducting temperature provided the following condition holds:
\begin{equation}
e^2\varkappa/t_0 \lesssim T_c.
\label{eq:cond-screen}
\end{equation}

When condition (\ref{eq:cond-screen}) is not fulfilled, the long-ranged nature of Coulomb interaction screened by high dielectric constant
becomes effective at large scales $L \gtrsim \varkappa^{-1}$ before the superconductivity occurs. While for scales
shorter than the screening radius the coupling constant $\gamma_c$ is enhanced by short-range interaction as compared to the BCS result,
at larger scales the Coulomb repulsion starts working in the opposite direction. As a result, one encounters the competition
between the enhancement and suppression of the superconductivity.
In this situation, a more general scheme of including Coulomb repulsion is necessary.

The simplest generalization of the RG procedure would be then a two-step RG.
At the first step, for $L\lesssim \varkappa^{-1}$, one uses the short-ranged RG with the initial values of all
interaction couplings determined by the short-range attraction (BCS line).
At the second step, for $L\gtrsim \varkappa^{-1}$, the RG equations are switched to the Coulomb case with $\gamma_s=-1$
and the initial values of other couplings given by the outcome of the first step.
However, within this two-step procedure the singlet amplitude $\gamma_s$ is instantly switched at $L\sim \varkappa^{-1}$
from the value dominated by the phonon-induced attraction, $\tilde{\gamma}_s(L)$, to the Coulomb dominated value $\gamma_s=-1$,
implying the change of its sign.

In order to smoothly describe the crossover regime,
an interpolating flow equation for the coupling $\gamma_s$ defined in Eq.~(\ref{eq:gammasq}) can be derived by replacing
the momentum by $L^{-1}$. In particular, for the case of preserved time and spin-rotational symmetries this yields [cf. Eq.~(\ref{eq:rg:final:gs})]
\begin{eqnarray}
\frac{\partial\gamma_s}{\partial y}  &=& - \frac{t}{2} (1+\gamma_s)\bigl ( \gamma_s+3\gamma_t+2\gamma_c+4 \gamma_c^2\bigr ) \frac{1}{Z_L+1}
\nonumber
\\
&-& (1+\gamma_s) \frac{Z_L}{Z_L+1}.
\label{eq:RGs}
\end{eqnarray}
In this flow equation we have introduced the new coupling $Z_L=\varkappa L$ satisfying
\begin{equation}
\frac{\partial Z_L}{\partial y} = Z_L.
\label{eq:RGZL}
\end{equation}
In Eq.~(\ref{eq:RGZL}) we have used the fact that $\varkappa$ is not renormalized by interactions, since it is determined by
electron charge and the thermodynamic density of states $\partial n/\partial \mu$ [\onlinecite{Fin}].
If the background medium is characterized by a momentum-dependent dielectric function $\epsilon(q)$, this would modify Eq.~(\ref{eq:RGZL})
accordingly. The RG flow governed by Eqs.~(\ref{eq:RGs}) and (\ref{eq:RGZL}) can be viewed as a two-step RG procedure with a
short-ranged singlet amplitude $\gamma_s \simeq \tilde{\gamma}_s(1+Z_L)-Z_L$ at the first step.

\begin{figure}[t]
\centerline{\includegraphics[width=7cm]{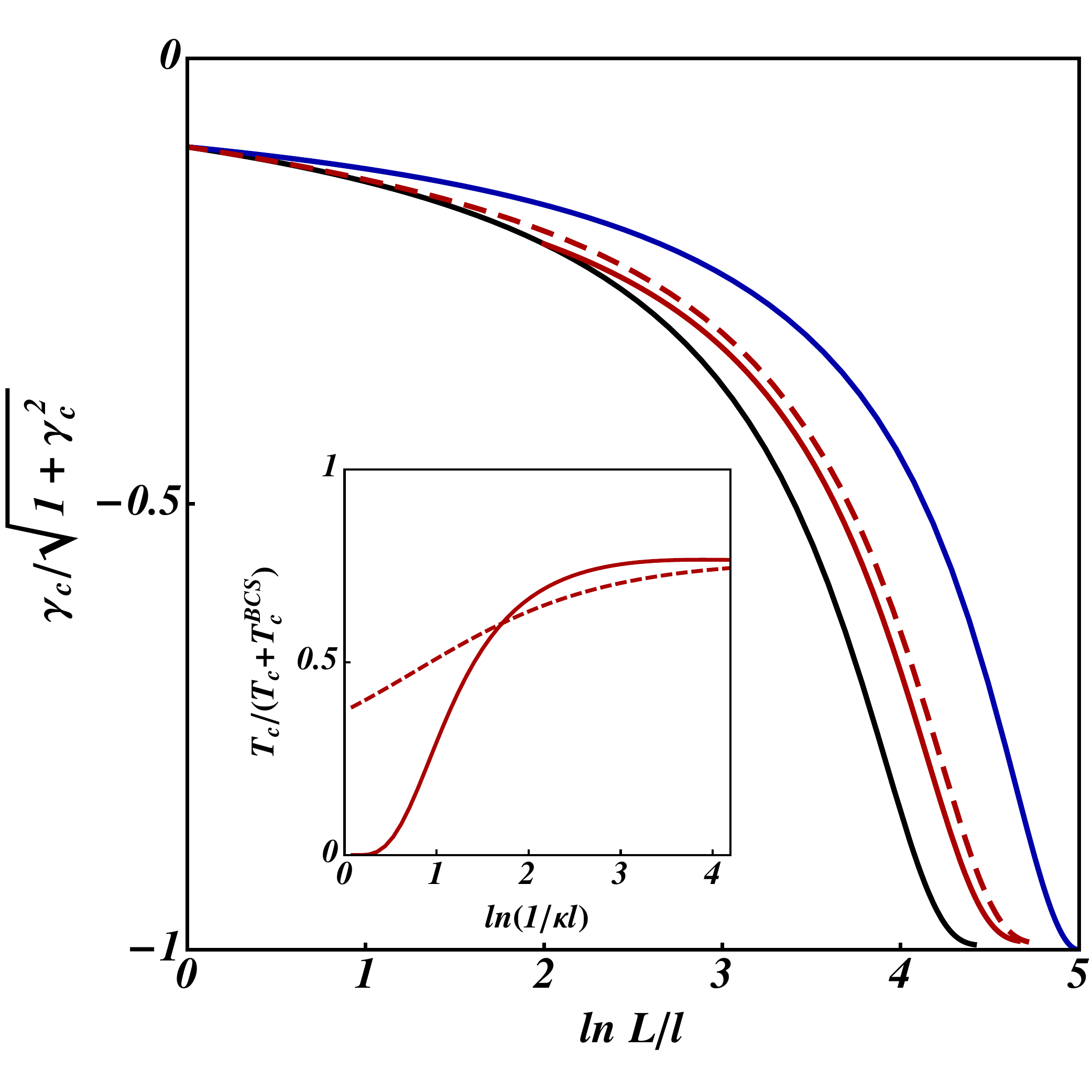}}
\caption{(Color online) Effect of a high dielectric constant on the renormalization of Cooper-channel coupling $\gamma_c$. Black curve corresponds to solution of RG Eqs. \eqref{eq:rg:final:t:SR} - \eqref{eq:rg:final:gc:SR}
for $\gamma_{c0}=\gamma_{t0}=-\gamma_{s0}=-0.1$, and $t_0=0.05$. Blue curve corresponds to the behavior of $\gamma_c$ on $L$ for clean BCS case. Red curve is  obtained from the numerical solution of the two-step RG equations for $\ln (1/\varkappa l)=2$. Red dashed curve is obtained from the numerical solution of the crossover RG Eqs. \eqref{eq:RGs} and \eqref{eq:RGZL}. Inset: Dependence of $T_c/(T_c+T_c^{BCS})$ on $\ln (1/\varkappa l)$ for two-step (solid) and crossover (dashed) RG equations.
}
\label{fig:RG-screen}
\end{figure}

We plot the results of the numerical evaluation of the renormalization of the Cooper-channel coupling $\gamma_c$ for $\ln (1/\varkappa l)=2$
using the two-step and interpolating RG procedures in Fig.~\ref{fig:RG-screen}. One can see that the enhancement of the superconductivity
at the first (short-ranged) step of the RG is more important than the suppression at the second (long-ranged) step. As shown in the inset, the overall enhancement of the superconductivity takes place (for chosen values of the bare interactions and resistivity) for $(\varkappa l)^{-1}\agt 3 \div 4$. For larger screening lengths both the two-step and interpolating RG procedures yield close results for the enhancement of $T_c$.

The long-range Coulomb repulsion can also be screened by a nearby metallic layer.
Specifically, the electron-electron repulsion can be considered as short-ranged on scales $L$ larger than
the spacer width $w_s$. When the mean-free path is larger than $w_s$, we have the short-range case from the very
beginning. In the opposite case $w_s\gg l$, without additional screening by the dielectric medium (i.e. for $\varkappa L\gg 1$)
the RG procedure corresponds to the
Coulomb case $\gamma_s=-1$ (up to small corrections of the order of $(\varkappa w_s)^{-1}\ll 1$)
for scales $L\lesssim w_s$. For larger scales the interaction becomes of the dipole-dipole type,
but the singlet interaction constant inherited from the first step remains Coulomb-like, $\gamma_s\simeq -1$. Therefore,
a metallic layer placed at the distance $w_s\gg l$ is not sufficient to screen the Coulomb repulsion such that
the superconductivity would be enhanced. However, the combination of the screening by a medium with large dielectric
constant (see above) and by the metallic layer does lead to the enhancement of the superconductivity as
compared to the cases when these screening mechanisms are considered separately. Indeed, these two mechanisms
make the interaction effectively short-ranged (with $|\gamma_s|<1$) for short and large scales, respectively.
In particular, for $\varkappa w_s \ll 1$ there is no room for the Coulomb regime at all.

\subsection{Enhancement of superconductivity for short-range repulsion}

Most of experiments on the superconducting transition in 2D films have been performed without
screening the long-range component of the interaction.
It is desirable to explore whether the mechanism of the enhancement of superconductivity addressed
in the present work may be employed in practice to obtain structures with substantially
enhanced $T_c$. The key condition is a suppression of the
long-range component of the Coulomb interaction [\onlinecite{PRL2012,Kravtsov12}].
This opens a new way for searching novel materials
exhibiting high-temperature superconductivity: one
needs the combination of a large dielectric
background constant and disorder in layered
structures.

As mentioned in Introduction, 2D superconductivity
has been recently realized in interfaces between two oxides, in particular, in  LaAlO$_3$/SrTiO$_3$
interfaces [\onlinecite{Caviglia2008,Ilani2014}].
These systems possess unique electrostatic properties owing to the giant
dielectric constant of SrTiO$_3$. In particular, the long-range component of the Coulomb
interaction is expected to be strongly screened in such materials.
Although currently, the highest $T_c$ reached in such materials is rather low as compared to
high-$T_c$ materials, the dependence of $T_c$ on the conductivity of a normal state is non-monotonic,
which agrees with the localization-induced mechanism of the superconductivity enhancement.
Further investigations are required to identify the ways for increasing $T_c$ in strongly screened oxide interfaces,
and to analyze optimal materials, structure design, and operation regimes, depending on
the microscopic details.

A possible route for increasing the superconducting transition temperature
in these materials is based on further suppression of the long-range
Coulomb interaction by designing a double-interface structure
with a LaAlO$_3$ layer sandwiched between two SrTiO$_3$ oxides.
In such a setup, already ten atomic layers of LaAlO$_3$ are sufficient,
so that the screening properties of the sample would be fully determined
by the giant dielectric constant of SrTiO$_3$.
At the same time, the two interfaces would be coupled by the interlayer
interaction, similarly to the Coulomb drag problem in double-layer structures.
The corresponding generalization of the sigma-model would include an additional
degree of freedom (a pseudospin in the interface space).
Furthermore, the doping of SrTiO$_3$ layers away from the interfaces can produce
an effective metallic gate made of the same material.

A simpler setup would involve an amorphous superconducting film placed on
a SrTiO$_3$ substrate (again possibly doped away from the interface) with high
dielectric constant instead of more conventional
SiO$_2$ or Al$_2$O$_3$ insulating substrates typically used in experiments on the SIT.
On the other side of the substrate one can place a metallic gate, thus realizing
both mechanisms of screening discussed above.
An interesting possibility of arranging a closely located metallic layer is provided by BN-Graphene
heterostructures [\onlinecite{BN}] with gated graphene layer serving as a metallic gate
and BN playing a role of a thin spacer. In this situation, a generalization of
the RG equations \eqref{eq:rg:final:t:G} - \eqref{eq:rg:final:z:G} to the case of two layers (similar to Ref. [\onlinecite{elio13}]) needs to be done.

\begin{figure}[t]
\centerline{\includegraphics[width=7cm]{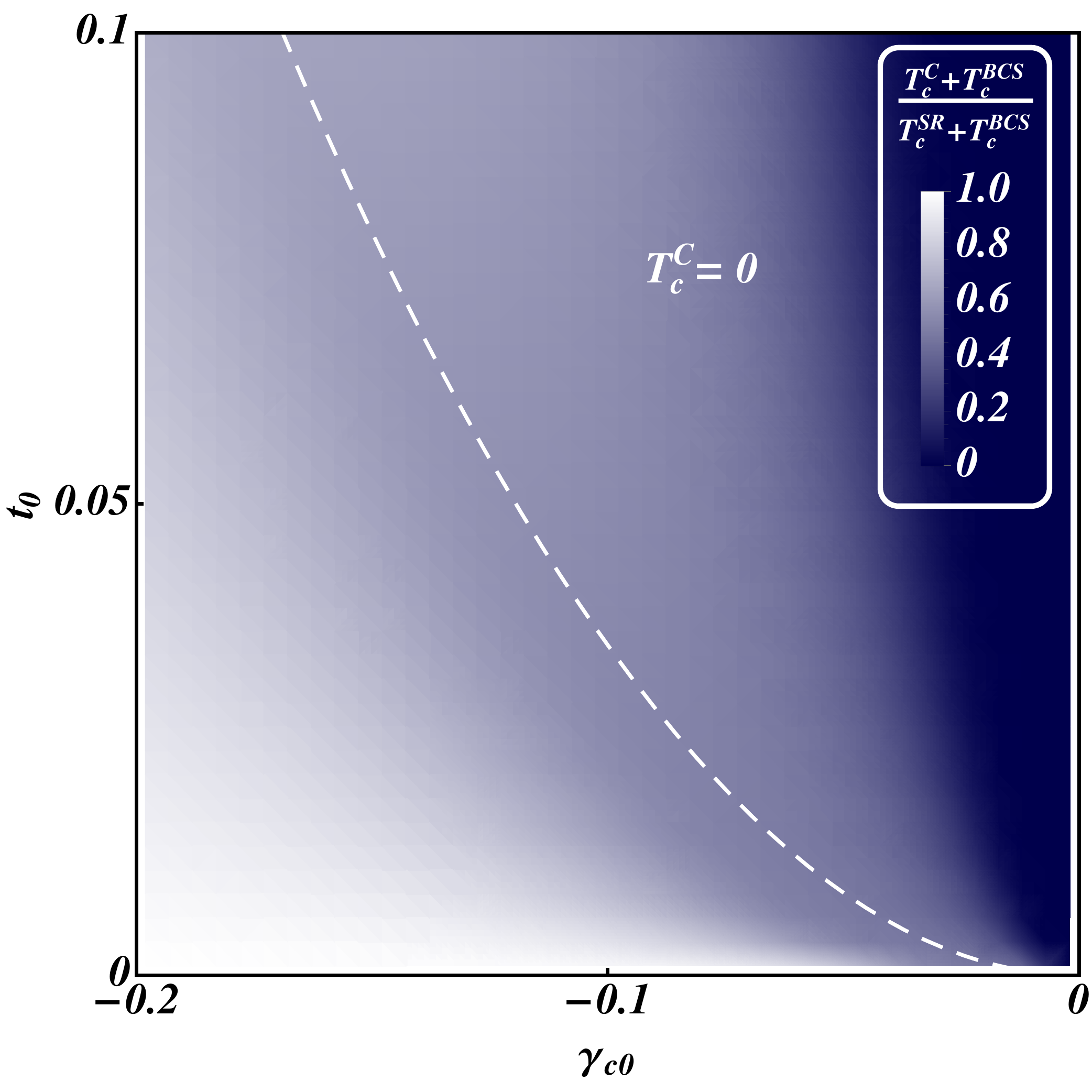}}
\caption{(Color online) The case of broken spin rotational symmetry. Comparison of superconducting transition temperature for the Coulomb interaction ($T_c^C$) and for the short-ranged interaction on the BCS line ($T_c^{SR}$). The color indicates the ratio $(T_c^C+T_c^{BCS})/(T_c^{SR}+T_c^{BCS})$. The dashed curve corresponds to the boundary of SC phase in the case of Coulomb interaction.}
\label{fig:Tc:CvsSR}
\end{figure}

Recently, the superconductivity has been studied in layered material Li$_x$ZrNCl
[\onlinecite{Exp-LiZrNCl}]. It was found that with increase of doping level $x$ the transition from insulator
to superconductor occurs at $x \approx 0.05$. The critical resistance is close to $h/2e^2$. The temperature dependence of the resistivity,
 $\rho(T)$, measured across  the transition
is qualitatively similar to one shown in Fig. \ref{fig:RG:Res:vs:Cond}d.
Near the SIT  the superconducting transition temperature increases with decrease of doping level: from $T_c(x\approx 0.12) \approx 11$ K to $T_c(x\approx 0.05) \approx 16$ K. Such behavior of superconducting temperature is in suit with dependence $T_c(t_0)$ on BCS line predicted by our theory (see the inset to Fig. \ref{fig:RGO:PD2:SIT} and Fig. \ref{fig:RGO:PD2:SIT:BCSLine}).
Similar nonmonotonous dependence of $T_c$ on disorder was measured in W and Mo based films
(for an overview, see Ref. [\onlinecite{Osofsky}]).

In the case of repulsion in the particle-hole channel,
the enhancement of superconductivity by Anderson localization (in comparison with the corresponding clean system)
occurs in a certain range
of (not too strong) interaction and (not too weak) disorder, see Figs. \ref{fig:RGO:PD2}, \ref{fig:RGO:PD2:SIT:BCSLine}
and \ref{fig:RGSO+SRI+Tc}.
It should be stressed, however, that the critical temperature for short-range interaction is predicted to be always higher
than $T_c$ for unscreened Coulomb interaction when other parameters are kept fixed,
see e.g. the inset to Fig.~\ref{fig:RG-screen}.
Therefore, we propose to perform benchmarking experiments, measuring $T_c$ in the same superconducting
film placed on the substrate with high dielectric constant (say, STO-material that screens long-range Coulomb
repulsion) and on the reference substrate with a not too high $\varepsilon$
(say, SiO$_2$ or Al$_2$O$_3$). The experiments should be performed for
sufficiently dirty samples (but still on superconducting side of the SIT),
since the stronger disorder leads to a stronger difference between the
critical temperatures in the two cases, see Fig. \ref{fig:Tc:CvsSR}.

\subsection{Relation to numerical results}

Recent numerical calculations [\onlinecite{Sondhi,Garcia}] demonstrate that disorder may indeed
enhance the superconductivity in a certain range of parameters.
These results should be contrasted with numerical simulations of a two-dimensional
disordered Hubbard model with strong on-site attraction
in small-size systems that yielded a monotonous suppression of $T_c$ with
increasing disorder [\onlinecite{Trivedi}].
The physics behind the results of our work develops
for not too strong disorder and interaction,
whereas Ref. [\onlinecite{Trivedi}] focussed on the opposite limit.
Specifically,
we predict  the enhancement of $T_c$ by Anderson localization in 2D when both
disorder and interaction are weak: $|\gamma_{c0}| \ll t_0\ll \sqrt{|\gamma_{c0}|} \ll 1$.
In terms of the disordered Hubbard model used in numerical simulations~[\onlinecite{Trivedi}]
this regime corresponds to the
following range of parameters: $|U|\ll V \ll \sqrt{|U|}\ll 1$ where $U$ and $V$ stand
for dimensionless interaction and disorder.

It is the strong interaction $U$ which allowed the authors of Ref.~[\onlinecite{Trivedi}] to
extract the information on superconducting properties from the simulation on a
rather small system of $8\times 8$ sites.
As seen from Fig. \ref{fig:RGO:PD2:SIT:BCSLine}, in the strong-coupling regime, our theory agrees with
the numerics of Ref.~[\onlinecite{Trivedi}].
Indeed, for strong attraction and strong disorder our theory predicts a suppression
of the mean-field $T_c$ (and hence a suppression of the true critical temperature).
The point is that, in realistic systems, the attraction is normally considerably weaker
(otherwise $T_c$ would be given by the Debye energy and no challenge of obtaining
high-temperature superconductivity existed) and this is precisely the range of $\gamma_c$ where the enhancement of
superconductivity is expected according to our predictions.
The results of Ref.~[\onlinecite{Trivedi}] may in addition reflect the difference between
$T_c$ and $T_{\rm BKT}$ in the strongly disordered case, see Sec. \ref{s5_0}. If one extrapolates the Beasley-Mooij-Orlando
estimate [\onlinecite{BKT}] to the regime of strong disorder, $t_m\sim 1$, one gets $(T_{c}-T_{\rm BKT})/T_c\sim 1$.
This correlates with the numerical findings of Ref.~[\onlinecite{Trivedi}]: for strong disorder
$T_{\rm BKT}$ may be significantly lower than the mean-field $T_c$ (this difference might be important near the SIT).

In fact, disordered Hubbard model contains all ingredients required for
the enhancement of superconductivity by multifractality, but the range of optimal parameters
requires large system sizes.
To verify our prediction numerically within the disordered Hubbard model, one has to use weaker interaction and disorder
and hence larger system sizes of $N\times N$ sites,
where due to the logarithmic renormalization of couplings in 2D
$N$ depends exponentially on the inverse disorder strength.
Rough estimates yield at least $N\sim 30 \div 50$ for the minimal system size
where the enhancement can be detected.
Indeed, in Ref.~[\onlinecite{Sondhi}], where an enhancement of the superconductivity
by disorder in a honeycomb lattice was detected for a certain range of parameters,
the number of sites in the attractive Hubbard model was $900 \div 1600$, in consistency with the above estimate.
At the same time, such system sizes are still much smaller than the
sizes of real macroscopic systems where the regime required for a strong enhancement of $T_c$ by our mechanism
can be realized.

\subsection{Beyond one-loop RG: Structure of the phase diagram}

In Sec. \ref{s4}, we have analyzed the one-loop RG equations for
the cases of preserved and broken spin-rotational symmetry. As we have shown,
the one-loop precision is applicable for $t\, \text{max}\{1,|\gamma_c|\}\lesssim 1$,
and therefore some fixed points of the full phase diagram remain unaccessible
at this level, see gray areas in Figs. \ref{fig:RGO:PD00} and \ref{fig:RGSO+Coulomb}.
Here we discuss an expected structure of the full phase diagram, going
beyond the one-loop RG equations. We will focus on the SIT part of the phase diagram,
first disregarding the complications related to the appearance of additional
phases such as ferromagnetic (FM) phase for the case of preserved spin-rotational symmetry
and the critical metal (CM) for the spin-orbit case. Furthermore, for simplicity,
we consider the Coulomb case, $\gamma_s=-1$, which allows us to reduce the parameter
space for the RG flow. We expect that the superconductor-insulator quantum phase transition
is not sensitive to details of interactions and, therefore, concentrate on the
simplest case of broken spin-rotational symmetry, where we have a two-parameter RG flow
(for couplings $t$ and $\gamma_c$).

\begin{figure}[t]
\centerline{(a) \includegraphics[width=7cm]{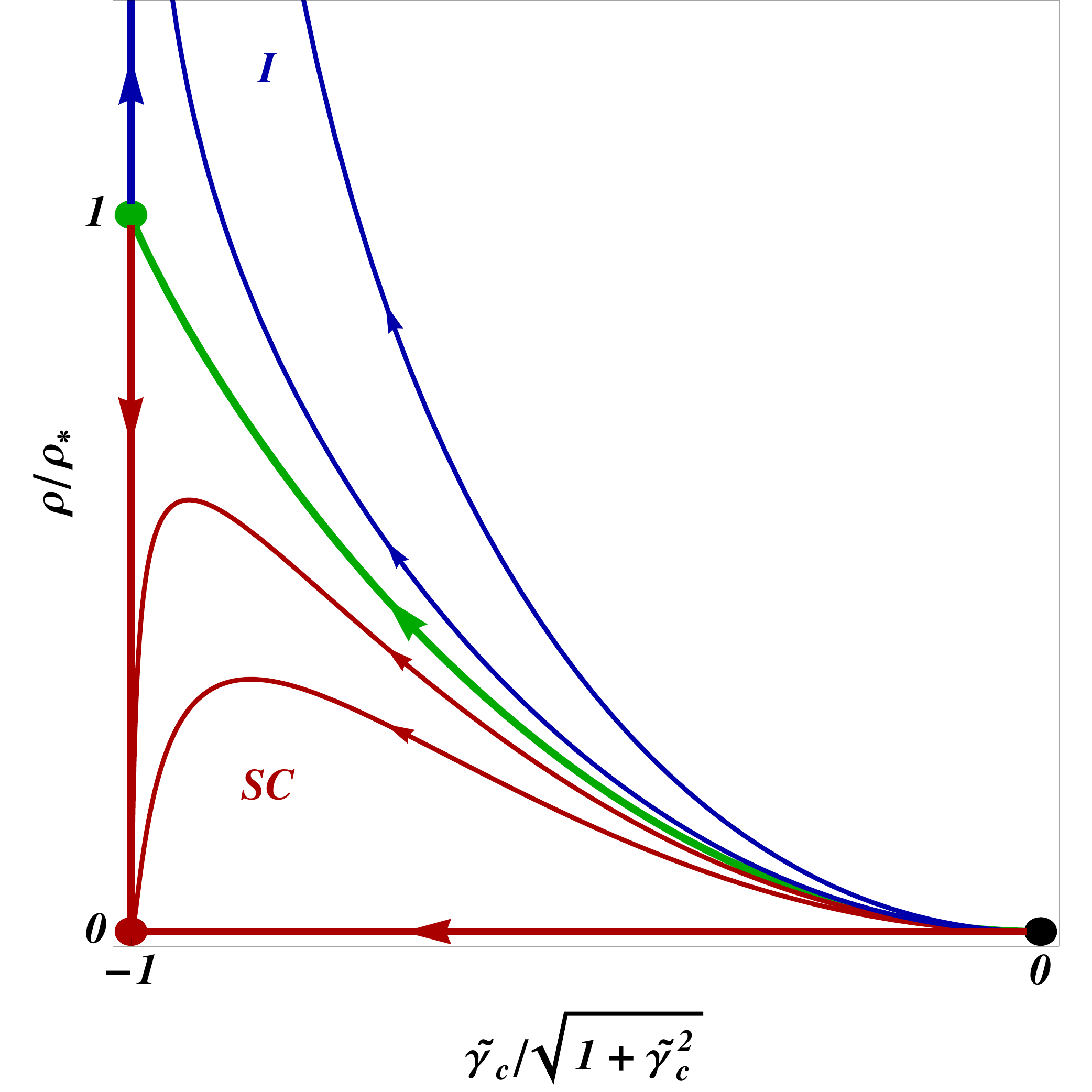}}
\centerline{(b) \includegraphics[width=7cm]{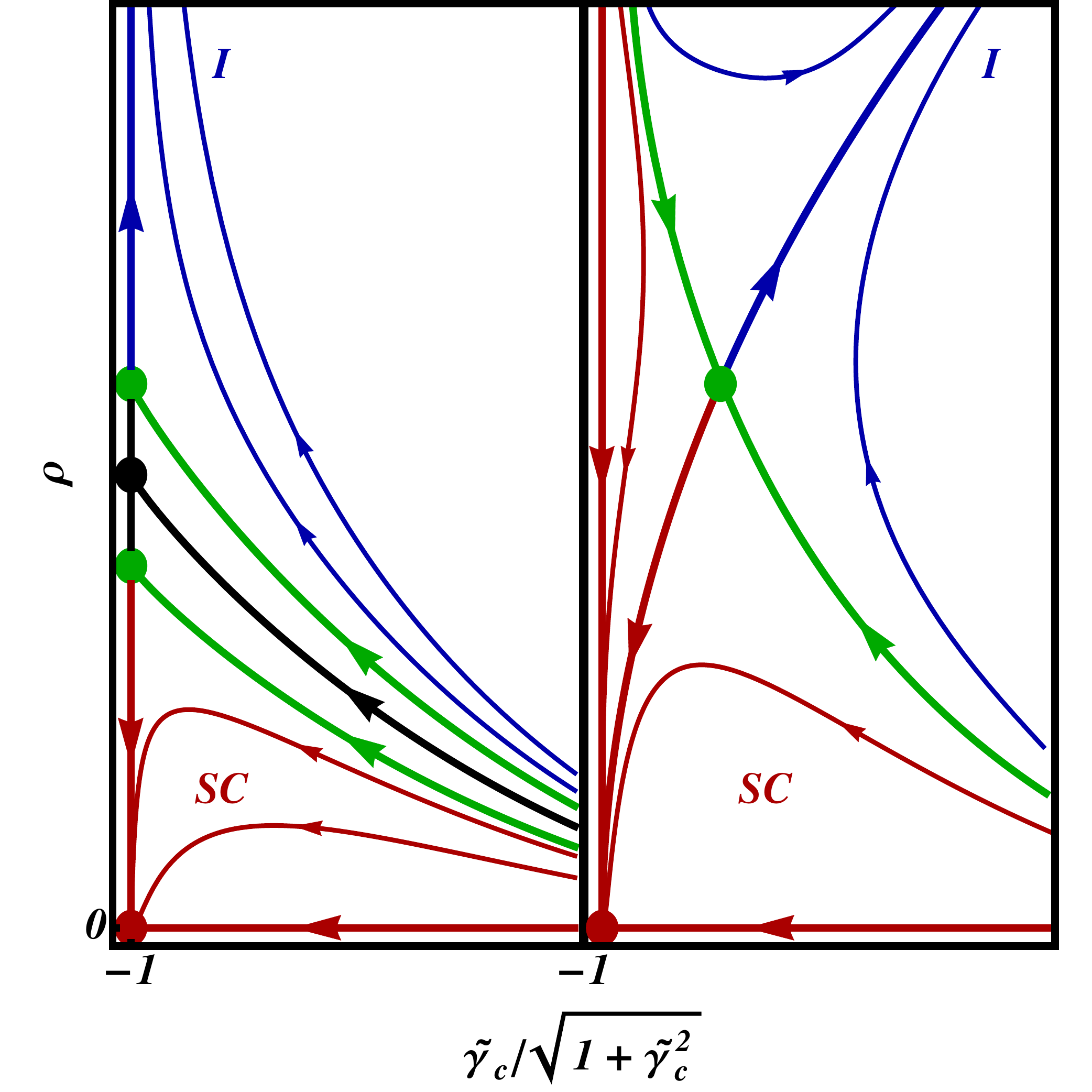}}
\caption{(Color online) Sketch of possible RG flows for the SIT  with the unstable fixed point at $\rho=\rho_*$ and $|\tilde\gamma_c|=\infty$ (a) and
with the metallic phase at $\tilde\gamma_c=\infty$ (b, left panel) or the fixed point at
$\rho\sim 1$ and $|\tilde\gamma_c|\sim 1$ (b, right panel). }
\label{fig:ExpRG}
\end{figure}

In Fig. \ref{fig:ExpRG}a we plot schematically the expected phase diagram for physical electrical
resistance $\rho$ (that can differ from the NLSM coupling $t$ as discussed in Sec.~\ref{s5})
and a parameter $\tilde{\gamma}_c$ (``generalized superconducting interaction'') characterizing the superconducting correlations in the system.
On the mean field level, this parameter is just equal to $\gamma_c$, but beyond the mean-field description it also reflects order-parameter fluctuations and thus characterizes the overall superconducting coherence (hence -- tilde), diverging when the true 2D superconductivity is established.
The superconducting fixed point is then located at $\rho=0$ and $\tilde{\gamma}_c=-\infty$.
As mentioned in Sec. \ref{s5_0} above, the SIT is most likely governed
by the fixed point at $\rho=\rho_*\sim 1$ and $\tilde{\gamma}_c=-\infty$, see Fig. \ref{fig:ExpRG}(a). The existence of such fixed point is compatible with the RG flows shown in Figs. \ref{fig:RGO:PD00} and \ref{fig:RGSO+Coulomb}. The SIT occurs through the separatrix connecting this fixed point with the trivial
clean non-interacting fixed point $\rho=\tilde\gamma_c=0$.

Two more possibilities compatible with Figs. \ref{fig:RGO:PD00} and \ref{fig:RGSO+Coulomb}
are as follows:
\begin{itemize}
\item[(i)] the fixed point at $\rho = \infty$ and $\tilde{\gamma}_c=-\infty$ is unstable and
the SIT fixed point is located at $\rho\sim 1$ and $|\tilde{\gamma}_c| \sim 1$, see the right panel in Fig. \ref{fig:ExpRG} (b);
\item[(ii)] the SIT fixed point is located at $\rho = \infty$ and $\tilde{\gamma}_c=-\infty$.
\end{itemize}
In both these cases the flow towards the superconducting fixed point would occur in the presence of strong superconducting correlations for an arbitrary high resistivity. The existence of the SIT in granular systems (or 2D Josephson-junction arrays) provides a strong evidence against such a scenario. Indeed, in such strongly inhomogeneous systems the superconductivity is established locally, but sufficiently strong disorder prevents vanishing of the total
resistance.

At the same time, our one-loop RG analysis does not exclude the possibility of existence of an intermediate
metallic phase around $\rho \sim 1$ at $\tilde{\gamma}_c=-\infty$, e.g. as shown in the left panel of Fig. \ref{fig:ExpRG} (b).
This metallic phase might be governed by either a segment of fixed points or by an attractive metallic fixed point.
In both cases, this intermediate metallic phase would resemble a so-called ``Bose metal'' mentioned in the Introduction.
Experimental evidence for existence of such a phase was reported in literature.
In the case of preserved spin-rotation invariance, we do not see any indications for such a scenario. For the spin-orbit class, our theory does suggest a critical-metal phase with resistivity of the order of resistance quantum, somewhat similar to the proposed ``Bose metal''. However, in the phase diagram this phase is separated from the superconductor by a narrow insulating region, so that the expected sequence of quantum phase transitions is SC -- I -- CM -- I.
More work is needed to prove or disprove the possibility of the intermediate metallic phase both within the NLSM
formalism and experimentally.

\begin{figure}[t]
\centerline{\includegraphics[width=7cm]{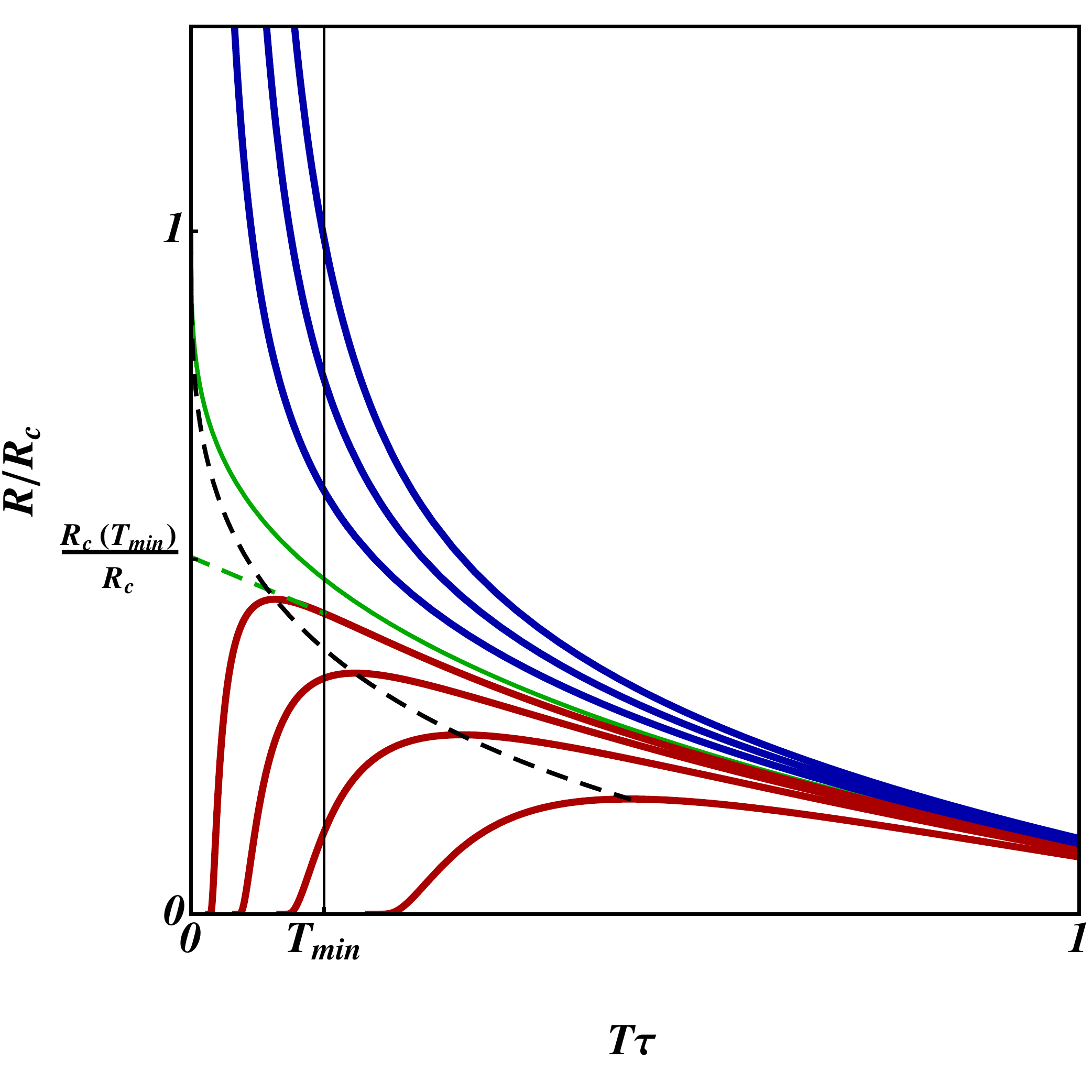}}
\caption{(Color online) Sketch of a typical dependence $R(T)$ near the SIT. The black dashed curve demonstrates how the maximal resistance approaches the critical one with decrease of $T$. The solid green curve indicates the separatrix. The dashed green line demonstrates approximation to the critical resistance from the curves available for $T>T_{\rm min}$ (see text). }
\label{fig:ExpRc}
\end{figure}

Returning to the phase diagram in Fig. \ref{fig:ExpRG} (a), we emphasize that the SIT
fixed point at $\rho_*\sim 1$ and $\tilde{\gamma}_c=-\infty$ is reached only at infinite
RG scale, corresponding to exactly zero $T$. This implies that in realistic experiments
(performed at finite temperature) the flow along the separatrix might seem as a flow towards
an insulator. As a result, the critical value of resistance $R_c(T_\text{min})$ inferred from the temperature dependence
of the resistivity measured down to finite $T_\text{min}$ might be lower than the true critical value $R_c=(h/e^2) \rho_*$, see Fig. \ref{fig:ExpRc}. Moreover, the `critical' values $R_c(T_{\rm min})$ extracted in such a manner from experimental data may significantly differ for different values of $T_\text{min}$.
This example is typical for the two- (and more) parameter scaling
and demonstrates that the ``non-universality'' of the critical resistance of the SIT might be an artifact of
the interpretation of the data obtained for finite temperatures.

It is also worth noting that in the case of preserved spin-rotational symmetry (realized in many SIT experiments),
the RG flow includes an extra dimension corresponding to the triplet coupling $\gamma_t$.
In this situation, depending on the initial parameters, the SIT fixed point at $\rho_*\sim 1$ in the three-dimensional parameter space can be reached both from above and from below, thus leading to different conclusions on the value of $R_c$ obtained at finite $T_\text{min}$
in different samples or settings. Moreover, the extra dimension in the phase diagram in this case
might result in a non-monotonic temperature dependence of the resistivity at the SIT, as observed
in some experiments.

Let us emphasize that the phase diagram in Fig. \ref{fig:ExpRG} is obtained within the framework of the so-called ``fermionic
mechanism'' of the SIT (associated to the works by Finkel'stein [\onlinecite{FinSC,Fin,FinPhysica}]).
There is a popular misconception in literature stating that, contrary to the
``bosonic mechanism'' [\onlinecite{Fisher90}] yielding $\rho_*\sim 1$, the fermionic mechanism
predicts $\rho_*\ll 1$ for a system with Coulomb interaction.
Our analysis demonstrates that this is not so: the SIT within the ``fermionic mechanism'' is governed by a fixed point with the critical resistivity
$\rho_*\sim 1$.  The confusion might arise if one neglects localization effects (due to interference and interactions)
in the analysis of $T_c$ as in Refs. [\onlinecite{Fin,FinPhysica}].
Indeed, for
very weak disorder ($t_0\ll \gamma_{c0}^2 $) the renormalization of $t$ is negligible, see
Figs. \ref{fig:RGO:PD00} and \ref{fig:RGSO+Coulomb}, where the red SC curves appear to be almost horizontal for small $t_0$.
However, with increasing disorder and approaching the separatrix ($t_0 \propto \gamma_{c0}^2 \ll 1$) the red SC curves become more and more pushed into the region of $\rho \sim 1$ at low temperatures, and the critical resistance $R_c$ is of the order of  $R_q$. In other words,
temperature dependence of resistivity for SC curves close to the transition reaches at the maximum a value $\sim R_q$ before dropping down with further lowering of the temperature. This is very well illustrated by the resistivity plots in the present paper.
We thus reiterate that both fermionic and bosonic mechanisms predict the SIT governed by a fixed point at $\rho_*\sim 1$.

\section{Summary and conclusions}
\label{s8}

To summarize, in this paper we have explored by means of the RG approach the interplay of
superconductivity, interaction, and localization
in 2D quantum systems. The focus has been put on the SIT in thin films.
Our main results are as follows.

\begin{enumerate}

\item
Within the non-linear sigma model formalism, we have derived the \textit{full set of one-loop} (in disorder strength $t\ll 1$)
\textit{RG equations} for a 2D disordered interacting system, Eqs.~\eqref{eq:rg:final:t:G} - \eqref{eq:rg:final:gc:G}.
Formally, these RG equations are valid for arbitrary interaction
couplings, including an arbitrary strong amplitude $\gamma_c$ in the Cooper channel.
The range of the applicability of one-loop RG equations has been identified
as the domain $t \max \{1,|\gamma_c|\} \lesssim 1$: beyond this range
higher loops become important when approaching to the insulator $t =\infty$, or else, to the superconducting instability, $\gamma_c=-\infty$.

\item
We have employed the RG framework to explore the structure
of the \textit{phase diagram} at zero magnetic field. The analysis of RG equations has been performed
for systems both with and without spin-orbit interaction (see Figs. \ref{fig:RGO:PD00}, \ref{fig:RGO:PD}, \ref{fig:RGO:PD2}, \ref{fig:RGO:PD2:SIT:BCSLine}, \ref{fig:RGSO+Coulomb}, \ref{fig:RGSO+SRI}, and \ref{fig:RGSO+SRI+Tc}). Furthermore,
the cases of short-ranged and long-ranged Coulomb interaction have been
investigated. In general, the phase diagram of a 2D disordered interacting system
is determined by multi-parameter scaling and depends on the symmetry of the problem.

\item
The \textit{enhancement of 2D superconductivity by Anderson localization} [\onlinecite{PRL2012}]
has been confirmed for the \textit{short-range} case.
We have identified the parameter
regions where the superconductivity is enhanced by localization,
both for the cases of preserved and broken spin-rotational symmetry (see Figs. \ref{fig:RGO:PD2}, \ref{fig:RGO:PD2:SIT:BCSLine}, and \ref{fig:RGSO+SRI+Tc}).

\item
In the case of preserved spin-rotational symmetry, the RG flow describes the three-parameter (for Coulomb
repulsion) or four-parameter (for short-range repulsion) scaling, thus rendering the phase
diagram multidimensional, with nontrivial fixed points appearing. In particular, a \textit{ferromagnetic
phase} develops with the metallic temperature behavior of the resistivity in a range of temperatures above the ferromagnetic instability (see Fig. \ref{fig:RGO:PD:Res}). This behaviour of resistivity in the ferromagnetic phase
may be confused with a tendency to superconductivity in experiments.

\item
The presence of spin-orbit coupling (which removes  the triplet interaction channel and converts weak localization into antilocalization) strongly affects the overall phase diagram. Two of this changes are fully expected. First, the spin-orbit coupling eliminates the ferromagnetic phase. Second, in the case of short-range interaction, a super metal phase emerges. What is much more intriguing,
our results indicate that a  \textit{critical-metal phase} with resistivity of the order of resistance quantum $R_q$ may arise (see Figs. \ref{fig:RGSO+Coulomb} and \ref{fig:RGSO+Coulomb+Res}). This phase bears certain similarity with a ``Bose metal'' phase, evidence for which has been found in some experiments.

\item
We have evaluated the \textit{temperature dependence of the electrical resistivity}, $\rho(T)$, for
given bare (high-temperature) couplings down to the lowest temperatures of the applicability
of the one-loop RG approach. In this temperature range the resistivity is dominated by the NLSM
coupling $t$ taken at the length scale $L_T$. At lower temperatures, in the close vicinity of $T_c$,
the electric resistivity is controlled by contributions due to inelastic processes (which are not automatically included by the RG procedure but rather require an additional calculation once the RG has been stopped by temperature).
These corrections lead to a strong suppression of $\rho(T)$ (see Fig. \ref{fig:RG:Res:vs:Cond}).

\item
We have studied the \textit{magnetoresistance near the SIT} (Sec.~\ref{s5}).
Both orbital and Zeeman effects of the magnetic field have been included in the
unifying RG scheme complemented by the analysis of fluctuation corrections
near the superconducting transition. A \textit{non-monotonous magnetoresistance},
see Fig.~\ref{fig:RG:Res:vs:Cond:H} has been predicted, with a maximum near the
critical field $H_c$, in agreement with experimental observations. The magnetoresistance becomes progressively stronger with lowering temperature and becomes giant as $T\to 0$, as also seen in experiments.

\item
We have further discussed in Sec.~\ref{s6} the \textit{limitations and generalizations}
of our approach as well as \textit{comparison between our theory and experiments}.
In particular, we have analyzed in detail how the screening of Coulomb interaction
by substrates with high dielectric constant and by external
metallic gates can be taken into account within our framework (see Fig. \ref{fig:RG-screen}).

\item
This consideration
allowed us to \textit{propose specific sample designs for experimental
observation of the superconductivity enhancement}.
Also, we argued that our results for enhancement of the superconductivity are
in qualitative  agreement with the experimental observations in layered material Li$_x$ZrNCl.

\item
We have discussed the possible \textit{overall structure of a generic SIT phase diagram} (see Fig. \ref{fig:ExpRG}) and the implications
of our findings for the experimental verification of the universality of this quantum
phase transition (see Fig. \ref{fig:ExpRc}). In particular, we have shown that, in contrast to a popular belief,
both ``fermionic'' and ``bosonic'' mechanisms
of the SIT have to do with the \textit{fixed point characterized by} $R\sim R_q$.

\end{enumerate}

Our findings are in a qualitative agreement with most of the experiments on 2D superconductivity in
disordered films. In particular, our approach explains a seemingly non-universal behavior of the critical
resistance $R_c$ found experimentally in different systems. Further, the analysis of the RG equations in a magnetic
field is in accord with the experimentally observed non-monotonic behavior
of the magnetoresistance near the SIT.

The detailed analysis of the temperature dependence of the resistivity in the close vicinity of the
``classical'' (finite-$T$) superconducting phase transition governed by the Berezinskii-Kosterlitz-Thouless physics
is a subject of future work [\onlinecite{future-Elio}]. The effect of disorder and interaction on the topological
sector of the theory (vortices in the order-parameter phase) is expected to be increasingly more pronounced
upon approaching the SIT. A related line of future research [\onlinecite{future-Elio}] is devoted to the
study of the zero-$T$ resistance of a disordered superconducting film as a function of the sample size.
Further, the RG approach developed here will be employed for studying the local tunneling
density of states (including its mesoscopic fluctuations) in the presence of superconducting
correlations [\onlinecite{LDOSSC}] as well as the fluctuations of the
superconducting order parameter near the SIT, as measured in experiments.
Finally, the role of strong local superconducting fluctuations in physics on the insulating side of the SIT still
requires further study.

\begin{acknowledgments}

We thank T.~Baturina, L.~Dell'Anna, M.~Feigel'man, A.~Finkelstein, A.~Frydman, V.~Gantmakher, E.~K\"onig, V.~Kravtsov, A.~Levchenko, J.~Mannhart, I.~Protopopov, R.~Schneider, M.~Skvortsov, C.~Strunk, and K.~Tikhonov for discussions. The work was funded by the Russian Science Foundation under the grant No. 14-42-00044.

\end{acknowledgments}

\appendix

\section{Background field renormalization of the nonlinear sigma model action \label{App:Der:BFM}}

In this appendix  we present details of the derivation of one-loop renormalization of the NLSM action \eqref{eq:NLSM} with the help of the background field renormalization. Let us separate the matrix field $Q$ into the ``fast'' ($Q$) and ``slow'' ($Q_0=T_0^{-1} \Lambda T_0$) modes as
\begin{equation}
Q \to  T^{-1}_{0} Q T_{0} . \label{app:Qhat}
\end{equation}
We assume that the ``fast'' matrix field $Q_{nm}$ has non-trivial structure in the Matsubara space for frequencies $|n|,|m| < N_{\rm max}$ whereas the ``slow" matrix field $T_0$ has non-trivial structure only for smaller frequencies:
\begin{equation}
(T_0)_{nm} = \begin{cases}
(T_0)_{nm},\quad & |n|\leqslant n_{\rm max} \,\, {\rm and}\, |m|\leqslant n_{\rm max} ,\\
\delta_{nm}, \quad & |n|\leqslant n_{\rm max} \,\, {\rm or}\, |m|\leqslant n_{\rm max} .
\end{cases}
\end{equation}
In what follows we assume that $N_{\rm max}\gg n_{\rm max} \gg 1$.

It is convenient to rewrite the sum of interacting terms $S_{\rm int}^{(\rho)}$, $S_{\rm int}^{(\sigma)}$ and  $S_{\rm int}^{(c)}$ in NLSM \eqref{eq:NLSM} as
\begin{equation}
S_{\rm int} = - \frac{\pi T}{4} \sum_{\alpha n} \sum_{r,j} \int d \bm{r} \Gamma_{rj} \Tr \Bigl [
J_{n,rj}^\alpha Q \Bigr ] \tr \Bigl [ J_{n,rj}^\alpha Q \Bigr ] ,
\end{equation}
where
\begin{equation}
J_{n,rj} = \begin{cases}
I_n t_{rj}, \quad r=0,3, \, j=0,1,2,3, \\
L_n t_{rj}, \quad r=1,2, \, j=0,1,2,3.
\end{cases} ,
\end{equation}
and
\begin{gather}
\Gamma_{r0}  =   (-1)^r \Gamma_s, \, \Gamma_{rj}  =  - (-1)^r \Gamma_t, \,  r=0,3, \, j=1,2,3 ,
\notag \\
\Gamma_{r0}  =   \Gamma_c, \, \Gamma_{rj}  =  0, \,  r=1,2, \, j=1,2,3 .
\end{gather}

The effective action for the slow $Q_0$ field is given by
\begin{equation}
{S}_{\rm eff}[Q_0] = \ln \int \mathcal{D}[Q] \exp {S}[T_0^{-1} Q T_{0}]
\label{SeffA}
\end{equation}
with
\begin{gather}
{S}[T_0^{-1} Q T_{0}]  = {S}[Q_0] +{S}[Q] +
O_\sigma +O_{\rm int}+ O_\eta ,
\end{gather}
where
\begin{align}
O_\sigma & = O_\sigma^{(1)} + O_\sigma^{(2),1}  + O_\sigma^{(2),2} \notag \\
O_{\rm int}& = O_{\rm int}^{(1),1}+O_{\rm int}^{(1),2}+O_{\rm int}^{(2),1}+O_{\rm int}^{(2),2} .
\end{align}
Here we introduce the following terms ($\delta Q = Q-\Lambda$)
\begin{align}
O_\sigma^{(1)}  & = - \frac{g}{8} \int\!\! d \bm{r} \Tr \bm{A} \delta Q \nabla \delta Q \notag \\
O_\sigma^{(2),1}  & = - \frac{g}{8} \int\!\! d \bm{r} \Tr \bm{A} \delta Q \bm{A} \Lambda \notag \\
O_\sigma^{(2),2}  & =  - \frac{g}{16} \int\!\! d \bm{r} \Tr \bm{A} \delta Q \bm{A} \delta Q \notag \\
O_{\rm int}^{(1),1} & =  - \frac{\pi T}{2} \sum_{\alpha n} \sum_{r,j} \int\!\! d \bm{r} \Gamma_{rj} \Tr \Bigl [
J_{n,rj}^\alpha \delta Q \Bigr ] \tr \Bigl [ J_{n,rj}^\alpha Q_0 \Bigr ] \notag\\
O_{\rm int}^{(1),2} &= - \frac{\pi T}{2} \sum_{\alpha n} \sum_{r,j} \int\!\! d \bm{r} \Gamma_{rj} \Tr \Bigl [
J_{n,rj}^\alpha \delta Q \Bigr ] \tr \Bigl [ A_{n,rj}^\alpha \delta Q \Bigr ], \notag \\
O_{\rm int}^{(2),1} &= - \frac{\pi T}{2} \sum_{\alpha n} \sum_{r,j} \int\!\! d \bm{r} \Gamma_{rj} \Tr \Bigl [
J_{n,rj}^\alpha Q_0 \Bigr ] \tr \Bigl [ A_{n,rj}^\alpha \delta Q \Bigr ], \notag \\
O_{\rm int}^{(2),2} &= - \frac{\pi T}{4} \sum_{\alpha n} \sum_{r,j} \int\!\! d \bm{r} \Gamma_{rj} \Tr \Bigl [
A_{n,rj}^\alpha \delta Q \Bigr ] \tr \Bigl [ A_{n,rj}^\alpha \delta Q \Bigr ], \notag\\
O_\eta &= 4\pi T Z_\omega \int\!\! d \bm{r}\tr A_\eta \delta Q  ,
\end{align}
where
\begin{equation}
\bm{A} = T_0 \nabla T_0^{-1}, \,
A_\eta = T_0 [\eta, T_0^{-1}], \, A_{n,rj}^\alpha = T_0 [ J_{n,rj}^\alpha, T_0^{-1}] .
\end{equation}

In the one-loop approximation the effective action ${S}_{\rm eff}[Q_{0}]$ can be obtained by expansion of ${S}[T_0^{-1} Q T_{0}]$ to the second order in $A_\eta$ and $A_{n;rj}^\alpha$. Then, we find
\begin{gather}
{S}_{\rm eff}[Q_{0}] = {S}[Q_0] + \langle O_\sigma \rangle + \langle O_{\rm int} \rangle  +  \langle O_\eta \rangle \notag \\
+ \frac{1}{2} \langle (O_\sigma +O_{\rm int}+ O_\eta )^2 \rangle , \label{app:SeffA2}
\end{gather}
where the average $\langle\dots\rangle$ is taken with respect to action~\eqref{eq:NLSM}.

For the perturbative (in $1/g$) treatment of the correlations of the fast fields in Eq. \eqref{app:SeffA2}
we shall use the square-root parametrization of the fast fields
\begin{gather}
Q = W +\Lambda \sqrt{1-W^2}, \qquad W= \begin{pmatrix}
0 & w\\
\bar{w} & 0
\end{pmatrix} .
\label{eq:Q-W}
\end{gather}
We adopt the following notations: $W_{n_1n_2} = w_{n_1n_2}$ and $W_{n_2n_1} = \bar{w}_{n_2n_1}$ with $n_1\geqslant 0$ and $n_2< 0$.
The blocks $w$ and $\bar{w}$ (in Matsubara space) obey
\begin{gather}
\bar{w} = -C w^T C,\qquad w = - C w^* C .
\end{gather}
The second equality here implies that in the expansion $w^{\alpha\beta}_{n_1n_2}= \sum_{rj} (w^{\alpha\beta}_{n_1n_2})_{rj} t_{rj}$ some of the elements
$(w^{\alpha\beta}_{n_1n_2})_{rj}$ are real and some are purely imaginary.

Expanding the NLSM action \eqref{eq:NLSM} to the second order in $W$, we find the following propagators for diffusive modes. The propagators of diffusons ($r=0,3$ and $j=0,1,2,3$) read
\begin{gather}
\Bigl \langle [w_{rj}(\bm{q})]^{\alpha_1\beta_1}_{n_1n_2} [\bar{w}_{rj}(-\bm{q})]^{\beta_2\alpha_2}_{n_4n_3} \Bigr \rangle =  \frac{2}{g} \delta^{\alpha_1\alpha_2} \delta^{\beta_1\beta_2}\delta_{n_{12},n_{34}}\notag \\
\times  \mathcal{D}_q(i\Omega_{12}^\varepsilon)\Bigl [\delta_{n_1n_3} - \frac{32 \pi T \Gamma_j}{g}\delta^{\alpha_1\beta_1}  \mathcal{D}_q^{(j)}(i\Omega_{12}^\varepsilon) \Bigr ] ,
\label{eq:prop:PH}
\end{gather}
where $\Omega_{12}^\varepsilon = \varepsilon_{n_1}-\varepsilon_{n_2}=2\pi T n_{12}=2\pi T (n_1-n_2)$. The standard diffuson propagator is given as
\begin{equation}
\mathcal{D}^{-1}_q(i\omega_n) =q^2+{16 Z_\omega |\omega_n|}/{g} .
\label{app:prop:free}
\end{equation}
The diffusons renormalized by interaction in the singlet  ($\mathcal{D}_q^{(0)}(\omega) \equiv \mathcal{D}_q^{s}(\omega)$) and triplet  ($\mathcal{D}_q^{(1)}(\omega)=\mathcal{D}_q^{(2)}(\omega)=\mathcal{D}_q^{(3)}(\omega) \equiv \mathcal{D}_q^{t}(\omega)$)   particle-hole channels are as follows
\begin{align}
[\mathcal{D}^s_q(i\omega_n)]^{-1} & =  q^2+{16 (Z_\omega+\Gamma_s) |\omega_n|}/{g},\notag \\
 [\mathcal{D}^t_q(i\omega_n)]^{-1} & =  q^2+{16 (Z_\omega+\Gamma_t) |\omega_n|}/{g} .
\end{align}
The propagators of singlet cooperon modes ($r=1,2$ and $j=0$) can be written as
\begin{gather}
\Bigl \langle [w_{r0}(\bm{q})]^{\alpha_1\beta_1}_{n_1n_2} [\bar{w}_{r0}(-\bm{q})]^{\beta_2\alpha_2}_{n_4n_3} \Bigr \rangle =  \frac{2}{g} \delta^{\alpha_1\alpha_2} \delta^{\beta_1\beta_2}\delta_{n_{14},n_{32}}\notag \\
\times  \mathcal{C}_q(i\Omega_{12}^\varepsilon)\Bigl [\delta_{n_1n_3} - \frac{64 \pi T z}{g}\delta^{\alpha_1\beta_1}  \mathcal{C}_q(i\Omega_{34}^\varepsilon)  \mathcal{L}_q(i\mathcal{E}_{12}) \Bigr ] ,
\label{eq:prop:PPS}
\end{gather}
where $\mathcal{E}_{12} = \varepsilon_{n_1}+\varepsilon_{n_2}$, $\mathcal{C}_q(i\omega_n) \equiv \mathcal{D}_q(i\omega_n)$, and the fluctuation propagator ($\gamma_c= \Gamma_c/Z_\omega$)
\begin{align}
\mathcal{L}^{-1}_q(i\omega_n) = \gamma_c^{-1}
+ \ln \frac{\Lambda}{4\pi T} & - \psi\left (\frac{D q^2 + |\omega_n|+\Lambda^\prime}{4\pi T}+\frac{1}{2} \right ) \notag \\
& + \psi\left (\frac{1}{2} \right ) .
\end{align}
Here $D = g/(16 Z_\omega)$ stands for the diffusion coefficient, $\psi(z)$ denotes the di-gamma function, and $\Lambda = 4\pi T N_{\rm max}$ ($\Lambda^\prime = 4\pi T n_{\rm max}$) determines the ultra-violet (infra-red) for the fast modes.
The propagators of triplet cooperons ($r=1,2$ and $j=1,2,3$) are insensitive to the interaction and are as follows
\begin{align}
\Bigl \langle [w_{rj}(\bm{q})]^{\alpha_1\beta_1}_{n_1n_2} [\bar{w}_{rj}(-\bm{q})]^{\beta_2\alpha_2}_{n_4n_3} \Bigr \rangle & =  \frac{2}{g} \delta^{\alpha_1\alpha_2} \delta^{\beta_1\beta_2} \delta_{n_1n_3}
\notag \\
&\times
 \delta_{n_2n_4}\mathcal{C}_q(i\Omega_{12}^\varepsilon) .
 \label{eq:prop:PPT}
\end{align}

In general, each term in the right hand side of Eq. \eqref{app:SeffA2} produce contributions which cannot be expressed in terms of $Q_0$ only.  However, all such contributions cancel in the total expression~\eqref{app:SeffA2}. Therefore, we will not list them below.  Expanding $\delta Q$ in series of W according to Eq.~\eqref{eq:Q-W} and performing averaging with the help of Eqs.~\eqref{eq:prop:PH} and \eqref{eq:prop:PPS}, we obtain the action \eqref{eq:NLSM} for the slow field $Q_0$  but with
\begin{align}
g(\Lambda)  & \to g(\Lambda^\prime) = g(\Lambda) + \delta g, \notag \\
\Gamma_{s,t,c}(\Lambda)  & \to  \Gamma_{s,t,c}(\Lambda^\prime) = \Gamma_{s,t,c}(\Lambda) + \delta \Gamma_{s,t,c} ,\notag \\
Z_\omega(\Lambda)  & \to Z_\omega(\Lambda^\prime) = Z_\omega(\Lambda) + \delta Z_\omega .
\end{align}
The contributions $\delta g, \delta \Gamma_{s,t,c}$, and $\delta Z_\omega$ take into account the effect of integration over the fast modes from the ultra-violet energy scale $\Lambda$ down to new ultra-violet scale $\Lambda^\prime$.

Below we list the different non-zero contributions to $\delta g, \delta \Gamma_{s,t,c}$, and $\delta Z_\omega$ from each term in the right hand side of Eq.~\eqref{app:SeffA2}. We start from corrections to the conductance:
\begin{gather}
\langle O_\sigma^{(2),1} \rangle \to \delta g_\sigma^{(2),1} = 2  \int_{q,\omega_n}
\Bigl [2 \mathcal{C}_q^2(i\omega_n) \mathcal{L}_q(i\omega_n) \notag \\
 + \sum_{j=0}^3   \gamma_j \mathcal{D}_q(i\omega_n)\mathcal{D}^{(j)}_q(i\omega_n) \Bigr ] ,
\end{gather}
and
\begin{gather}
\langle O_\sigma^{(2),2} \rangle \to \delta g_\sigma^{(2),2} = - 4 \int_q \mathcal{C}_q(0) .
\end{gather}
Here we use the following notation,
\begin{gather}
 \int_{q,\omega_n} \equiv \frac{2\pi T}{D}\sum_{n>0} \int \frac{d^2\bm{q}}{(2\pi)^2} \Theta\bigl (\Lambda-Dq^2-|\omega_n| \bigr ) \notag \\
 \times \Theta\bigl (Dq^2+|\omega_n|-\Lambda^\prime \bigr ) ,
 \end{gather}
 and
\begin{gather}
 \int_{q} \equiv  \int \frac{d^2\bm{q}}{(2\pi)^2} \Theta\bigl (\Lambda-Dq^2\bigr )\Theta\bigl (Dq^2-\Lambda^\prime \bigr ) ,
 \end{gather}
 where $\Theta(x)$ stands for the Heaviside step function.
 Next,

\begin{widetext}
\begin{gather}
\left \langle O_{\rm int}^{(2),2} + \frac{1}{2} \left [ O_{\rm int}^{(1),2} \right ]^2 \right \rangle
\to \delta g_{\rm int}^{(2),2}
= -2  \int_{q,\omega_n} \sum_{j=0}^3   \gamma_j  \mathcal{D}_q(i\omega_n)\mathcal{D}^{(j)}_q(i\omega_n)
\left [ 1 - 2 q^2 \mathcal{D}_q(i\omega_n) \right ] \notag \\
- 4  \int_{q,\omega_n} \mathcal{L}_{q}(i\omega_n) \mathcal{C}^{2}_q(i\omega_n)
\left [ 1 - 2 q^2 \mathcal{C}_q(i\omega_n) \right ] .
\end{gather}
In total, we find
\begin{gather}
g(\Lambda^\prime) = g(\Lambda) -4 \int_q \mathcal{C}_q(0) + 4 \int_{q,\omega_n} q^2 \Biggl [
2  \mathcal{C}^{3}_q(i\omega_n)\mathcal{L}_q(i\omega_n)
+
\sum_{j=0}^3 \gamma_j \mathcal{D}^{2}_q(i\omega_n)\mathcal{D}^{(j)}_q(i\omega_n)
\Biggr ] .
\label{eq:app:dg}
\end{gather}

The renormalization of $Z_\omega$ is described by the following terms:
\begin{gather}
\langle O_\eta \rangle \to \delta Z_\omega^\eta= \frac{2}{g} Z_\omega  \int_{q,\omega_n}
\Bigl [2 Z_\omega\mathcal{C}_q^2(i\omega_n) \mathcal{L}_q(i\omega_n)
 + \sum_{j=0}^3   \Gamma_j \mathcal{D}_q(i\omega_n)\mathcal{D}^{(j)}_q(i\omega_n) \Bigr ] ,
\label{eq:app:z:1}
\end{gather}
\begin{gather}
\left \langle O_{\rm int}^{(2),2} + \frac{1}{2} \left [ O_{\rm int}^{(1),2} \right ]^2 \right \rangle
\to \delta Z_\omega^{(2),2} = \frac{2}{g} \int_{q,\omega_n}  \Biggl \{
\sum_{j=0}^3   \Gamma_j \bigl [
\mathcal{D}^{-1}_q(i\omega_n)\mathcal{D}^{(j)}_q(i\omega_n) - 1
\bigr ] \partial_n D_q(i\omega_n)+2 Z_\omega  \mathcal{C}_q(i\omega_n) \partial_n
  \mathcal{L}_{q}(i\omega_n) \Biggr \} ,
 \label{eq:app:z:2}
\end{gather}
where $\partial_n f(i\omega_n)  \equiv (g/16) \partial f/\partial \omega_n$. Combining Eqs. \eqref{eq:app:z:1} and \eqref{eq:app:z:2} together, we obtain
\begin{gather}
Z_\omega(\Lambda^\prime) = Z_\omega(\Lambda)+ \frac{2}{g}  (\Gamma_s+3\Gamma_t) \int_{q,\omega_n} \mathcal{D}_q^2(i\omega_n) + \frac{4}{g} Z_\omega
\int_{q,\omega_n}  \Bigl [ \mathcal{C}_q(i\omega_n) \partial_n
  \mathcal{L}_{q}(i\omega_n)   - \mathcal{L}_{q}(i\omega_n)  \partial_n \mathcal{C}_q(i\omega_n)
  \Bigr ] .
  \label{eq:app:dz}
 \end{gather}
The corrections to the interaction amplitude $\Gamma_s$ are as follows:
\begin{gather}
\langle O_{\rm int}^{(2),1} \rangle \to \delta \Gamma_{s, {\rm int}}^{(2),1} = \frac{4}{g} \Gamma_{s} \int_{q,\omega_n}
\Bigl [2 Z_\omega\mathcal{C}_q^2(i\omega_n) \mathcal{L}_q(i\omega_n)   + \sum_{j=0}^3  \Gamma_j \mathcal{D}_q(i\omega_n)\mathcal{D}^{(j)}_q(i\omega_n)\Bigr ] ,
\end{gather}
\begin{align}
\left \langle O_{\rm int}^{(2),2} + \frac{1}{2} \left [ O_{\rm int}^{(1),2} \right ]^2 \right \rangle \to
\delta \Gamma_{s,{\rm int}}^{(2),2} = &-\frac{2}{g} (\Gamma_s+3\Gamma_t)  \int_q \mathcal{D}_q(0)
\notag
\\
&- \frac{4}{g} Z_\omega  \int_q \mathcal{C}_q(0) \mathcal{L}_q(0) -\frac{8}{g} Z_\omega  \int_{q,\omega_n}
\mathcal{C}^2_q(i\omega_n) \mathcal{L}^2_q(i\omega_n) ,
\end{align}
\begin{gather}
\left \langle O_{\rm int}^{(1),1} O_{\rm int}^{(1),2} \right \rangle \to
\delta \Gamma_{s,{\rm int}}^{(1),1;(1),2} = -  \frac{4}{g}  \Gamma_{s}  \int_{q,\omega_n}
 \Bigl [2  Z_\omega\mathcal{C}_q^2(i\omega_n) \mathcal{L}_q(i\omega_n)
 + \sum_{j=0}^3  \Gamma_j \mathcal{D}_q(i\omega_n)\mathcal{D}^{(j)}_q(i\omega_n)\Bigr ] .
 \end{gather}
In total, we find
\begin{gather}
\Gamma_s(\Lambda^\prime) = \Gamma_s(\Lambda)-\frac{2}{g} (\Gamma_s+3\Gamma_t) \int_q \mathcal{D}_q(0)
- \frac{4}{g} Z_\omega  \int_q \mathcal{C}_q(0) \mathcal{L}_q(0)
-\frac{8}{g} Z_\omega^2  \int_{q,\omega_n}
\mathcal{C}^2_q(i\omega_n) \mathcal{L}^2_q(i\omega_n)  .
\label{eq:app:dGs}
\end{gather}

The corrections to the interaction amplitude $\Gamma_t$ can be listed as follows:
\begin{gather}
\langle O_{\rm int}^{(2),1} \rangle \to \delta \Gamma_{t, {\rm int}}^{(2),1} = \frac{4}{g} \Gamma_{s} \int_{q,\omega_n}
\Bigl [2  Z_\omega\mathcal{C}_q^2(i\omega_n) \mathcal{L}_q(i\omega_n)   + \sum_{j=0}^3  \Gamma_j \mathcal{D}_q(i\omega_n)\mathcal{D}^{(j)}_q(i\omega_n)\Bigr ] ,
\end{gather}
\begin{gather}
\left \langle O_{\rm int}^{(2),2} + \frac{1}{2} \left [ O_{\rm int}^{(1),2} \right ]^2 \right \rangle \to
\delta \Gamma_{t,{\rm int}}^{(2),2}
= -\frac{2}{g} (\Gamma_s-\Gamma_t)  \int_q \mathcal{D}_q(0)
+ \frac{4}{g}  Z_\omega  \int_q \mathcal{C}_q(0) \mathcal{L}_q(0) -\frac{8}{g}\Gamma_t^2 \int_{q,\omega_n}
\mathcal{D}^{t2}_q(i\omega_n) ,
\end{gather}
\begin{gather}
\frac{1}{2}\left \langle \left [O_{\rm int}^{(1),1}\right ]^2  \right \rangle \to
\delta \Gamma_{t,{\rm int}}^{(1),1;(1),1}
= \frac{8}{g} \Gamma_{t}^2 \int_{q,\omega_n}
 \Bigl [\mathcal{D}^{2}_q(i\omega_n)-\mathcal{D}^{t2}_q(i\omega_n)\Bigr ] -
 \frac{8}{g}  \Gamma_{t}^2 \left (\frac{D}{4\pi T}\right )^2\int_{q,\omega_n}
 \mathcal{L}_q(i\omega_n)\psi^{\prime\prime}\left ( \mathcal{X}_{q,i|\omega_n|}\right )
 ,
 \end{gather}
\begin{gather}
\left \langle O_{\rm int}^{(1),1} O_{\rm int}^{(1),2} \right \rangle \to
\delta \Gamma_{t,{\rm int}}^{(1),1;(1),2} = \frac{4}{g}  \Gamma_{t} \int_{q,\omega_n}
 \Bigl [2   Z_\omega\mathcal{C}_q^2(i\omega_n) \mathcal{L}_q(i\omega_n)
 - \sum_{j=0}^3  \Gamma_j \mathcal{D}_q(i\omega_n)\mathcal{D}^{(j)}_q(i\omega_n) + 4 \Gamma_t \mathcal{D}^{t2}_q(i\omega_n)\Bigr ] .
 \end{gather}
 Here $\mathcal{X}_{q,i|\omega_n|} = (Dq^2+|\omega_n|+\Lambda^\prime)/(4\pi T) +1/2$.In total, we obtain
\begin{gather}
\Gamma_t(\Lambda^\prime) = \Gamma_t(\Lambda) -\frac{2}{g} (\Gamma_s-\Gamma_t) \int_q \mathcal{D}_q(0) + \frac{8}{g}  \Gamma_t^2 \int_{q,\omega_n} \mathcal{D}^2_q(i\omega_n)
+ \frac{4}{g}  Z_\omega  \int_q \mathcal{C}_q(0) \mathcal{L}_q(0)
\notag \\
+ \frac{16}{g} \Gamma_{t}  Z_\omega \int_{q,\omega_n}  \mathcal{L}_q(i\omega_n) \Biggl [
 \mathcal{C}_q^2(i\omega_n)
 -
\frac{\Gamma_{t}}{2z} \left (\frac{D}{4\pi T}\right )^2\psi^{\prime\prime}\left ( \mathcal{X}_{q,i|\omega_n|}\right )\Biggr ]
  .
\label{eq:app:dGt}
\end{gather}

Finally, the corrections to the Cooper channel interaction amplitude $\Gamma_c$ are given as
\begin{gather}
\langle O_{\rm int}^{(2),1} \rangle \to \delta \Gamma_{c, {\rm int}}^{(2),1}
= \frac{4}{g} \Gamma_{c} \int_{q,\omega_n}
\Bigl [2   Z_\omega\mathcal{C}_q^2(i\omega_n) \mathcal{L}_q(i\omega_n)   + \sum_{j=0}^3  \Gamma_j \mathcal{D}_q(i\omega_n)\mathcal{D}^{(j)}_q(i\omega_n)\Bigr ] ,
\end{gather}
\begin{align}
\left \langle O_{\rm int}^{(2),2} + \frac{1}{2} \left [ O_{\rm int}^{(1),2} \right ]^2 \right \rangle \to
\delta \Gamma_{c,{\rm int}}^{(2),2} =& -\frac{2}{g} (\Gamma_s-3\Gamma_t)  \int_q \mathcal{D}_q(0)
\notag \\
 &-\frac{2}{g} \Gamma_s  Z_\omega \int_{q,\omega_n}
\mathcal{D}^{-1}_q(i\omega_n)\mathcal{D}^{(s)}_q(i\omega_n) \mathcal{L}_q(i\omega_n)
\Biggl [
\mathcal{D}^{2}_q(i\omega_n)+\mathcal{C}^{2}_q(i\omega_n)\Biggr ],
\end{align}
\begin{gather}
\frac{1}{2}\left \langle \left [O_{\rm int}^{(1),1}\right ]^2  \right \rangle \to
\delta \Gamma_{c,{\rm int}}^{(1),1;(1),1} = - \Gamma_c \mathcal{L}_{q=0}(i\omega_n=0) \int_{\omega_m} \mathcal{C}_{q=0}(i\omega_{m})
 ,
 \end{gather}
\begin{gather}
\left \langle O_{\rm int}^{(1),1} O_{\rm int}^{(1),2} \right \rangle \to
\delta \Gamma_{c,{\rm int}}^{(1),1;(1),2} = \frac{4}{g} \Gamma_{c}
\Biggl [-\mathcal{L}_{q=0}(i\omega_n=0) \int_{\omega_m} \mathcal{C}_{q=0}(i\omega_{m}) \Biggr ]
  \int_{q,\omega_n}
\Bigl [2  Z_\omega \mathcal{C}_q^2(i\omega_n) \mathcal{L}_q(i\omega_n)   \notag \\
+ \sum_{j=0}^3  \Gamma_j \mathcal{D}_q(i\omega_n)\mathcal{D}^{(j)}_q(i\omega_n)\Bigr ] .
 \end{gather}
 Here we introduce
 \begin{equation}
 \int_{\omega_n} \equiv \frac{2\pi T }{D}\sum_{n>0}\Theta\bigl (\Lambda-Dq^2-|\omega_n| \bigr )
 \Theta\bigl (Dq^2+|\omega_n|-\Lambda^\prime \bigr )  .
 \end{equation}
In total, we obtain
\begin{gather}
\Gamma_c(\Lambda^\prime) = \Gamma_c(\Lambda) + \Gamma_c \Biggl [-\mathcal{L}_{q=0}(i\omega_n=0) \int_{\omega_m} \mathcal{C}_{q=0}(i\omega_{m}) \Biggr ] \frac{4}{g} \int_{q,\omega_n}
\Bigl [2  Z_\omega \mathcal{C}_q^2(i\omega_n) \mathcal{L}_q(i\omega_n)   + \sum_{j=0}^3  \Gamma_j \mathcal{D}_q(i\omega_n)\mathcal{D}^{(j)}_q(i\omega_n)\Bigr ]
\notag \\
-\frac{2}{g} (\Gamma_s-3\Gamma_t)  \int_q \mathcal{D}_q(0)-\frac{2}{g} \Gamma_s Z_\omega \int_{q,\omega_n}
\mathcal{D}^{-1}_q(i\omega_n)\mathcal{D}^{(s)}_q(i\omega_n) \mathcal{L}_q(i\omega_n)  \Biggl [
\mathcal{D}^{2}_q(i\omega_n)+\mathcal{C}^{2}_q(i\omega_n)\Biggr ] .
\label{eq:app:dGc}
\end{gather}

We emphasize that Eqs. \eqref{eq:app:dz} and \eqref{eq:app:dGs} implies that
\begin{gather}
Z_\omega(\Lambda^\prime)+\Gamma_s(\Lambda^\prime) = Z_\omega(\Lambda)+ \Gamma_s(\Lambda) +\frac{4}{g} Z_\omega \int_{q,\omega_n}\mathcal{C}_q(i\omega_n)  \mathcal{L}^2_q(i\omega_n)\Bigl [
\psi^\prime(\mathcal{X}_{q,i|\omega_n|}) -\frac{4\pi T}{D} \mathcal{C}_q(i\omega_n)
\Bigr ] = Z_\omega(\Lambda)+ \Gamma_s(\Lambda) .
\end{gather}
Here we employ the following relation: $\partial_n \mathcal{L}_q(i\omega_n) = [D/(4\pi T)]\mathcal{L}^2_q(i\omega_n) \psi^\prime(\mathcal{X}_{q,i|\omega_n|})$. Also since $ \Lambda^\prime \gg 4\pi T$ we use that
$\psi^\prime \left ( \mathcal{X}_{q,i|\omega_n|}\right ) \approx ({4\pi T}/{D}) \mathcal{C}_q(i\omega_n)$.

The one-loop renormalization of the parameters of the NLSM action obtained from the background field procedure can be summarized as follows:
\begin{gather}
g(\Lambda^\prime)  = g(\Lambda) - 4 \int_q \mathcal{C}_q(0) + 4 \int_{q,\omega_n} q^2 \Biggl [
2 \mathcal{C}^{3}_q(i\omega_n)\mathcal{L}_q(i\omega_n) +
\sum_{j=0}^3 \gamma_j \mathcal{D}^{2}_q(i\omega_n)\mathcal{D}^{(j)}_q(i\omega_n)
\Biggr ] ,
\label{eq:app:dg-f}
\end{gather}
\begin{gather}
Z_\omega(\Lambda^\prime)  = Z_\omega(\Lambda)+ \frac{2}{g}  (\Gamma_s+3\Gamma_t) \int_{q,\omega_n} \mathcal{D}^2_q(i\omega)  + \frac{4 Z_\omega}{g}
\int_{q,\omega_n}  \mathcal{C}^2_q(i\omega_n) \Bigl [
  \mathcal{L}^2_{q}(i\omega_n)   + \mathcal{L}_{q}(i\omega_n)  \Bigr ] ,
  \label{eq:app:dz-f}
 \end{gather}
\begin{gather}
\Gamma_s(\Lambda^\prime)  = \Gamma_s(\Lambda)- \frac{2}{g}  (\Gamma_s+3\Gamma_t) \int_{q,\omega_n} \mathcal{D}^2_q(i\omega)  - \frac{4 Z_\omega}{g}
\int_{q,\omega_n}  \mathcal{C}^2_q(i\omega_n) \Bigl [
  \mathcal{L}^2_{q}(i\omega_n)   + \mathcal{L}_{q}(i\omega_n)  \Bigr ] ,
  \label{eq:app:dGs-f}
 \end{gather}
\begin{gather}
\Gamma_t(\Lambda^\prime)  = \Gamma_t(\Lambda) -\frac{2}{g} \left (\Gamma_s-\Gamma_t-  4 \frac{\Gamma_t^2}{Z_\omega}\right ) \int_{q,\omega_n}\mathcal{D}^2_q(i\omega_n)
 + \frac{4}{g} \left (Z_\omega + 4 \Gamma_t + \frac{2\Gamma_t^2}{Z_\omega} \right ) \int_{q,\omega_n} \mathcal{C}^2_q(i\omega+n) \mathcal{L}_q(i\omega_n)  \notag \\
 -  \frac{4}{g D}  \int_{q,\omega_n} \mathcal{C}^2_q(i\omega_n) \mathcal{L}^2_q(i\omega_n) ,
\label{eq:app:dGt-f}
\end{gather}
and
\begin{gather}
\Gamma_c(\Lambda^\prime)  =  \Gamma_c(\Lambda) - \frac{\Gamma_c^2}{Z_\omega} \int_{\omega_n} \mathcal{C}_{q=0}(i\omega_n)  -\frac{2}{g} (\Gamma_s-3\Gamma_t)  \int_{q,\omega_n} \mathcal{D}^2_q(i\omega_n)
+ \frac{8 \Gamma_c}{g} \int_{q,\omega_n} \mathcal{C}^2_q(i\omega_n) \mathcal{L}_q(i\omega_n)\notag \\
 +\frac{12 \Gamma_t \Gamma_c}{g Z_\omega}  \int_{q,\omega_n} \mathcal{D}_q(i\omega_n) \mathcal{D}^t_q(i\omega_n)
.
\label{eq:app:dGc-f}
\end{gather}
\end{widetext}
To derive RG equations from Eqs. \eqref{eq:app:dg-f} - \eqref{eq:app:dGc-f} we choose $\Lambda^\prime = \Lambda + d\Lambda$. Then, provided $d\Lambda/\Lambda$ corresponds to $- 2 dL/L$ we obtain RG equations \eqref{eq:rg:final:t} - \eqref{eq:rg:final:z}.

\section{Enhancement of $T_c$ in the case of weak short ranged interaction \label{AppTcEnh}}

\subsection{Orthogonal symmetry class}

Expanding RG Eqs. \eqref{eq:rg:final:t:G} - \eqref{eq:rg:final:gc:G} with $n=3$ to the lowest order in interactions
$\gamma_{s,t,c}$ we find
\begin{equation}
\frac{d t}{d y} =  t^2, \qquad
\frac{d}{dy} \begin{pmatrix} \gamma_s \\ \gamma_t \\ \gamma_c
\end{pmatrix} = -\frac{t}{2} \mathcal{R}_o
\begin{pmatrix} \gamma_s \\ \gamma_t \\ \gamma_c
\end{pmatrix} -
\begin{pmatrix} 0 \\ 0 \\ 2\gamma_c^2
\end{pmatrix}  .
\label{e2.6}
\end{equation}
Here within our accuracy we neglect the interaction corrections to $t$ in comparison with weak-localization correction. The matrix
\begin{equation}
\mathcal{R}_o=\left( \begin{array}{ccc}
1 & 3 & 2 \\
1 & -1 & -2 \\
1 & -3 & 0\\
\end{array} \right)
\end{equation}
has the following eigenvalues and eigenvectors:
\begin{equation}
\label{e2.8}
\lambda=-4\ : \ \  \begin{pmatrix} -1 \\ 1 \\ 1
\end{pmatrix}
\ ;\quad
\lambda^\prime=2\ :
\begin{pmatrix} 1 \\ 1 \\ -1
\end{pmatrix}
\ {\rm and} \
\begin{pmatrix} 1  \\ -1 \\ 2
\end{pmatrix} .
\end{equation}
If the $\gamma_c^2$ term is neglected, the solution of the linear
system (\ref{e2.6}) approaches the eigenvector with $\lambda=-4$,
i.e., interaction parameters tends to the BCS line $-\gamma_s = \gamma_t = \gamma_c$.
Let us expand the vector formed by $\gamma_{s,t,c}$ in eigenvectors
(\ref{e2.8}):
\begin{equation}
\label{e2.9}
\begin{pmatrix} \gamma_s \\ \gamma_t \\ \gamma_c
\end{pmatrix} =
\begin{pmatrix} -1 & 1 & 1 \\
                           1 & 1 & -1 \\
                           1 & -1 & 2
\end{pmatrix}
\begin{pmatrix} a \\ b \\ c \end{pmatrix} .
\end{equation}
Transforming the set of equations \eqref{e2.6} to the new variables
$a,b$, and $c$, we get
\begin{align}
\frac{dt}{dy} &= t^2 , \notag \\
\frac{da}{dy} &= 2 t a - \frac{2}{3} (a-b+2c)^2  , \notag \\
\frac{db}{dy} &= - t b ,
\notag \\
\frac{dc}{dy} &= - t c - \frac{2}{3} (a-b+2c)^2 .
\label{e2.13}
\end{align}
Equations \eqref{e2.13} are supplemented by the following initial conditions: $t(0)=t_0$, $a(0)=a_0$, $b(0)=b_0$, and $c(0)=c_0$ where
\begin{equation}
\begin{pmatrix}
 a_0 \\ b_0 \\ c_0
\end{pmatrix} =
\begin{pmatrix}
-{1/6} & {1/2} & {1/3} \\
                           {1/2} &{1/2}  & 0 \\
                           {1/3} & 0 & {1/3}
\end{pmatrix}
\begin{pmatrix}
 \gamma_{s,0} \\ \gamma_{t,0} \\ \gamma_{c,0}
\end{pmatrix} .
\end{equation}
From Eqs. \eqref{e2.13} it is easy to find
\begin{equation}
t(y) = \frac{t_0}{1 -t_0y} , \qquad b(y) = b_0(1 -t_0y) \equiv b_0 \frac{t_0}{t} .
\label{e2.15}
\end{equation}
Since $b$ decreases upon RG flow, it is not important and we neglect it in
the future analysis.

Equations for the remaining two variables, $a$ and $c$ are coupled. If
the quadratic term is neglected, then $a$ increases and $c$
decreases. This suggests that $c$ can be neglected. This is confirmed
by a more careful analysis which shows that, although on the very last
interval of RG ``time'' $y$ the variable $c$ starts to increase and
becomes of the same order as $a$ (i.e. of order unity), this
weakly affect the RG scale at which this happens (i.e. the temperature of the
superconducting transition). Thus, we neglect $c$ in
what follows.

We can now easily solve the remaining equation for $a$. We assume the
starting value $a_0$ to be negative (which means that there is
attraction in the Cooper channel that is supposed to lead to the
superconductivity), $a = - |a|$. This is in particular the case when
$\gamma_{c,0}$ is the dominant coupling and $\gamma_{c,0}<0$. Then the
equation reads
\begin{equation}
\label{e2.16}
\frac{d|a|}{dy}  = 2t|a| + \frac{2}{3} a^2 \,.
\end{equation}
Solving this equation, we obtain
\begin{equation}
\label{e2.18}
a(y) = - \left( \frac{t_0^2}{|a_0|t^2} +  \frac{2 t_0}{3 t^2}
- \frac{2}{3t}\right)^{-1} \, .
\end{equation}

Let us analyze the result obtained. Let us first assume that
$|a_0| \ll t_0$. Then the second term in brackets in the right hand side of
(\ref{e2.18}) is small compared to the first one and can be neglected,
\begin{equation}
\label{e2.19}
a^{-1}(y) = - \frac{1}{t} \left( \frac{t_0^2}{|a_0|t} -
  \frac{2}{3}\right) \,.
\end{equation}
With increasing RG scale $y$ the
resistance $t$
increases
together with the
interaction $a$.
If $t$ reaches first unity, we get an
insulator; if $a \sim 1$ happens first, we get a superconductor. It is
easy to see that the second possibility (superconductivity) is
realized if  $|a_0| \gg t_0^2$.  Then at the point of divergence of $a$ we have a resistance
$t_* \simeq 3t_0^2/(2 |a_0|) \ll 1$. This occurs at $y_* \simeq \frac{1}{t_0} - \frac{1}{t_*}$, i.e. we can estimate the temperature of superconducting transition as $T_c \sim \exp(-2y_*)$ yielding
\begin{equation}
\label{e2.22}
T_c \sim \frac{1}{\tau} \exp\left\{- \frac{2}{t_0}\left [1-\frac{t_0}{t_*}\right ]
\right\} \, .
\end{equation}
Here the factor $2$ in the exponent originates from a translation
of the length scale into energy (temperature).
Under the above assumption  $|a_0| \ll t_0$ the second term in square
brackets in the exponential of \eqref{e2.22} is just a small correction to the first
one.

The transition temperature~\eqref{e2.22} is much
higher than the BCS temperature $T_c^{\rm BCS} = (1/\tau)
e^{-1/|\gamma_{c,0}|}$, so that the superconductivity is strongly enhanced by
disorder. The origin of the enhanced superconductivity is in the
increase of $|a|$ governed by the eigenvalue $\lambda =-4$ of matrix $\mathcal{R}_o$ which
yields the eigenvalue $-(t/2) \times (-4) = 2t$ of the linear
part of the system in Eqs. \eqref{e2.6}. This is nothing but the anomalous
multifractal exponent $-\Delta_2$ for this symmetry class [\onlinecite{Wegner}]. (We have in
mind the ``weak multifractality'' in 2D.) Therefore, the (multi)fractality is the source of the enhancement of the
superconductivity.

By solving \eqref{e2.13} with $b=0$ and $a$ given by Eq.~\eqref{e2.19},
one finds that although $|c|$ decreases initially, eventually with
increasing RG scale towards $y_*$ it becomes of the order of unity: $|c(y_*)|\sim 1$.
Therefore, to determine precise value of $t_*$ one has to solve coupled equations
for $a$ and $c$ [with $(b=0)$].

If $|a_0| \ll t_0^2$, the resistance reaches unity before the
interaction becomes strong, and the system is an insulator. Finally,
if $|a_0| \gg t_0$, the disorder is not particularly important, and
the transition temperature is given by usual clean BCS $T_c^{\rm BCS}$. In the latter case neglecting $b$ and $c$ is not parametrically justified and leads to an incorrect numerical factor in the exponent.

For the initial values of interaction parameters used in the inset of Fig. \ref{fig:RGO:PD2:SIT} we find that $a_0 = -0.01$. Thus, we expect the regime with $T_c \approx T_c^{BCS}$ for $t_0 \ll |a_0| = 0.01$. In the range of $t_0$ between $|a_0|=0.01$ and $\sqrt{|a_0|} = 0.1$ the transition occurs at $T_c\gg T_c^{BCS}$. At $t_0 \sim \sqrt{|a_0|} = 0.1$ the superconductor-insulator transition is expected. The above crude analysis of RG equations linearized in interactions is in good agreement with numerical solutions of full RG equations \eqref{eq:rg:final:t:G} - \eqref{eq:rg:final:gc:G}.

\subsection{Symplectic symmetry class}

Now we consider the symplectic symmetry class, i.e. assume that the spin
symmetry is completely broken. Expanding Eqs. \eqref{eq:rg:final:t:G} - \eqref{eq:rg:final:gc:G} with $n=0$ to the lowest order in interactions $\gamma_{s,t,c}$ we obtain
\begin{equation}
\label{e3.5}
\frac{d t}{d y} = - \frac{t^2}{2}, \quad
\frac{d}{dy} \begin{pmatrix} \gamma_s \\  \gamma_c
\end{pmatrix} =
 -\frac{t}{2} \mathcal{R}_s
\begin{pmatrix} \gamma_s \\ \gamma_c
\end{pmatrix} -
\begin{pmatrix} 0 \\ 2\gamma_c^2
\end{pmatrix} .
\end{equation}
Here we neglect interaction corrections to $t$ in comparison with weak anti localization.
The matrix
\begin{equation}
\mathcal{R}_s = \begin{pmatrix}
1 &  2 \\
1 & 0\\
\end{pmatrix}
\end{equation}
has the following eigenvalues and eigenvectors:
\begin{equation}
\label{e3.6}
\lambda=-1\, : \quad  \begin{pmatrix} -1 \\ 1
\end{pmatrix}
\ ;\qquad
\lambda^\prime=2\, :
\quad \begin{pmatrix} 2 \\ 1
\end{pmatrix} .
\end{equation}
If the $\gamma_c^2$ term is neglected, the solution of the linear
system in Eqs. \eqref{e3.5} approaches the eigenvector with $\lambda=-1$,
i.e., interaction amplitudes approach the BCS (in the absence of $\gamma_t$) line $\gamma_s = - \gamma_c$.
As in the orthogonal case, we can expand the vector formed by $\gamma_{s,c}$ in the eigenvectors
\begin{equation}
\label{e3.7}
\begin{pmatrix}
 \gamma_s \\ \gamma_c
\end{pmatrix} =
\begin{pmatrix} -1 & 2 \\
                          1 & 1
\end{pmatrix}
\begin{pmatrix} a \\ c \end{pmatrix} .
\end{equation}
Transforming the set of equations \eqref{e3.5} to the new variables $a$ and $c$, we find
\begin{align}
\frac{dt}{dy} = & - \frac{t^2}{2} , \notag\\
\frac{da}{dy}  =& \frac{t}{2} a - \frac{4}{ 3} (a+c)^2, \notag \\
\frac{dc}{dy}  =& - t c - \frac{2}{ 3} (a+c)^2 .
\label{e3.8b}
 \end{align}
Equations \eqref{e3.8b} are supplemented by the initial conditions: $t(0)=t_0$, $a(0)=a_0$ and $c(0)=c_0$, where
\begin{equation}
\label{e3.8}
\begin{pmatrix} a_0 \\ c_0
\end{pmatrix} =
\begin{pmatrix} - {1/3} &  {2/3}  \\
                           {1/3} &   {1/3}
\end{pmatrix}
\begin{pmatrix} \gamma_{s,0} \\ \gamma_{c,0}
\end{pmatrix} .
\end{equation}

From the first of Eqs. \eqref{e3.8b} we find
 \begin{equation}
\label{e3.4}
t(y) = \frac{t_0}{1+ y t_0/2} .
\end{equation}
Equations for two variables, $a$ and $c$ are coupled. If the quadratic term is neglected, then $a$ increases
and $c$ decreases. At the later stage of RG the quadratic terms leads to enhancement of $c$.
This suggests that $c$ can be neglected for qualitative analysis of RG equations \eqref{e3.8b}.
We thus neglect $c$ and keep only $a$ (fully analogously to what we have done in the
orthogonal case). The resulting equation for $a$ reads
\begin{equation}
\label{e3.9}
\frac{d a}{dy}  = \frac{t}{ 2} a - \frac{4}{ 3} a^2 \,.
\end{equation}
We solve this equation with the result
\begin{equation}
\label{e3.10}
a = \frac{1}{t} \left ( \frac{1}{a_0 t_0} + \frac{4}{ 3 t^2} - \frac{4}{3 t_0^2}\right )^{-1} \,.
\end{equation}
Provided $a_0<0$, the superconducting instability is possible.  The new, different from standard clean BCS behavior emerges under the condition $|a_0| \ll t_0$. Then the condition $a \sim 1$ yields
\begin{equation}
\label{e3.11}
t_* \simeq 2 \left(|a_0| t_0/3\right)^{1/2} \ll t_0  .
\end{equation}
By solving Eq.~\eqref{e3.8b} with $a$ given by Eq.~\eqref{e3.10}, we
find that although $|c|$ decreases initially, eventually with increasing RG scale it reaches $a$: $c\sim a$ at $t=t_*$. Therefore, to determine precise value of $t_*$ one has to solve coupled equations for $a$ and $c$.

Equation~\eqref{e3.11} yields the following estimate for the transition temperature
\begin{equation}
\label{e3.12}
T_c \sim \frac{1}{\tau} e^{-2y_*}\sim e^{-4/t_*} \sim \exp\left ( - \frac{\mathcal{C}}{\sqrt{|a_0| t_0}} \right ) \gg T_c^{\rm BCS} .
\end{equation}
The constant $\mathcal{C}$ is of the order of unity and depends on ratio $c_0/a_0$.
The clean BCS result is restored (up to small corrections) at  $|a_0| \gg t_0$. For $a_0 < 0$ (or, equivalently, $\gamma_{c,0}<\gamma_{s,0}/2$) there is no transition to the supermetallic phase with increase of $t_0$.

As in the orthogonal case, the source of the enhancement of the
superconducting temperature is in the first term on the right hand side of
Eq.~\eqref{e3.9}. The eigenvalue $t/2$ is  the anomalous
multifractal exponent $-\Delta_2$ for the symplectic symmetry
class. Therefore, also in this case
the (multi)fractality is the source of the enhancement of the
superconductivity. This enhancement is less efficient than in the
orthogonal case for two reasons, because of antilocalizing behavior
that leads to decrease of $t$ and therefore weakening of
multifractality.


\begin{thebibliography}{100}
\bibitem{footnote-T0}
There is a subtle difference between the scaling with temperature and with the system size.
At zero $T$, RG equations \eqref{eq:rg:final:t:G} - \eqref{eq:rg:final:gc:G} describe the behavior of the system
upon changing the system size $L$.
The divergence of $\gamma_c$ indicates establishing of the absolute value of
superconducting order parameter on the length scale $L_c$.
We note that both the total electrical resistance $\rho(L_c)$ and the NLSM coupling $t(L_c)$ which, in general,
are different from each other due to the contribution of inelastic processes
(see discussion in Sec. \ref{s5_0}), are finite.
At scales $L>L_c$, the length-scale dependence of the (four-contact) resistance
of a square sample is governed by the contribution of non-trivial configurations of
the order parameter [\onlinecite{future-Elio}]
(similar to the ``quantum phase slips'' in a quasi-one-dimensional systems).
The resistance becomes zero only in the thermodynamic limit, $L=\infty$.
\bibitem{footnoteA} In addition to fixed points that describe phases and transitions between them
there is a non-trivial fixed point at $\gamma^{*}_c \approx -0.3$, $\gamma^{*}_t \approx 2.5$, $\gamma_s^*=-1$ and $t^{*} \approx 0.03$
(stable with respect to the $\gamma_s$ direction). Since $t^*$ is a parameterless number, the one-loop approximation is, strictly speaking, not sufficient for determining the couplings at this fixed point. However, the numerical smallness of the obtained value of $t^*$ suggests that the one-loop RG is still reasonably well controlled near this fixed point. This fixed point is unstable in two directions and is located on the critical surface separating the superconducting and ferromagnetic phases. It resembles the Nishimori point in 2D random-bond Ising model [\onlinecite{Nishimori}]. The analysis of RG Eqs. \eqref{eq:rg:final:t:Coulomb} - \eqref{eq:rg:final:gc:Coulomb} in the vicinity of this fixed point suggests that it separates two fixed points on the critical surface  which control the transitions between SC and FM phases. We expect that one such fixed point is situated in the strong coupling region
whereas the other corresponds to
a certain
clean limit (cf. discussion in Ref. [\onlinecite{KB}]). We will investigate this interesting possibility elsewhere.
\bibitem{footnote-TITS}
Two-dimensional systems with strong spin-orbit interaction can be topologically protected from localization,
when they appear as surface states of 3D topological insulators and superconductors [\onlinecite{TI-reviews}].
In terms of the NLSM, such metallic states are described by an action with an additional topological term,
see, e.g., Refs.~[\onlinecite{elio13}]. The combined effect of disorder, Coulomb interaction, and topology
in such system may lead to the emergence of a novel critical point [\onlinecite{OGM10}] with $t\sim 1$, while no insulating phase
appears. In this work, we do not consider 2D surfaces of 3D topological insulators and superconductors and relegate the discussion
of the full phase diagrams for topologically protected systems (including graphene with superconducting islands [\onlinecite{Bouchiat}]) to future work.
\end{thebibliography}
\end{document}